\documentclass[a4paper,11pt]{article}
\pdfoutput=1 

\usepackage{jheppub} 

\usepackage[T1]{fontenc} 

\newcommand{\be}{\begin{equation}}
\newcommand{\ee}{\end{equation}}
\newcommand{\ba}{\begin{array}}
\newcommand{\ea}{\end{array}}
\newcommand{\bea}{\begin{eqnarray}}
\newcommand{\eea}{\end{eqnarray}}

\usepackage{natbib,bm}
\usepackage{enumerate}
\usepackage{graphicx}
\usepackage[usenames,dvipsnames]{color}
\usepackage{xcolor} 
\usepackage{amsmath}
\usepackage{amsfonts}
\usepackage{amssymb}
\usepackage{bbm}
\usepackage{latexsym}
\usepackage[english]{babel}
\usepackage{multirow}
\usepackage{float}
\usepackage{url}
\usepackage{slashed}
\usepackage{breakurl}
\usepackage{mathrsfs}

\usepackage{ulem,fancyvrb}
\usepackage{xcolor}

\title{Probing sub-GeV leptophilic dark matter at Belle II and NA64}

\author[a]{Jinhan Liang,}
\author[a,b]{Zuowei Liu}
\author[a]{and Lan Yang}

\affiliation[a]{Department of Physics, Nanjing University, \\Nanjing 210093, China}
\affiliation[b]{CAS Center for Excellence in Particle Physics, \\Beijing 100049, China}

\emailAdd{jinhanliang@smail.nju.edu.cn}
\emailAdd{zuoweiliu@nju.edu.cn}
\emailAdd{lanyang@smail.nju.edu.cn}

\abstract{

An analysis is given 
of the Belle II sensitivities and NA64 constraints on 
the sub-GeV Dirac dark matter that interacts with charged leptons. 
We consider two different types of interactions between sub-GeV Dirac dark matter and  the charged leptons: 
the EFT  operators  and the light  vector  mediators. 
We compute the Belle II mono-photon sensitivities 
on sub-GeV dark matter
with 50 ab$^{-1}$ data which are expected to be accumulated in the full Belle II runs. 
Although 
the Belle II mono-photon sensitivities 
on the EFT operators 
are of similar size as the LEP constraints, 
Belle II can probe new parameter space of the light 
vector mediator models that are unexplored by LEP. 
For both the EFT operators and the light vector mediator models, 
the Belle II mono-photon 
sensitivities can be several orders 
of magnitude stronger than the current dark 
matter direct detection limits, 
as well as the white dwarf limits. 
The light vector mediator  
can also be directly searched for 
by reconstructing the invariant mass of its 
di-lepton decay final states at Belle II, 
which is found to be complementary to the mono-photon channel.
We compute the NA64 constraints on the sub-GeV 
Dirac dark matter  
and provide analytic expressions of the dark matter 
cross section in  
the Weizs\"acker-Williams approximation, 
for the EFT operators, 
and for the light vector mediator models. 
We find that the current NA64 data 
(with $2.84 \times 10^{11}$ electron-on-target events)
provide strong  
constraints on sub-GeV dark matter. 
Although the NA64 constraints are found to be about one  
order of magnitude  smaller than the Belle II 
sensitivities for the EFT operators, 
NA64 can probe some regions of the parameter space in  
the light vector mediator models that are beyond 
the reach of Belle II. 
We also find that Belle II and NA64 
can probe the 
canonical dark matter annihilation cross section 
in thermal freeze-out in a significant portion of the parameter space of the models considered.
}

\arxivnumber{2111.15533}

\begin{document} 
\maketitle
\flushbottom

\section{Introduction}
\label{sec:intro}

Although dark matter (DM) 
makes up a quarter of the total energy density of the universe, 
its particle property remains unknown today 
{\cite{Bertone:2004pz,Feng:2010gw}. 
During the past decades, a great amount of 
theoretical and experimental efforts have been put 
into searches for the weakly interacting massive 
particles (WIMPs), 
 which have constrained 
the DM-nucleus cross section to 
an unprecedented level 
\cite{Roszkowski:2017nbc, Schumann:2019eaa}.  
Recently, dark matter direct detection (DMDD) 
experiments have also started to provide 
compelling limits on sub-GeV dark matter 
particles. 
For sub-GeV dark matter, electronic signals become important 
in DMDD experiments. 
Scattered by DM, electrons in the target can be either ionized 
or excited. 
The DMDD experiments with an ionization signal include 
XENON10 \cite{Essig:2017kqs}, 
XENON100 \cite{Essig:2017kqs}, 
XENON1T \cite{Aprile:2019xxb}, 
DarkSide-50 \cite{DarkSide:2018ppu}, 
and PandaX \cite{PandaX-II:2021nsg}; 
the experiments with an excitation signal 
include 
SENSEI \cite{Barak:2020fql}, 
DAMIC \cite{Aguilar-Arevalo:2019wdi}, 
EDELWEISS \cite{Arnaud:2020svb}, 
and SuperCDMS~\cite{Agnese:2018col}. 
The excitation signal can have a lower energy threshold 
than the ionization signal, leading to a better sensitivity 
for lighter dark matter. 
Currently, the xenon target experiments 
and SENSEI provide the leading DMDD constraints 
to sub-GeV DM.  
Astrophysical processes can also give competitive 
constraints to sub-GeV DM, for example, 
heating constraints in white dwarfs due to DM 
\cite{Bell:2021fye, Bertone:2007ae, 
McCullough:2010ai, Hooper:2010es, 
Amaro-Seoane:2015uny, Panotopoulos:2020kuo}. 
Furthermore, interactions between sub-GeV DM 
and cosmic rays 
\cite{Ema:2018bih,Dent:2020syp,Cao:2020bwd}, 
and Sun \cite{An:2017ojc, Emken:2021lgc}
can significantly alter the velocity of the DM particle, 
and thus enhance the sensitivity of the DMDD experiments.

In this paper, we  study the Belle II sensitivities and the NA64 constraints 
on the 
sub-GeV dark matter that interacts with charged leptons. 
Belle II is operated at SuperKEKB
which collides 7 GeV electrons with 4 GeV positrons \cite{Kou:2018nap}. 
In the 8-year data taking, 
Belle II is expected to accumulate 50 $\rm ab^{-1}$ data \cite{Kou:2018nap},  
which is much more than  
other low energy electron-positron colliders, 
such as BaBar and BESIII. 
Moreover, the calorimeter of Belle II is much more hermetic 
with non-projective barrel crystals, 
which makes it an ideal detector for DM searches  
\cite{Kou:2018nap, BaBar:2001yhh}.

Electron collider constraints on DM have been studied  
previously, including Belle II \cite{Kou:2018nap,Liang:2019zkb,Duerr:2019dmv,Duerr:2020muu,Kang:2021oes,Essig:2013vha,Izaguirre:2015zva,Filimonova:2019tuy,Izaguirre:2015zva,Izaguirre:2015yja,Boehm:2020wbt},
LEP \cite{Ellis:2001hv, Freitas:2014jla, Primulando:2020rdk,Richard:2014vfa,Fox:2011fx}, 
and other electron colliders
\cite{Essig:2009nc, Anastasi:2015qla, BaBar:2017tiz, BaBar:2008aby, Liu:2018jdi, Habermehl:2020njb, Chae:2012bq,Dev:2021jrg, Kalinowski:2021tyr, Barman:2021hhg, Profumo:2009tb, Liu:2019ogn, Xiang:2017yfs, Birkedal:2004xn, Yu:2014ula, Hochberg:2017khi, Liu:2017lpo, Alikhanov:2017cpy, Borodatchenkova:2005ct,Graham:2021ggy, Zhang:2019wnz}.
In this paper, we study the capability of the Belle II experiment 
in probing the parameter space of the sub-GeV dark matter models, 
including both the 
effective field theory (EFT) operators and the  
light vector mediator models. 
To our knowledge, Belle II constraints on various EFT operators 
between DM and charged leptons have not been thoroughly studied 
in the literature. 
Certain light mediator models, e.g., the dark photon model 
has been studied in Ref.\ \cite{Kou:2018nap}. 
Here we consider a more general light mediator model 
in which the light mediator has both vector 
and axial-vector couplings to fermions in the hidden sector 
and in the SM sector. 
Thus we carry out detailed Belle II analyses both for the 
EFT operators and for the light mediator models 
with different mass relations and different couplings. 
We compute the mono-photon constraints 
on the EFT operators and 
on the light mediator models, 
and further compare the limits 
to the DMDD constraints.  
We find that the Belle II mono-photon limits 
can be much stronger 
than current DMDD constraints, 
and can also constrain the proposed DM 
models to interpret the recent excess events in Xenon1T 
electron recoil data \cite{XENON:2020rca}. 
For the light mediator models, 
we further compute the Belle II limits 
due to 
the visible decay final states of the 
mediator, 
and find that the visible channel can be complementary 
to the mono-photon channel.

NA64 is an electron fixed target experiment operated at 
CERN with the incident electron energy of $\sim 100$  GeV 
and a lead target. 
NA64 has collected $2.84 \times 10^{11}$ electron-on-target (EOT) data
in the year 2016, 2017, and 2018 \cite{Banerjee:2019pds}. 
The DM signature at NA64 is a significant missing energy \cite{Banerjee:2019pds}.
NA64 constraints on DM 
have been analyzed recently, including 
DM with a dark photon mediator \cite{Banerjee:2019pds, Gninenko:2017yus}, 
millicharged DM \cite{Gninenko:2018ter}, 
DM with EM form factors \cite{Chu:2018qrm}, 
and pseudo-Dirac dark matter \cite{Berlin:2020uwy}. 

In this work, we carry out a systematic study on NA64 constraints 
for a number of DM models (EFT operators and 
the light vector mediator models), which, 
to our knowledge, has not been done in the literature. 
We also provide analytic expressions of the differential cross sections 
for various models 
in the Weizs\"acker-Williams approximation (WWA)  
\cite{Williams:1935dka,vonWeizsacker:1934nji}. 
We find that NA64 and Belle II can be complementary in probing 
sub-GeV DM models.

The rest of the paper is organized as follows. 
In Sec.\ \ref{sec:modsig}, 
we introduce two different types of dark matter models: 
fermionic DM interating with SM via EFT operators and 
via the light vector mediator models. 
We discuss both the signal events and the 
SM background events in the mono-photon channel 
for the Belle II analysis
in Sec.\ \ref{sec:monoBgSig}. 
We compute the Belle II mono-photon sensitivities  
for the EFT operators and 
for the light vector mediator models 
in Sec.\ \ref{sec:monoEFT} and Sec.\ \ref{sec:monoZ} 
respectively,  
and further compare them to the DMDD limits. 
We analyze the Belle II di-muon limits on  
the light vector mediator models in Sec.\ \ref{sec:visiZ}. 
We compute the NA64 constraints 
on the EFT operators 
and on the light vector mediator models 
in Sec.\ \ref{sec:NA64}. 
The analytic expressions of the DM cross sections in the WWA 
at the NA64 experiment 
are given in Appendix \ref{sec:dxsec}. 
We compute the Belle II sensitivities on the 
dark matter annihilation cross section
in Sec.\ \ref{sec:DMrelic}.
We summarize our findings in Sec.\ \ref{sec:summary}.

\section{Dark Matter Models and the mono-photon signal}
\label{sec:modsig}

In this paper, we consider two {different types} 
of DM models: 
(1) {fermionic} DM interacts with charged leptons via EFT 
operators; 
(2) {fermionic} DM interacts with charged leptons via a light 
{vector} mediator. 
There are a variety of EFT operators
between the SM and dark matter. 
Here we consider the fermionic dark matter 
that has four-fermion EFT interaction with charged leptons as follows \cite{Fox:2011fx, Chae:2012bq}
\bea
\cal{L}&=&\frac{ 1}{ \Lambda_V^2} O_V  \equiv \frac{1}{\Lambda_{V}^{2}} \bar{\chi} \gamma_{\mu} \chi \bar{\ell} \gamma^{\mu} \ell, 
\label{eq:EFT1} \\ 
\cal{L}&=&\frac{ 1}{ \Lambda_A^2} O_A \equiv  \frac{1}{\Lambda_{A}^{2}} \bar{\chi} \gamma_{\mu} \gamma_{5} 
\chi \bar{\ell} \gamma^{\mu} \gamma_{5} \ell, 
\label{eq:EFT2}\\
\cal{L}&=&\frac{ 1}{ \Lambda_s^2} O_s \equiv  \frac{1}{\Lambda_{s}^{2}} \bar{\chi} \chi \bar{\ell} \ell, 
\label{eq:EFT3}\\
\cal{L}&=&\frac{ 1}{ \Lambda_t^2} O_t \equiv \frac{1}{\Lambda_{t}^{2}} \bar{\chi} \ell \bar{\ell} \chi,
\label{eq:EFT4}
\eea
where $\chi$ is the Dirac DM, 
$\ell$ is the SM charged lepton, 
and $\Lambda$ is the new physics scale. 
{The first three EFT operators can be obtained by integrating 
out an $s$-channel mediator in a UV complete model; 
the last EFT operator can be obtained by integrating out 
a $t$-channel mediator \cite{Fox:2011fx}. 
Thus, we use $\Lambda_{V}$ 
($\Lambda_{A}$) to denote the vector (axial-vector) case; 
for the two scalar operators we use $\Lambda_{s}$ and $\Lambda_{t}$
to refer to the possible UV-completions. 
\footnote{For simplicity, 
we have assumed universal couplings for different lepton flavors.}}
{The production cross section of 
$e^+ e^- \to \bar \chi \chi \gamma$
at electron colliders
for the above 
four EFT operators 
are computed} in
Ref.\ \cite{Chae:2012bq}; 
we collect these cross section 
formulas 
in Appendix \ref{sec:xsecEFT}.

We consider a  
light mediator model 
in which the light mediator is a spin one particle 
with couplings to both hidden sector dark matter and 
charged leptons; 
the interaction 
Lagrangian is given by 
\begin{equation}
\mathcal{L}=Z_{\mu}^{\prime} \bar{\chi} \gamma^{\mu}
\left(g_{v}^{\chi}-g_{a}^{\chi}\gamma_{5}\right)
\chi+Z_{\mu}^{\prime} \bar{\ell} \gamma^{\mu}
\left(g_{v}^{\ell}-g_{a}^{\ell} \gamma_{5}\right) \ell,
\label{eq:Z_model}
\end{equation}
where $Z'$ denotes the light mediator, 
$\chi$ is the dark matter, 
$\ell$ is the SM charged lepton, 
$g_{v}^{\chi,\ell}$ ($g_{a}^{\chi,\ell}$) 
is the vector (axial-vector) coupling.
The mono-photon cross section 
at the electron colliders for the process $e^+e^-\to\gamma 
 Z^\prime\to\gamma\chi\bar{\chi}$ is given by \cite{Liu:2019ogn}
\bea
\frac{d \sigma}{d E_{\gamma} d z_{\gamma}} & = &
\frac{\alpha
s_{\gamma}^{2} 
\left[\left(g_{v}^{\ell}\right)^{2}
+\left(g_{a}^{\ell}\right)^{2}\right]
}{6 \pi^{2} s E_{\gamma}
\left[\left(s_{\gamma}-m_{Z^{\prime}}^{2}\right)^{2}+m_{Z^{\prime}}^{2}
\Gamma_{Z^{\prime}}^{2}\right] }
\sqrt{1-4\frac{m_{\chi}^2}{s_\gamma}}
\left[1+\frac{E^2_\gamma}{s_\gamma}
\left(1+z_{\gamma}^{2}\right)\right]\frac{1}{{1-z_{\gamma}^{2}}}
\nonumber \\
& \times & 
\left[\left(g_{v}^{\chi}\right)^{2}\left(1+2 \frac{m_{\chi}^2}{s_\gamma}\right)+\left(g_{a}^{\chi}\right)^{2}\left(1-4 \frac{m_{\chi}^2}{s_\gamma}\right)\right],
\label{eq:monoZp}
\eea
where $E_\gamma$ and $\theta_\gamma$ are the 
photon energy and polar angle respectively 
in the center of mass frame, 
$s$ is the square of the center of mass energy, 
$z_\gamma=\cos\theta_\gamma$, 
$s_\gamma=s-2\sqrt{s}E_\gamma$, 
and $m_{Z^\prime}$ and $\Gamma_{Z^\prime}$
are the mass and the total decay width of the 
$Z'$ boson. 
The $Z'$ total decay width is given by 
\begin{equation}
\Gamma_{Z^{\prime}}=\Gamma\left(Z^{\prime} \rightarrow \chi \bar{\chi}\right)+\sum_{\ell} \Gamma\left(Z^{\prime} \rightarrow \ell \bar{\ell}\right),
\label{eq: Zbosondecaywidth}
\end{equation}
where $\Gamma\left(Z^{\prime} \rightarrow \chi \bar{\chi}\right)$ 
is the invisible decay width with DM in the final state, 
and $\Gamma\left(Z^{\prime} \rightarrow \ell \bar{\ell}\right)$ is the 
decay width with SM particles in the final state. 
The invisible decay width is given by 
\begin{equation}
\Gamma\left(Z^{\prime} \rightarrow \chi \bar{\chi}\right)=
\frac{m_{Z^{\prime}}}{12 \pi} \sqrt{1-4 \frac{m_{\chi}^{2}}{m_{Z^{\prime}}^{2}}}
\left[\left(g_{v}^{\chi}\right)^{2}\left(1+2 \frac{m_{\chi}^{2}}{m_{Z^{\prime}}^{2}}\right)
+\left(g_{a}^{\chi}\right)^{2}\left(1-4 \frac{m_{\chi}^{2}}{m_{Z^{\prime}}^{2}}\right)\right],
\label{eq: ZtoDMwidth}
\end{equation}
$\Gamma\left(Z^{\prime} \rightarrow \ell \bar{\ell}\right)$ can be computed 
similarly by substituting the couplings and mass for lepton.

\section{Mono-photon searches at Belle II}
\label{sec:monoBgSig}
In this section,
we use the mono-photon final state, 
$e^+ e^- \to \chi \bar{\chi} \gamma$, to 
probe the DM models at Belle II.
For each of the DM models,
the number of signal events
is calculated by  the analytic expressions of the differential cross sections
offered in section \ref{sec:modsig}.
In our analysis, we consider 
both the reducible background 
and the irreducible background for the 
mono-photon process. 

The mono-photon irreducible background is due to   
the $e^+e^-\to
\gamma\nu\bar{\nu}$ process in the SM; 
the differential cross 
section of {the} $e^+ e^- \rightarrow 
\gamma \nu \bar{\nu}$ {process in the SM} is given by 
\cite{Ma:1978zm, Gaemers:1978fe, Liu:2018jdi}  
\be
\frac{d \sigma_{\nu\bar{\nu}\gamma}}{d E_{\gamma} d z_{\gamma}}=\frac{\alpha G_{F}^{2} s_{\gamma}^{2}}{4 \pi^{2} s E_{\gamma}\left(1-z_{\gamma}^{2}\right)} 
\left[ 8 s_{W}^{4}-\frac{4}{3} s_{W}^{2} +1 \right] 
\left[1+\frac{E_{\gamma}^{2}}{s_{\gamma}}\left(1+z_{\gamma}^{2}\right)\right],
\label{eq:nuBG}
\ee
where $G_F$ is the Fermi constant, 
$s_W \equiv \sin \theta_{W}$ with $\theta_W$ being
the weak mixing angle.

Photons at Belle II are detected in the ECL and KLM sub-detectors, 
both of which consist of three segments: 
the forward detector, 
the backward detector, 
and the barrel detector \cite{Kou:2018nap}. 
The mono-photon 
reducible backgrounds at the 
Belle II detector come from 
the SM processes in which one 
or more SM final state particles 
are not detected by the detector.
The main reducible background in our analysis
is due to the 
$e^+ e^- \to \gamma \slashed{\gamma} \slashed{\gamma}$ 
process,\footnote{We use ``slash'' to denote 
a particle that is not detected by the detector.} 
where two of the final state photons are not detected 
because one photon escapes in the beam direction and 
the other escapes in the region where the 
detector has no coverage or a very low 
detection efficiency, 
for example, the 
gaps between different segments of the 
ECL and KLM sub-detectors, 
and the gap located
at $90^\circ$ of the ECL barrel \cite{Kou:2018nap}.
\footnote{The reducible BG due to 
$e^+ e^- \to \gamma \slashed{\ell}^+ \slashed{\ell}^-$ 
is subdominant, 
because charged leptons can  
either {be} detected by tracking detectors if emitted in the central region \cite{Kou:2018nap}, 
or {be} effectively removed by kinematic conditions if emitted along the beam directions \cite{Liang:2019zkb}.}

The reducible BG {at the Belle II detector} 
has been analyzed by 
Ref.\ \cite{Kou:2018nap}. 
For the sub-GeV DM particles, 
we adopt the low-mass 
region given in Ref.\ \cite{Kou:2018nap} 
as the signal region in our analysis;
recently a fitting function for the boundary of
this region is given in Ref.\ \cite{Duerr:2019dmv} 
\bea
\label{eq:SR1}
\theta_{\mathrm{min}}^{\mathrm{low}} &=& 5.399^{\circ} E_{\mathrm{CMS}}(\gamma)^{2} / \mathrm{GeV}^{2}-58.82^{\circ} E_{\mathrm{CMS}}(\gamma) / \mathrm{GeV}+195.71^{\circ}, \\
\theta_{\mathrm{max}}^{\mathrm{low}}  &=& -7.982^{\circ} E_{\mathrm{CMS}}(\gamma)^{2} / \mathrm{GeV}^{2}+87.77^{\circ} E_{\mathrm{CMS}}(\gamma) / \mathrm{GeV}-120.6^{\circ}, 
\label{eq:SR}
\eea 
where $\theta_{\rm min}^{\rm low}$ 
and $\theta_{\rm max}^{\rm low}$ 
are the minimum and maximum 
angles for the photon in the 
lab frame, namely 
$\theta_{\min }^{\text {low }}<
\theta_\gamma^{\rm lab}<
\theta_{\max }^{\text {low }}$.\footnote{We use
``lab'' to denote the variable in the lab frame.} 
In the signal region, 
about 300 mono-photon events from the 
reducible backgrounds are 
expected with 20 $\rm fb^{-1}$ data \cite{Kou:2018nap}, 
corresponding to 
$\sim$$7.5\times 10^5$ mono-photon events 
with 50 ab$^{-1}$ data; 
there are about $1.9\times 10^3$ mono-photon events 
from the irreducible background process 
$e^+ e^- 
\to \nu \bar{\nu} \gamma$ with  50 ab$^{-1}$ data.

\section{Mono-photon constraints on EFT operators} 
\label{sec:monoEFT}

We compute the Belle II 
90\% C.L.\ limits on the EFT operators, 
by using the criterion ${N_s / \sqrt{N_b}} =\sqrt{2.71}$, 
where $N_s$ ($N_b$) is the number of signal 
(background) events in the signal region. 
Fig.\ (\ref{fig:monoEFT}) shows the Belle II 
90\% C.L.\ lower bounds 
on the new physics scale $\Lambda$ 
of the EFT operators, 
from the mono-photon channel 
with $50$ ab$^{-1}$ integrated luminosity. 
As shown in the left panel figure of Fig.\ (\ref{fig:monoEFT}), 
the Belle II 90\% C.L.\ lower bounds are 
$\sim${280} GeV, for $\Lambda_{V}$, 
$\Lambda_{A}$, and $\Lambda_{s}$,  
and are about $\sim$220 GeV for $\Lambda_{t}$. 
We further compare the Belle II limits 
to the LEP limits  
analyzed by Ref.\ \cite{Fox:2011fx}. 
Mono-photon data with 650 pb$^{-1}$ at various $\sqrt{s}$ 
from 180 GeV to 209 GeV have been collected by 
the DELPHI detector at LEP 
\cite{DELPHI:2003dlq, Fox:2011fx}.
The LEP mono-photon data 
are binned in 19 
$x_\gamma= E_\gamma/ E_{\rm beam}$ bins \cite{Fox:2011fx}, 
where $E_\gamma$ and $E_{\rm beam}$ 
are the energy of photon and the beam energy respectively.
The LEP 
90\% C.L.\ lower limits on EFT operator with sub-GeV mass,
are about 480 GeV for
$\Lambda_{V}$ and $\Lambda_{A}$,
440 GeV for $\Lambda_{s}$, 
and 340 GeV for $\Lambda_{t}$ \cite{Fox:2011fx}.

Although the expected integrated luminosity of 
Belle II is about five orders of magnitude larger 
than LEP, their limits on the EFT operators 
turn out to be of similar size. This is largely 
due to the fact that EFT operators 
and the SM processes depend on $\sqrt{s}$ 
in different ways.  
For the four-fermion EFT operators, the cross section is 
proportional to $s$ (to compensate the $\Lambda^4$ factor in the 
denominator), whereas for the QED process (responsible for 
the reducible background at Belle II), the cross 
section is inversely proportional to $s$. 
For that reason, the dominant reducible background 
at Belle II becomes totally negligible at LEP, 
whereas the cross sections of EFT operators at LEP 
are enhanced by a factor of $\sim$400 as compared to Belle II. 
The weak processes that lead to the irreducible mono-photon backgrounds 
have a similar proportionality on $s$ as the EFT operators 
up to the $Z/W$ mass scale. 
Taking these effects into consideration, we find that 
LEP is expected to have similar constraints 
on the four-fermion EFT operators as Belle II.

\begin{figure}[htbp]
\begin{centering}
\includegraphics[width=0.49 \textwidth]{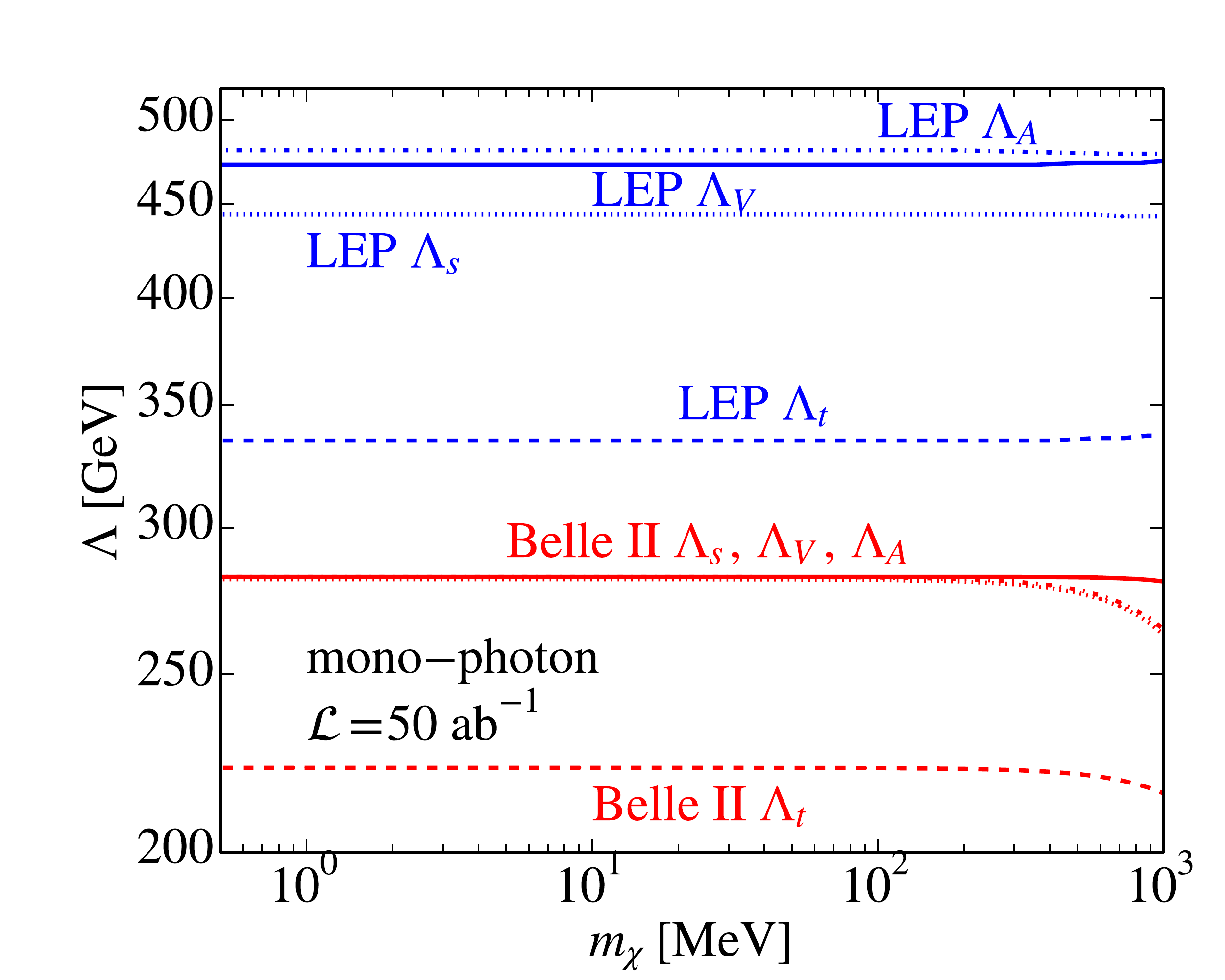}
\includegraphics[width=0.49 \textwidth]{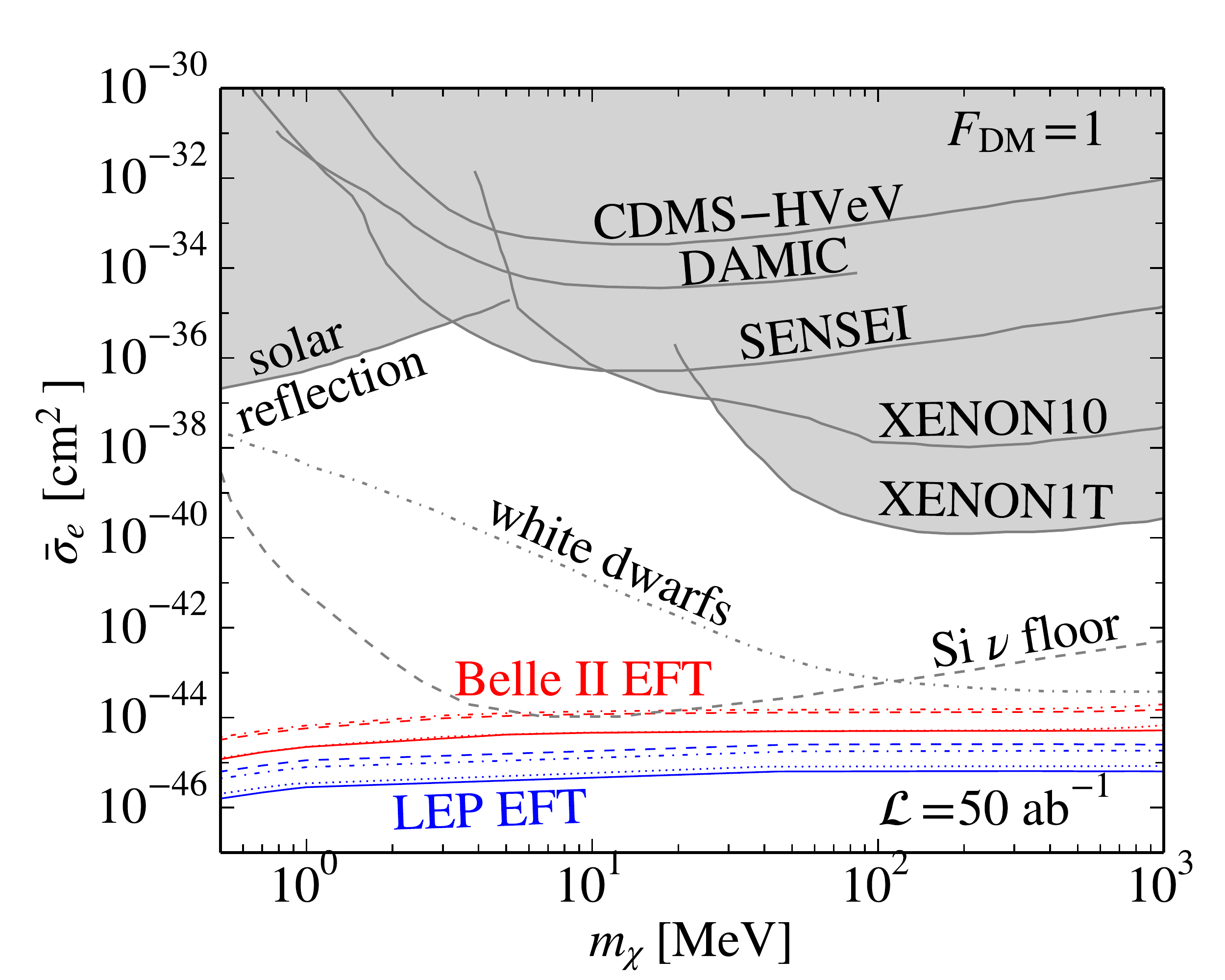}
\caption{Left panel: Belle II {90\%} C.L.\ 
lower bounds (red curves) on the new physics scale 
$\Lambda_A$ (dot-dashed), $\Lambda_V$ (solid), 
$\Lambda_s$ (dotted), and  $\Lambda_t$ (dashed), 
in the EFT operators as a function of the DM mass, 
via the mono-photon 
channel with the integrated luminosity 
 $50$ ab$^{-1}$.    
The LEP constraints (blue curves) \cite{Fox:2011fx} 
with the luminosity of $650 ~\rm pb^{-1}$
are also shown.
Right panel: 
Belle II and LEP constraints on the DM-electron cross section 
at the reference momentum $q=\alpha m_e$ for 
the EFT operators, 
as well as constraints from DMDD experiments with $F_{\rm DM}(q)=1$, 
including   
XENON10 \cite{Essig:2017kqs}, 
CDMS-HVeV \cite{SuperCDMS:2018mne}, 
XENON1T \cite{Aprile:2019xxb}, 
DAMIC \cite{DAMIC:2019dcn}, 
and SENSEI \cite{SENSEI:2020dpa}. 
Constraints for boosted DM from solar 
reflection \cite{An:2017ojc} and 
constraints on dark matter capture in white dwarfs 
are also shown. 
We compute the white dwarf limit 
at the reference momentum $q=\alpha m_e$ 
by using the lower bound on $\Lambda$ given 
in Ref.\ \cite{Bell:2021fye}.
The  dashed gray line shows the neutrino floor limit 
for silicon target detectors where the exposure 
is 1000 kg-year \cite{Essig:2018tss}.}
\label{fig:monoEFT}
\end{centering}
\end{figure}

We further compare the collider constraints 
on the EFT operators to other experimental constraints. 
We compute the DM-electron scattering cross section 
at the momentum transfer 
$q \equiv |{\bf q}|=\alpha m_{e}$ \cite{Essig:2011nj, Essig:2015cda}, 
by using the limit on $\Lambda$ in the EFT operators 
\be 
\label{eq:trans}
\bar{\sigma}_{e} \equiv \frac{\mu_{\chi e}^{2}}{16 \pi m_{\chi}^{2} m_{e}^{2}} 
 \left. \overline{\left|\mathcal{M}_{\chi e}(q)\right|^{2}}\right|_{q = \alpha m_{e}}, 
\ee 
where $m_\chi$ is the DM mass, 
$m_e$ is the electron mass, 
$\mu_{\chi e}$ is the reduced mass. 
Here both DM and electron are assumed to {be}
non-relativistic, 
and the dependence on $q$ 
is solely in the matrix element ${\cal M}_{\chi e}$, 
which can be factorized as 
$\overline{\left|\mathcal{M}_{\chi e}(q)\right|^{2}}=
\overline{\left|\mathcal{M}_{\chi e}(\alpha m_{e})\right|^{2}} 
\left|F_{\mathrm{DM}}(q)\right|^{2}$.
We have 
$\overline{\left|\mathcal{M}_{\chi e}(\alpha m_{e}) \right|^{2}} 
\simeq 16 m_e^2 m_\chi^2/\Lambda^4$ for all the four EFT 
operators except $\Lambda_A$ which 
is 3 times larger. 
The form factor for the EFT operators 
considered in our analysis 
is found to be $F_{\rm DM}(q) \simeq 1$.
\footnote{See appendix \ref{app:ME} for the expressions  
of $\overline{\left|\mathcal{M}_{\chi e}(q)\right|^{2}}$ for the EFT operators
and also the range of $F_{\rm DM}(q)$ for the momentum of interest.} 
For the EFT operators the collider 
constraints from Belle II and LEP on sub-GeV DM  
are found to be much stronger than the DMDD limits, 
including the constraints 
from 
SENSEI \cite{SENSEI:2020dpa}, 
CDMS-HVeV \cite{SuperCDMS:2018mne}, 
DAMIC \cite{DAMIC:2019dcn}, 
XENON10 \cite{Essig:2017kqs}, 
XENON1T \cite{Aprile:2019xxb}, 
{DMDD limits via solar reflection} \cite{An:2017ojc}, 
and white dwarfs 
\cite{Bell:2021fye, Bertone:2007ae, 
McCullough:2010ai, Hooper:2010es, 
Amaro-Seoane:2015uny, Panotopoulos:2020kuo}. 
The constraint on $\bar \sigma_e$ from white dwarfs,
as shown in Fig.\ (\ref{fig:monoEFT}),
is computed via Eq.\ (\ref{eq:trans}), by
using of the lower bound on $\Lambda_V \simeq 200$ GeV 
in Ref.\ \cite{Bell:2021fye}.\footnote{Our 
white dwarf constraint on $\bar \sigma_e$
is different from Ref.\ \cite{Bell:2021fye}
where the cross section is evaluated at the
momentum/energy scale relevant for
DM captures in white dwarfs
\cite{Sandra:Robles}. 
The white dwarfs limits are 
$\Lambda_V \simeq \Lambda_A \simeq 200 ~\rm GeV$ 
and $\Lambda_s \simeq 200 ~\rm MeV$
when $m_\chi > 100 ~\rm MeV$ as given in 
Ref.\ \cite{Bell:2021fye}.}
Signals due to neutrino-target scatterings  
are the irreducible background in DMDD, 
which are often referred to as the 
neutrino floor. 
The gray dashed line in Fig.\ (\ref{fig:monoEFT}) 
shows the neutrino floor for Si detectors 
with a 1000 kg-year exposure 
\cite{Essig:2018tss}.\footnote{The 
neutrino floors for Xe and Ge targets are 
higher than Si.} 
Thus, it is remarkable that LEP and Belle II 
can probe the parameter space beyond the neutrino floor,
especially in the sub-MeV mass region, 
as shown in Fig.\ (\ref{fig:monoEFT}). 
We note that the DMDD limits are 
the same for all the four EFT operators, 
since $F_{\rm DM}(q)=1$ is used, 
but the collider limits are slightly different 
for the four EFT operators.
Thus the electron collider constraints, from Belle II and LEP, 
can further extend to the sub-MeV DM region 
where many of the current direct detection experiments 
lose sensitivity due to the low recoil energy.

\section{Mono-photon constraints on light mediator model} 
\label{sec:monoZ}

We investigate the capability of the Belle II detector 
in probing the light mediator models in which 
the light mediator $Z'$ couples to both DM 
and charged leptons. 
Unlike the four-fermion EFT operators, 
the collider cross section in the light-mediator models 
is not proportional to $s$. 
For that reason, the Belle II is expected to 
explore some new parameter space in the 
light-mediator models that has not been 
probed by the LEP experiment.

In this analysis, we are interested in the 
$Z'$ mass below the Belle II $\sqrt{s} \simeq 10$ GeV.  
Thus we consider three $Z'$ masses in the MeV-GeV mass range:
10  MeV,  0.6  GeV, and 5 GeV. 
We note that for ultralight mediators, 
constraints from  
cosmic microwave background (CMB)
and baryon acoustic oscillations (BAO) 
are usually much more stringent than 
collider searches \cite{Buen-Abad:2021mvc}.

\begin{figure}[htbp]
\begin{centering}
\includegraphics[width=0.49 \textwidth]{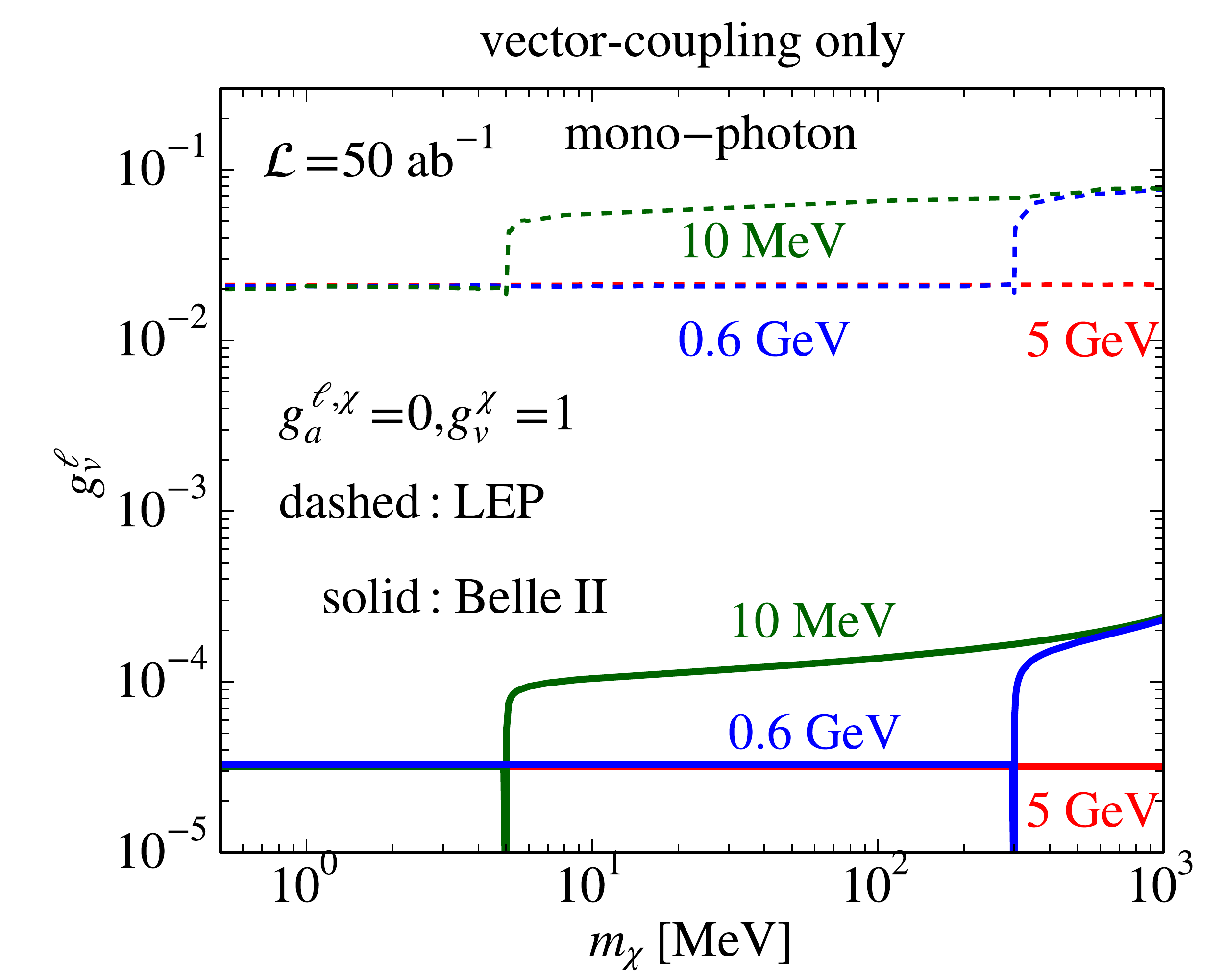}
\includegraphics[width=0.49 \textwidth]{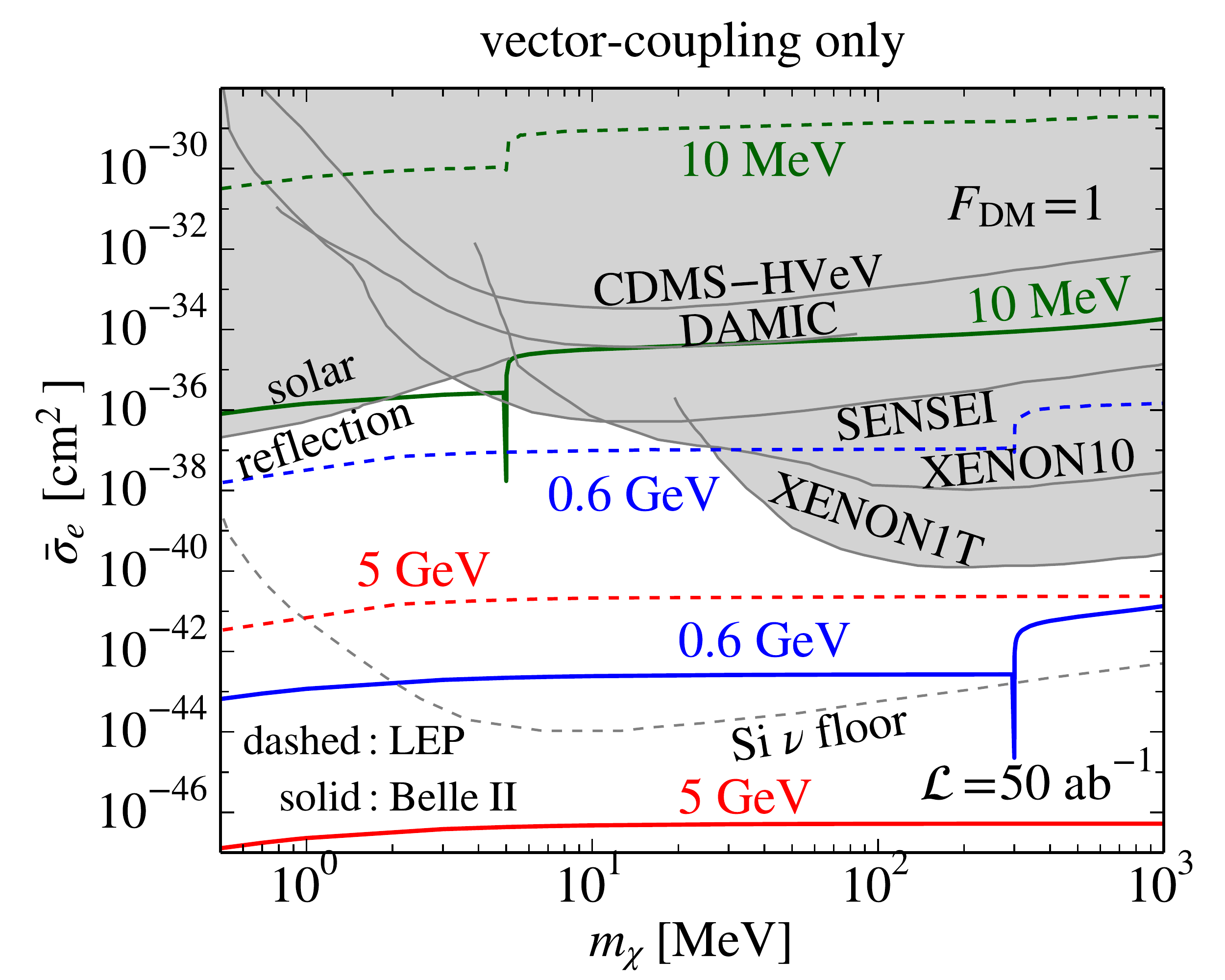}
\caption{The expected Belle II {90\%} upper bound (solid lines)
with 50 $\rm ab^{-1}$ integrated luminosity 
on $g_v^\ell$ (left panel) and $\bar\sigma_e$ (right panel)  
in light mediator models 
where only vector couplings are assumed. 
Three $Z'$ masses 
are considered for the Belle II analysis: 
$m_{Z^\prime}=5$ GeV (red), $m_{Z^\prime}=0.6$ 
GeV (blue), and $m_{Z^\prime}=10$ MeV (green). 
The LEP constraints (dashed lines) 
with 650  $\rm pb^{-1}$ integrated luminosity 
are also shown for two $Z'$ masses: 
$m_{Z^\prime}=5$ GeV (red), and $m_{Z^\prime}=0.6$ 
GeV (blue), and $m_{Z^\prime}=10$ MeV (green). 
Constraints from DMDD experiments with $F_{\rm DM}(q)=1$ 
(valid for $m_{Z'} \gg \alpha m_e$) 
are also shown:  
SENSEI \cite{SENSEI:2020dpa}, 
XENON10 \cite{Essig:2017kqs} 
and XENON1T \cite{Aprile:2019xxb}, 
CDMS-HVeV \cite{SuperCDMS:2018mne}, 
DAMIC \cite{DAMIC:2019dcn}, 
and solar reflection \cite{An:2017ojc}. 
The gray dashed line shows the neutrino floor limit for 
silicon detectors \cite{Essig:2018tss}.}
\label{fig:monoZV}
\end{centering}
\end{figure}

\begin{figure}[htbp]
\begin{centering}
\includegraphics[width=0.49 \textwidth]{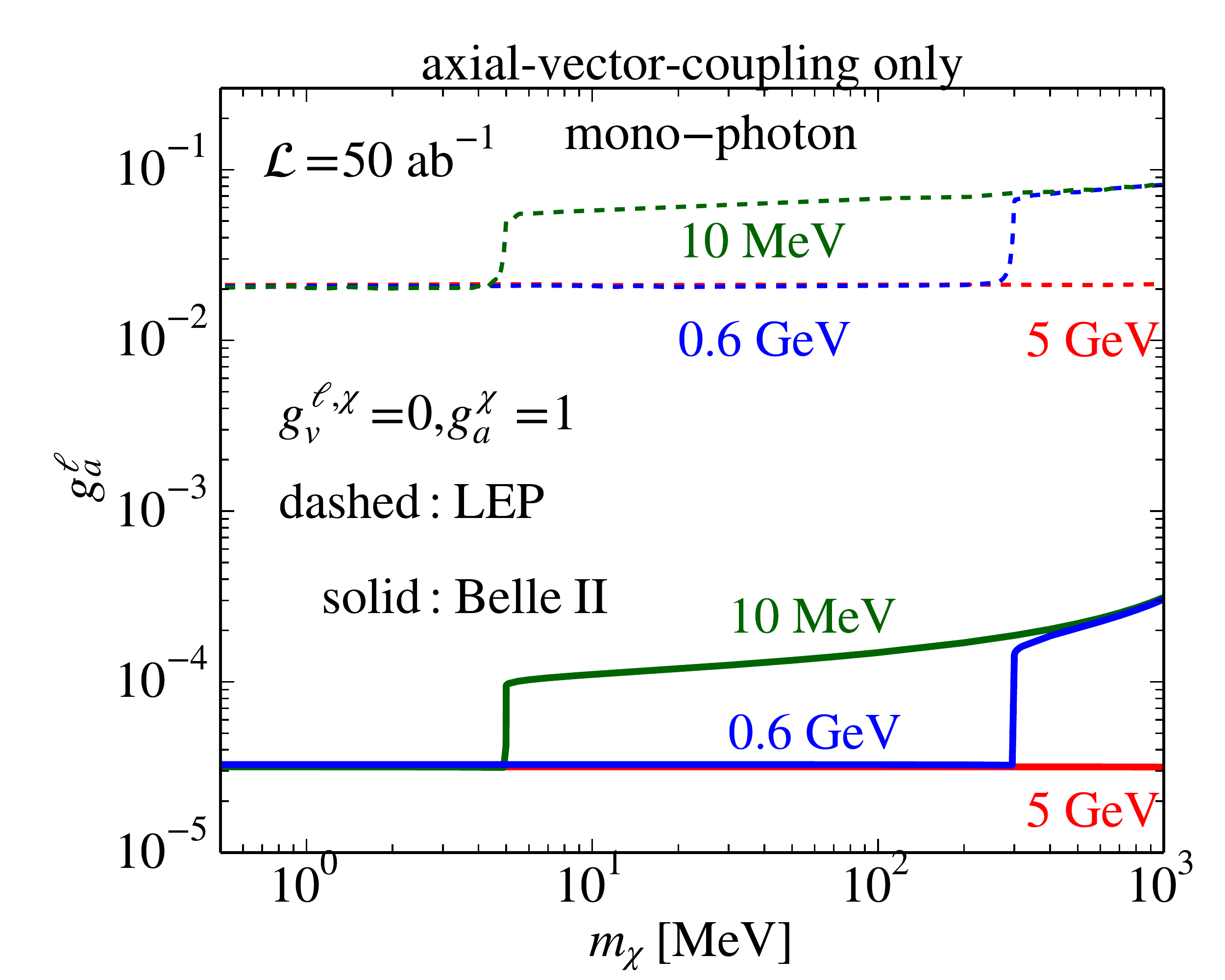}
\includegraphics[width=0.49 \textwidth]{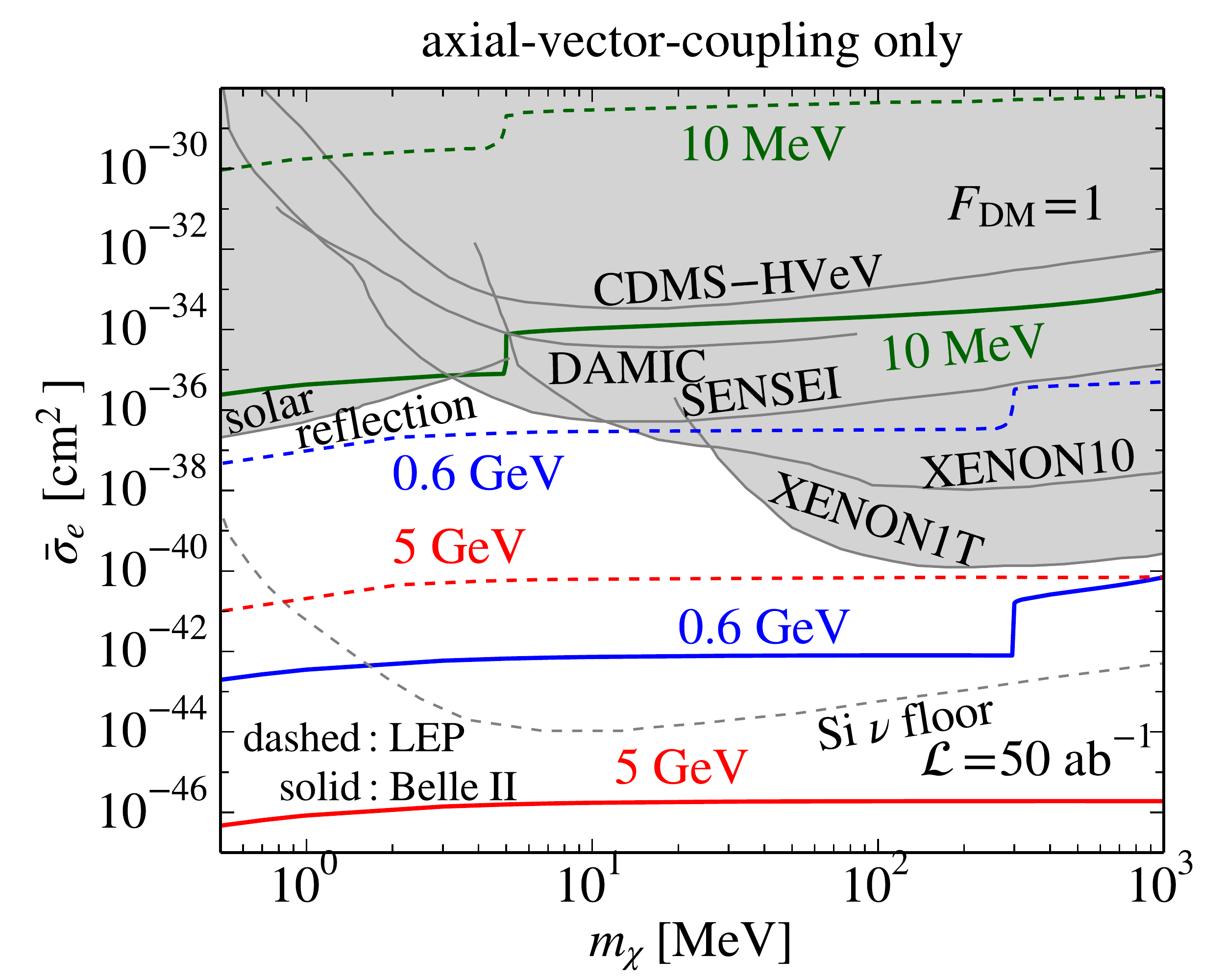}
\caption{Same as Fig.\ (\ref{fig:monoZV}) but for 
light mediator models 
where only axial-vector couplings are assumed.} 
\label{fig:monoZA}
\end{centering}
\end{figure}

We compute the Belle II 
{90\%} C.L.\ limits on the light-mediator models 
using the same criterion as the EFT operators, 
namely by setting ${N_s / \sqrt{N_b}} =\sqrt{2.71}$, 
where $N_s$ is obtained by integrating 
Eq.\ (\ref{eq:monoZp}) in the signal region. 
The expected Belle II 90\% C.L.\ upper bounds 
with $50$ ab$^{-1}$ data 
on the gauge coupling are shown on left panel figure of 
Fig.\ (\ref{fig:monoZV}) and Fig.\ (\ref{fig:monoZA}), 
where we only consider vector couplings 
and axial-vector couplings respectively. 
The collider signals depend  strongly on the 
mass relations between the light mediator 
and DM. 
There are two categories: 

\begin{itemize}

\item $m_{Z'} > 2 m_\chi$. 
The $Z'$ boson mainly decays into dark matter. 
Thus the $Z'$ boson can be produced on-shell in 
the $e^+ e^- \to \chi \bar{\chi} \gamma$  process 
and is exhibited as a resonance in the mono-photon 
energy spectrum 
(see e.g.\ \cite{BaBar:2008aby,BaBar:2017tiz,Hochberg:2017khi,Liu:2019ogn}).
The mono-photon cross section 
can be approximated by 
$\sigma_{\chi \bar{\chi} \gamma} \simeq 
\sigma_{\gamma Z'} \times {\rm BR} (Z' \to \chi \bar{\chi})$. 
Because the branching ratio ${\rm BR} (Z' \to \chi \bar{\chi}) \simeq 1$ 
in the parameter space of interest in this analysis, and 
the cross section $\sigma_{\gamma Z'}$ is proportional to $(g^\ell)^2$, 
the mono-photon cross section depends on $g^\ell$, but 
not on $m_\chi$ or $g^\chi$. 
The $m_\chi$ independence can be seen 
in Fig.\ (\ref{fig:monoZV}) and Fig.\ (\ref{fig:monoZA}) 
in the mass range $m_\chi < m_{Z'}/2$.

\item $m_{Z'} < 2 m_\chi$.
The $Z'$ boson can only decay into the SM particles. 
Thus the $Z'$ boson is produced off-shell in the 
$e^+ e^- \to \chi \bar{\chi} \gamma$ process 
without a resonance in the mono-photon energy spectrum. 
In this case, $\sigma_{\chi \bar{\chi} \gamma}$ is proportional to 
$(g_\ell g_\chi)^2$ and depends on $m_{\chi}$, 
which can seen in Fig.\ (\ref{fig:monoZV}) and Fig.\ (\ref{fig:monoZA}).

\end{itemize}

For both the vector-only case and 
the axial-vector-only case, 
the Belle II upper limits are about $g^\ell \sim 3 \times 10^{-5}$
when $m_{Z'} > 2 m_\chi$,  
as shown in Fig.\ (\ref{fig:monoZV}) and Fig.\ (\ref{fig:monoZA}). 
For the vector-only case,
the sensitivity is highly enhanced 
if $m_{Z'} = 2 m_\chi$. 
We further compare the Belle II limits to 
the LEP limits. 
LEP contraints on
light mediator models 
with $m_{Z'} \geqslant 10$ GeV 
have been studied in Ref.\ \cite{Fox:2011fx}. 
Here we analyze the LEP constraints 
to the region where $m_{Z'} < 10$ GeV, 
following the analysis of Ref.\ \cite{Fox:2011fx}; 
the details of our LEP analysis are given in section \ref{sec:LEP}. 
As shown in Fig.\ (\ref{fig:monoZV}) and Fig.\ (\ref{fig:monoZA}),
the LEP constraints are about $3 \times 10^{-2}$
for both vector and axial-vector couplings when $m_{Z'} > 2 m_\chi$,
which are three orders of magnitude 
weaker than Belle II. 
Unlike the EFT operators, 
there are resonance signals in the 
mono-photon energy spectrum,
which correspond to the Breit-Wigner 
resonance of the $Z'$ boson, 
in the light mediator models. 
One could select the events near the resonance 
to further improve the significance of the searches. 
We have not taken advantage of this, because such 
a study requires the detailed knowledge of the 
subdetectors to simulate the reducible background, 
which, however, is beyond the scope of this study.

Similar to the analysis for the EFT operators, 
we compare the collider limits on the light mediator models 
to DMDD limits on 
the right panel figures of Figs.\ (\ref{fig:monoZV}) and (\ref{fig:monoZA}).
In the non-relativistic limit, 
the amplitude for the light mediator models is given by 
\be
\overline{\left|\mathcal{M}_{\chi e}\right|^{2}}
\simeq 16 m_e^2 m_\chi^2  {  (g_v^\ell g_v^\chi)^2 + 3( g_a^\ell g_a^\chi)^2 \over  ( m_{Z'}^2+q^2)^2}.\label{eq:M2Zp}
\ee 
Thus we have  $F_{\rm DM} (q) \simeq 1$ for  
$m_{Z'} \gg \alpha m_e$, which is the case for the model 
points considered in our study. 
The reference amplitude 
$\overline{\left|\mathcal{M}_{\chi e} (q=\alpha m_e) \right|^{2}}$ 
is obtained by neglecting $q^2$ term in the denominator 
of Eq.\ \eqref{eq:M2Zp}, resulting in an $m_{Z'}^{-4}$ dependence 
in  the reference cross section $\bar{\sigma}_e$. 
We find that for the light mediator in the GeV scale, 
the Belle II limits can be 
several orders of magnitude stronger than 
the DMDD limits, including 
SENSEI \cite{SENSEI:2020dpa}, 
CDMS-HVeV \cite{SuperCDMS:2018mne}, 
DAMIC \cite{DAMIC:2019dcn}, 
XENON10 \cite{Essig:2017kqs}, 
XENON1T \cite{Aprile:2019xxb}, 
and {DMDD limits via solar reflection} \cite{An:2017ojc}. 
For example, 
Belle II can explore the parameter space well below the neutrino floor 
for silicon detectors \cite{Essig:2018tss}, 
for the case where $m_{Z'} \gtrsim 5 ~\rm GeV$. 
However, 
for the mediator mass at the MeV scale, the Belle II limits become 
somewhat weaker due to the $m_{Z'}^{-4}$ dependence. 
For example, the parameter space to be probed by Belle II 
for the $m_{Z'} \sim 10 ~\rm MeV$ case has 
already been excluded by the current DMDD limits, 
except the parameter space in the vicinity of $m_{Z'} \simeq 2 m_\chi$ in 
the vector-coupling-only case, where 
the Belle II limits are significantly enhanced.

A meta-stable particle that decays into 
SM particles are constrained by BBN bounds. 
Ref.\ \cite{Hufnagel:2018bjp} finds that 
the upper bound on the lifetime of the 
meta-stable particle that decays into electron 
and/or photon final states must be less 
than $\sim 10^{-1}\, (10^3)$ sec if the mass is about 
1 GeV (MeV), in order to satisfy the BBN bound. 
As shown in Figs.\ \eqref{fig:monoZV} and \eqref{fig:monoZA}, 
the smallest vector/axial-vector coupling probed by Belle II 
is about $g^\ell \simeq {\cal O}(10^{-5})$, leading to 
a lifetime of $\tau \simeq {\cal O} (10^{-13})$ sec 
for $m_{Z'} \sim 1$ GeV  
and $\tau \simeq {\cal O} (10^{-10})$ sec 
for $m_{Z'} \sim 1$ MeV, 
which are much smaller than the BBN bounds. 
Thus, the BBN bounds on the light vector mediator models 
considered in this study are 
much weaker than the Belle II sensitivities.

We further display the Belle II mono-photon constraints on $\bar{\sigma}_e$ 
with 50 ab$^{-1}$ for each model point 
in the $m_\chi-m_{Z'}$ plane for the vector-coupling-only case,
as shown in Fig.\ (\ref{fig:mzmxcontour}).
Because the $\bar \sigma_e$ value to be probed by Belle II 
is proportional to $m_{Z'}^{-4}$, 
the DM models with a smaller $m_{Z'}$ is less constrained, 
for example $\bar \sigma_e > 10^{-30}$ cm$^{2}$ is still allowed for 
a sub-MeV mediator. 
We also find that the constraint decreases with the DM mass $m_\chi$ 
and becomes very strong in the vicinity of the $m_{Z'} = 2 m_\chi$ line.

Recently, excess events in the electron recoil data 
are observed in the Xenon1T experiment 
\cite{XENON:2020rca}. 
A number of papers have used DM to explain such an excess, 
some of which require a sizable DM-electron interaction cross section 
\cite{Su:2020zny,Jho:2020sku,Chen:2020gcl, Du:2020ybt}. 
We note that for EFT operators between DM and electron, 
and for the GeV-mediator models, such strong 
DM-electron interaction cross sections are likely to be 
constrained by Belle II. 
However, for the models with a relatively light mediator, 
the DM-electron cross section can be significantly large, 
for example, $\bar \sigma_e \gtrsim {\cal O} (10^{-30})$ cm$^{2}$ 
for the mediator mass below MeV is likely to remain unconstrained 
with the Belle II data.  

\begin{figure}[htbp]
\begin{centering}
\includegraphics[width=0.49 \textwidth]{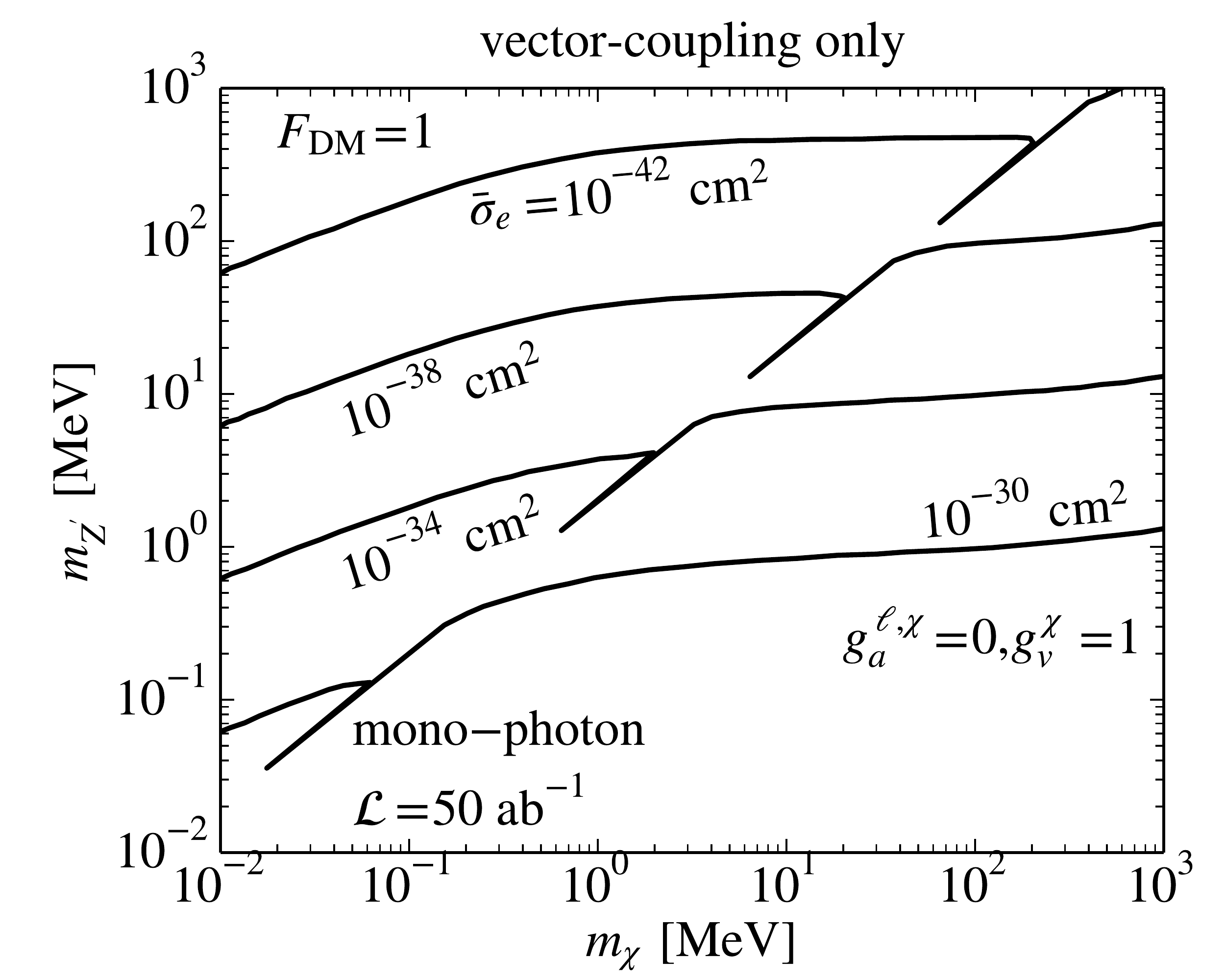}
\caption{The expected Belle II upper bound 
on $\bar \sigma_e$ for each model point in 
the $m_\chi-m_{Z^\prime}$ plane, 
in the mono-photon channel with 50 $\rm ab^{-1}$,
where only vector couplings are considered. 
We have estimated the limits in the vicinity of $m_{Z'}=2 m_\chi$, 
where the Belle II  sensitivity is highly enhanced. 
The limits along the $m_{Z'}=2 m_\chi$ line are shown for illustrative 
purposes only; the more accurate values require a detailed analysis 
which is beyond the scope of the current work.} 
\label{fig:mzmxcontour}
\end{centering}
\end{figure}

\section{Di-lepton constraints on light mediator models}
\label{sec:visiZ}

Because for the light mediator that couples to leptons, 
it is inevitable that the mediator can decay into 
a pair of final state leptons if kinematically allowed, 
one can search for the dark matter 
via the visible decay of the light mediator. 
Here we choose the process $e^+e^-
\to\gamma\mu^+\mu^-$
to search for the $Z^\prime$ resonance in 
the di-muon invariant mass spectrum.\footnote{Another di-lepton invariant mass 
channel is $e^+e^-\to \gamma e^+e^- $, 
which, however, has a much larger background 
due to an additional $t$-channel diagram and 
the photon conversion process in the low invariant mass region \cite{BaBar:2014zli}. Therefore, we do not consider 
$e^+e^-\to \gamma e^+e^- $ in this study.
}
We consider 
the following two $Z'$  masses: 
$m_{Z'} = 0.6$ GeV 
and $m_{Z'} = 5$ GeV.

We use {\sc Madgraph} \cite{Alwall:2014hca} to 
generate $10^5$ events for the 
$e^+e^- \to\gamma\mu^+\mu^-$ process 
for each new physics model point and for 
the SM. 
The main SM backgrounds
are from  $e^+e^-
\to\gamma\mu^+\mu^-$ mediated by photon, 
since  the center of mass energy is much
smaller than the mass of $Z$ boson. 
We use the following preselection cut for 
{\sc Madgraph} simulations: we select photons 
that are within the 
angle coverage of ECL such that 
$12.4^\circ$ $<\theta_\gamma^{\rm lab}<$ $155.1^\circ$,
and muons within 
the angle coverage of KLM 
such that 
$25^\circ$ $<\theta_{\mu^\pm}^{\rm lab}<$ $155^\circ$ 
\cite{Kou:2018nap}. 
We adopt the  ``three isolated clusters'' \cite{Duerr:2020muu}  as the trigger condition,
which requires that  
(i) at least three 
isolated calorimeter clusters with a 
minimum distance of $d_{\rm min} =30$ cm;\footnote{We 
use the incident position on the 
first layer of the ECL detectors 
to compute $d_{\rm min}$. 
The inner surface of the barrel region of the detector is 
$r=125$ cm away from the beam and with 
the polar angle 
32.2$^\circ < \theta< $128.7$^\circ$; 
the forward (backward) detector is placed at  
$z=+196\, (-102)$ cm with 12.4$^\circ < \theta< $31.4$^\circ$
 (130.7$^\circ < \theta< $155.1$^\circ$) 
\cite{Kou:2018nap}.}
(ii) at least one of the three 
clusters needs to have $E_{\rm lab} > $ 
0.5 GeV and the two additional clusters 
$E_{\rm lab} > $ 0.18 GeV;
(iii) all three 
clusters need to have 18.5$^\circ$
$\leq\theta_{\rm lab}\leq$ 139.3$^\circ$. 
We apply the isolation cuts 
to the photon and muon events 
that are simulated via {\sc Madgraph}. 
For the triggered events,
we follow Ref.\ \cite{Duerr:2019dmv} 
to apply the selection cuts for muons and 
photons as follows. 
We select a pair of muons such that 
(i) both $p_{\mathrm{T}}\left(\mu^{+}\right)$ and 
$p_{\mathrm{T}}\left(\mu^{-}\right)>0.05 ~\mathrm{GeV}$, 
(ii) the opening angle of the muon pair is larger than $0.1$ rad,  
and (iii) the invariant mass of the muon pair 
$m_{\mu \mu}>0.03 ~ \mathrm{GeV}$.\footnote{The
invariant mass cut $m_{\mu\mu}>30$ MeV 
hinders the di-muon sensitivities to very light mediators, 
e.g., the $m_{Z'} = 10$ MeV case as analyzed in the 
mono-photon channel.}
We select photons that satisfy
$E_{\rm lab} >0.5 ~ \rm GeV $
and $17^\circ  \leq \theta_{\rm lab} \leq 150^\circ$.

To search for the $Z'$ resonance, 
we further apply a detector cut 
of $|m_{\mu \mu}- m_{Z'}| < 2 \max \{\Gamma_{Z^\prime},  \sigma_{m_{\mu \mu}}\}$, 
{where $m_{\mu \mu}$ is the reconstructed di-muon invariant mass, 
and $ \sigma_{m_{\mu \mu}}$ is its uncertainty.} 
The resonant mass resolution is 
$\sim 0.2\%$ for charmonium and 
0.3\% for bottomonium resonances \cite{Kou:2018nap}. 
In our analysis, we adopt a 
constant resolution as $\sigma_{m_{\mu\mu}} = 0.01$ GeV 
for the di-muon invariant mass measurement, 
since we are primarily interested in the light mediators.\footnote{In 
the $e^+e^-\to\gamma\mu^+\mu^-$ process at the electron colliders, 
one can determine the di-muon invariant mass by using the measured 
photon energy via
$m^2_{\mu\mu} = s-2\sqrt{s}E_\gamma$, 
and the uncertainty is given by  
$
\sigma_{m_{\mu\mu}}=\sigma_{E_\gamma}\sqrt{s} /m_{\mu\mu},  
$ 
where $\sigma_{E_\gamma}/{E_\gamma} = 2\%$ \cite{Kou:2018nap}. 
In the $m_{Z'} = 5 \rm ~ GeV$ 
case, we have 
$\sigma_{m_{\mu\mu}}/m_{\mu\mu} \simeq 3.4\%$ 
from the photon energy measurement, 
which is much larger than 
the di-muon channel.
Hence, the di-muon channel has a better resolution 
for a narrow $Z'$ resonance than the mono-photon channel.} 
To our knowledge, the two $Z'$ masses considered here do not coincide 
with any significant di-muon backgrounds from hadron decays. 
Otherwise a more sophisticated study is in order.   
We note that taking into account the angular distributions of
the final state particles does not improve the sensitivities, 
because the new physics process has a similar 
di-muon angular distribution as the SM background.

\begin{figure}[htbp]
\begin{centering}
\includegraphics[width=0.49 \textwidth]{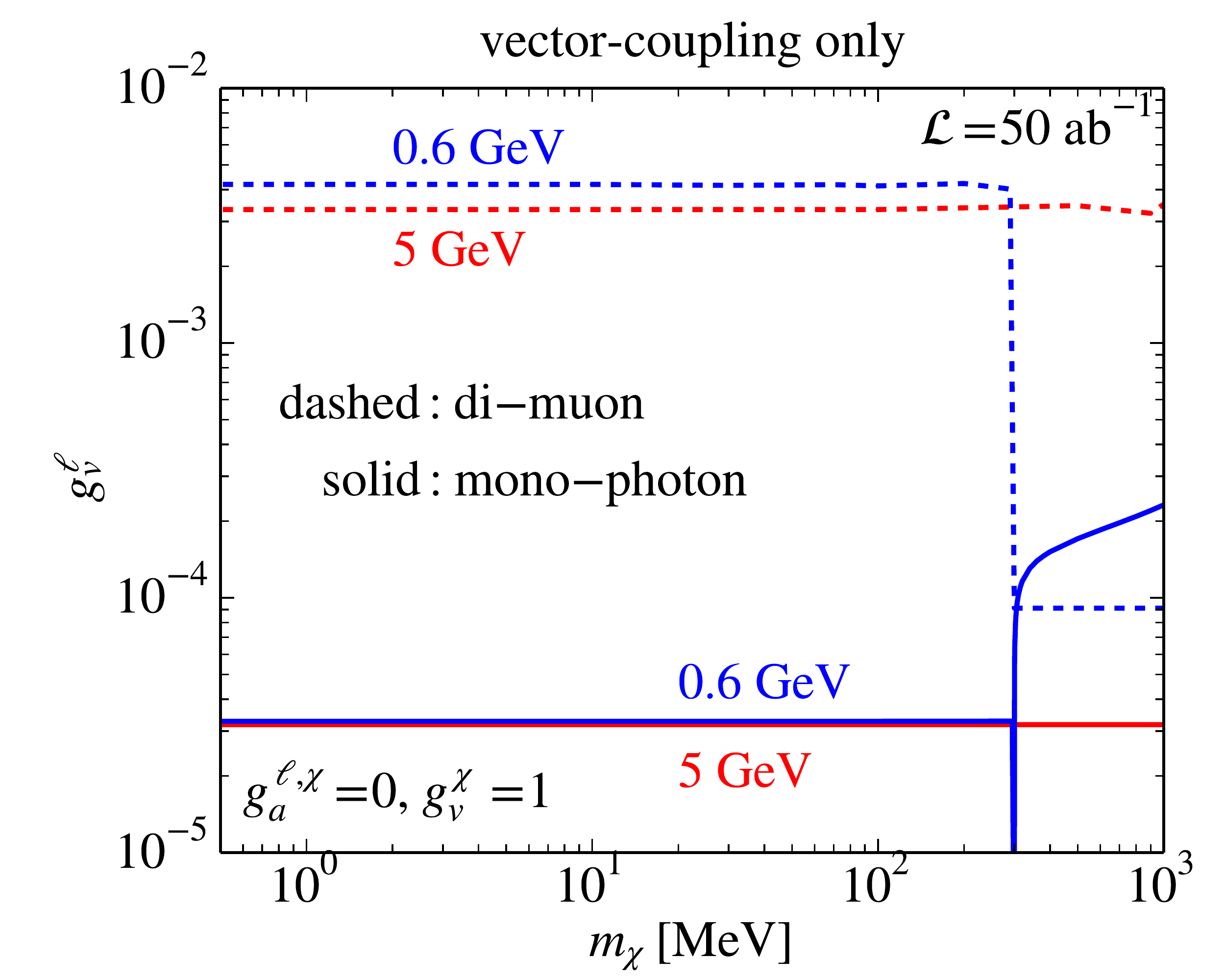}
\includegraphics[width=0.49 \textwidth]{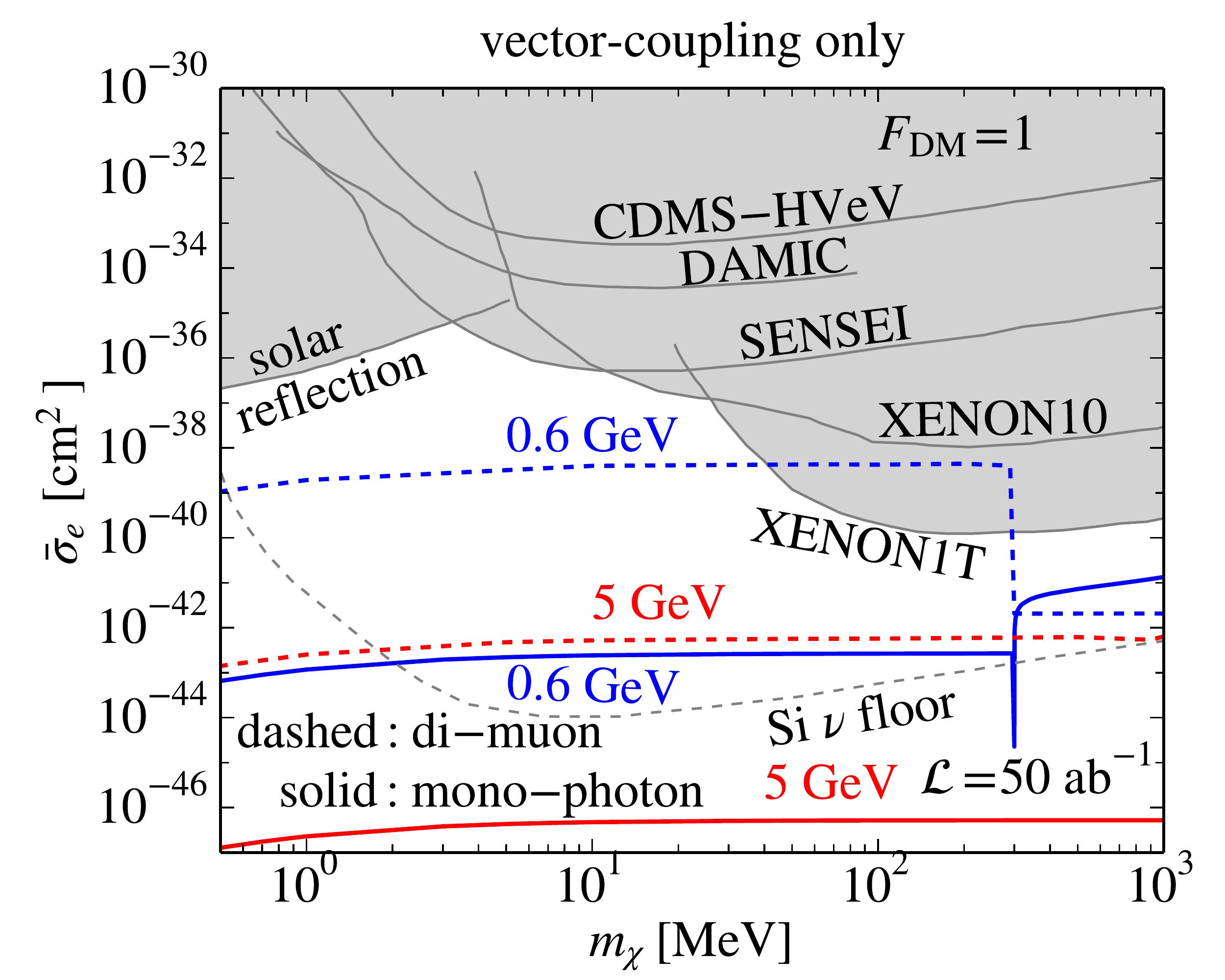}
\caption{The expected Belle II {90\%} C.L.\ upper bound via visible search (dashed)
with 50 $\rm ab^{-1}$ integrated luminosity 
on $g_v^\ell$ (left panel) and $\bar\sigma_e$ (right panel)  
in light mediator models 
where only vector couplings are assumed. 
Two $Z'$ masses 
are considered for the {di-muon} search analysis: 
$m_{Z^\prime}=5$ GeV (red) and $m_{Z^\prime}=0.6$ 
GeV (blue).
The Belle II {mono-photon} search constraints
(solid) 
are shown for the two $Z'$ masses. 
Constraints from DMDD experiments with $F_{\rm DM}(q)=1$ 
(valid for $m_{Z'} \gg \alpha m_e$) 
are also shown:  
SENSEI \cite{SENSEI:2020dpa}, 
XENON10 \cite{Essig:2017kqs} 
and XENON1T \cite{Aprile:2019xxb}, 
CDMS-HVeV \cite{SuperCDMS:2018mne}, 
DAMIC \cite{DAMIC:2019dcn}, 
and solar reflection \cite{An:2017ojc}. 
The gray dashed line shows the neutrino 
floor limit for 
silicon detectors \cite{Essig:2018tss}.}
\label{fig:zvisibleV}
\end{centering}
\end{figure}

We compute the Belle II sensitivity  
({90\%} C.L.\ upper bound) 
to the new physics model in the di-muon channel
with 50 $\rm ab^{-1}$ integrated luminosity, 
by using the condition 
${N_s / \sqrt{N_b}}=\sqrt{2.71}$, 
where $N_s$ is the number of new physics signal events, 
and $N_b$ is the number of  SM background events. 
The expected {90\%} C.L.\ upper bounds  
on $g_v^\ell$ are shown 
on the left panel figure of Fig.\ (\ref{fig:zvisibleV}) 
where we take $g^\ell_a = g^\chi_a = 0$ and $g_v^\chi =1$;  
the corresponding limits on $\bar\sigma_e$ 
are shown 
on the right panel figure of Fig.\ (\ref{fig:zvisibleV}). 
For the $m_{Z'} = 5$ GeV case, the 
expected 90\% C.L.\ upper bound on $g_v^\ell$ 
in the di-muon channel is $g_v^\ell \lesssim 3 \times 10^{-3}$ 
in the MeV-GeV DM mass range, which 
is about two orders of magnitude 
weaker than in the mono-photon channel. 
The $m_{Z'} = 0.6$ GeV case is similar 
to the $m_{Z'} = 5$ GeV except in the 
mass range $m_\chi > 0.3$ GeV, where 
the di-muon limit becomes stronger than the 
mono-photon limit. This is due to the fact that
for the case where $m_{Z'} = 0.6$ GeV 
and $m_\chi > 0.3$ GeV, 
di-leptons can be produced on the $Z'$ resonance, 
but DM can only be produced off the $Z'$ resonance. 
The di-muon limit for the $m_{Z'} = 5$ GeV case
is comparable to the neutrino floor limit 
of the silicon detectors \cite{Essig:2018tss} and 
is several orders of magnitude 
stronger than the current DMDD limits, 
which includes 
SENSEI \cite{SENSEI:2020dpa}, 
XENON10 \cite{Essig:2017kqs} 
and XENON1T \cite{Aprile:2019xxb}, 
CDMS-HVeV \cite{SuperCDMS:2018mne}, 
DAMIC \cite{DAMIC:2019dcn}, 
and solar reflection \cite{An:2017ojc}, 
as shown in the right panel figure of 
Fig.\ (\ref{fig:zvisibleV}). 
The di-muon limit for the $m_{Z'} = 0.6$ GeV case 
is stronger than the current DMDD limits 
except the mass range of 
20 MeV $\lesssim m_\chi \lesssim$ 300 MeV 
where XENON1T \cite{Aprile:2019xxb} 
becomes stronger.

The expected 90\% C.L.\ upper bounds  
on $g_a^\ell$ are shown 
on the left panel figure of Fig.\ (\ref{fig:zvisibleA}) 
where we take $g^\ell_v = g^\chi_v = 0$ and $g_a^\chi =1$;  
the corresponding limits on $\bar\sigma_e$ 
are shown 
on the right panel figure of Fig.\ (\ref{fig:zvisibleA}). 
The di-muon limit on $g_a^\ell$ is about 
three times weaker than $g_v^\ell$ in the 
low DM mass range.

The different limits on $g_a^\ell$ and $g_v^\ell$ are 
primarily due to different behaviors in the 
photon-$Z'$ interference terms in the cross section. 
The total amplitude square of 
the $e^+ e^- \to \mu^+ \mu^- \gamma$ process 
can be parameterized as 
{
$|\mathcal{M}|^2 = |\mathcal{M}_\gamma+ \mathcal{M}_{Z'}
+ \mathcal{M}_{Z}|^2$,
}
where 
$\mathcal{M}_\gamma$, 
$\mathcal{M}_{Z'}$, 
and $\mathcal{M}_{Z}$
denote 
the amplitudes mediated by the 
photon, $Z'$, and $Z$ respectively. 
Since $m_Z$ is much larger {than} $\sqrt{s}$ of Belle II, 
we neglect $\mathcal{M}_{Z}$ here.
Hence, the $e^+ e^- \to \mu^+ \mu^- \gamma$ cross section 
receives three contributions: 
$\sigma = \sigma_\gamma +  \sigma_{\gamma Z'} +  \sigma_{Z'}$, 
where $\sigma_\gamma$ denotes the SM background
mediated by the photon, 
$\sigma_{\gamma Z'}$ denotes the cross section due to the 
$\gamma - Z'$ interference term, 
and $\sigma_{Z'}$ denotes the cross section due to 
the $Z'$ term. 
The expressions of $\sigma_{Z'}$ are similar in the 
vector only case and in the axial-vector only case; 
the $\sigma_{\gamma Z'}$ contributions, however,  
are very different. 
For example, we 
have $\sigma_{\gamma Z'}=$ 2.36 (-0.379) fb 
and $\sigma_{Z'}=$ 0.472 (0.603) fb 
for the case where 
$g_v^\ell \, (g_a^\ell)=0.01$, 
$m_{Z'}=5$ GeV, and  
$m_\chi=1$ GeV. 
Therefore the total cross section 
in the vector case is about one 
order of magnitude larger than the axial-vector case.
Although we have imposed 
detector cuts to select events from the $Z'$ resonance, 
the contribution from $\sigma_{\gamma Z'}$ 
turns out to be comparable to that from $\sigma_{Z'}$, 
because the NP couplings 
($g_v^\ell$ and $g_a^\ell$) are much smaller than the QED  
coupling constant $e$.

\begin{figure}[htbp]
\begin{centering}
\includegraphics[width=0.49 \textwidth]{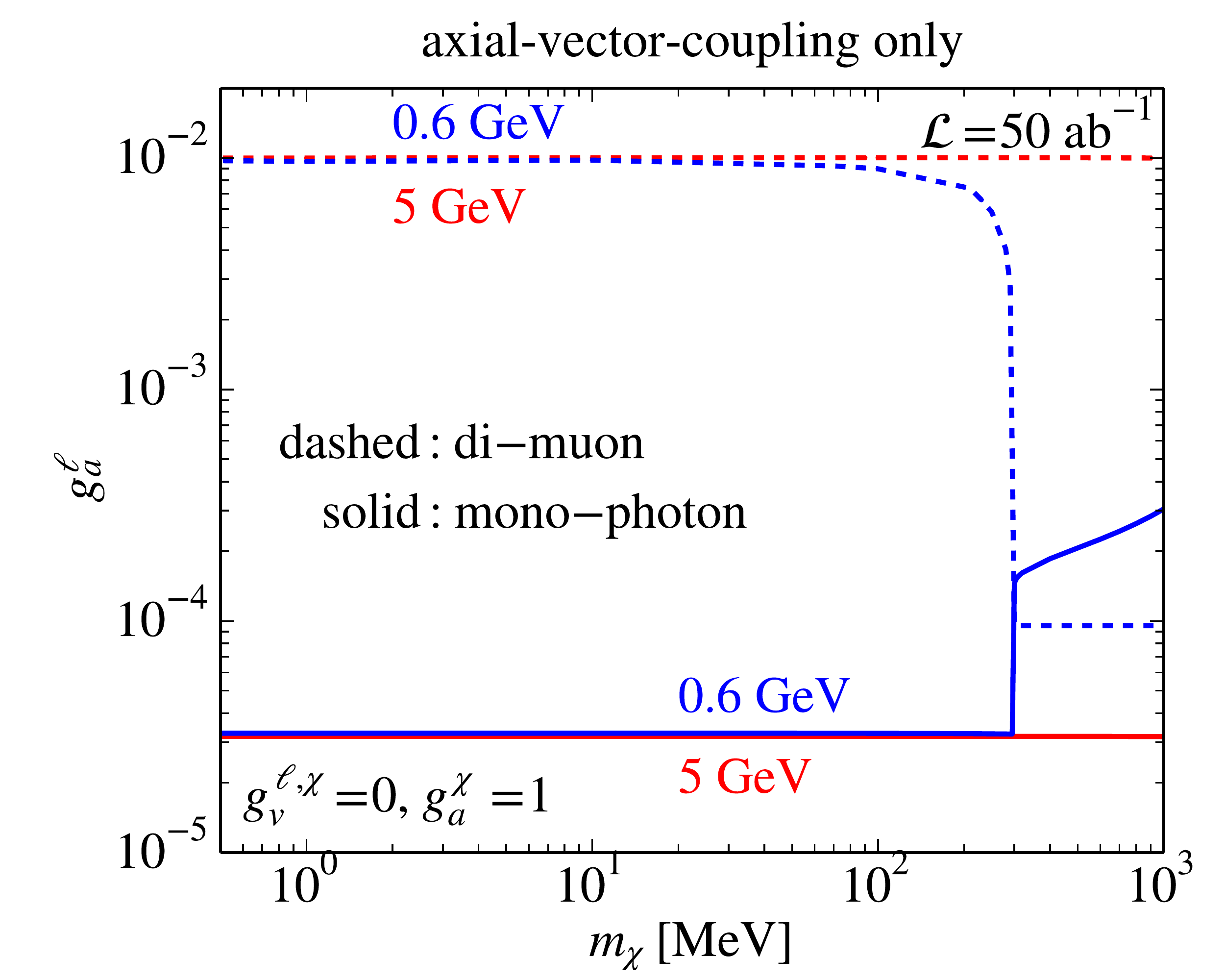}
\includegraphics[width=0.49 \textwidth]{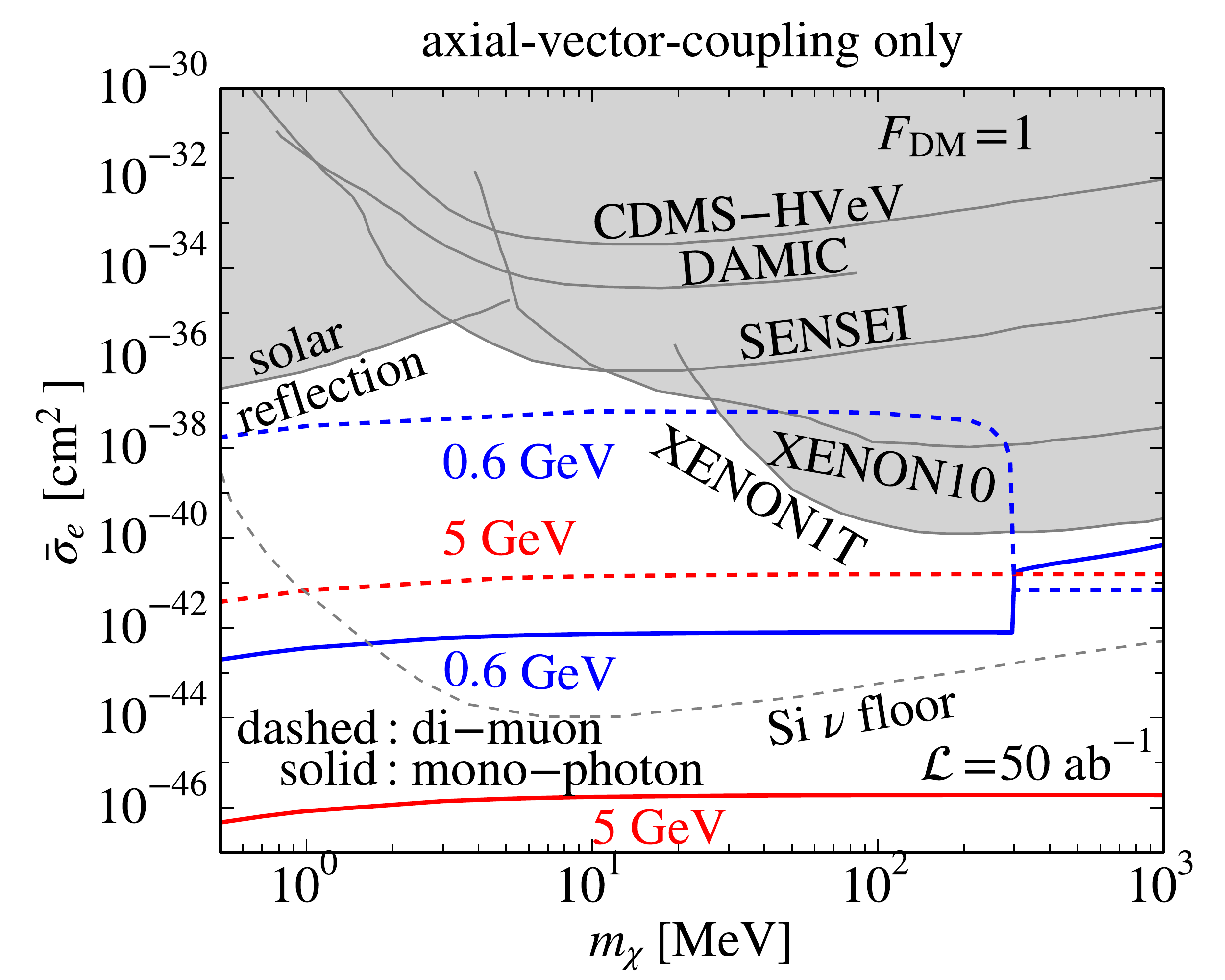}
\caption{Same as Fig.\ (\ref{fig:zvisibleV}) but 
where only axial-vector coupling are assumed.}
\label{fig:zvisibleA}
\end{centering}
\end{figure}

\begin{figure}[htbp]
\begin{centering}
\includegraphics[width=0.49 \textwidth]{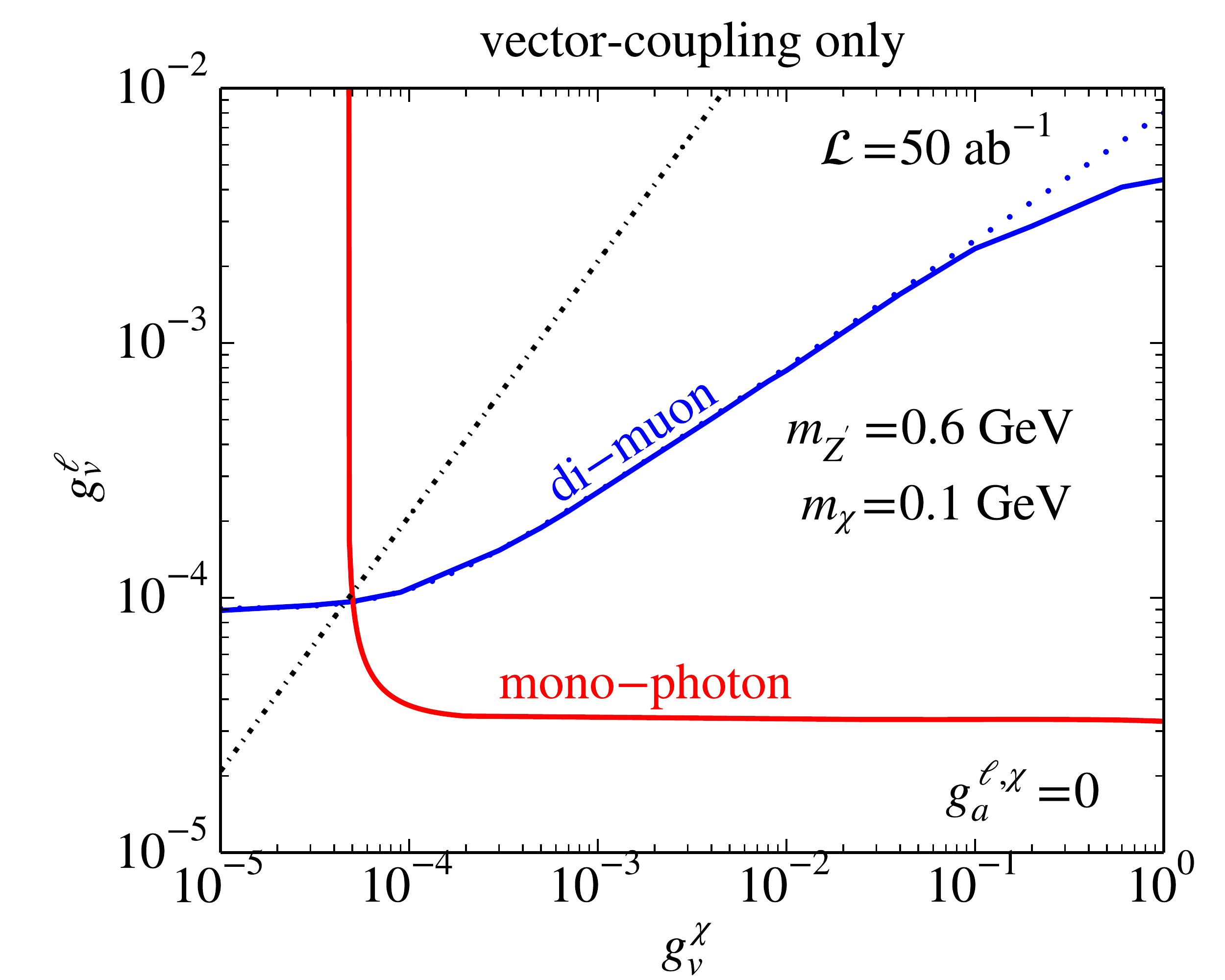}
\includegraphics[width=0.49 \textwidth]{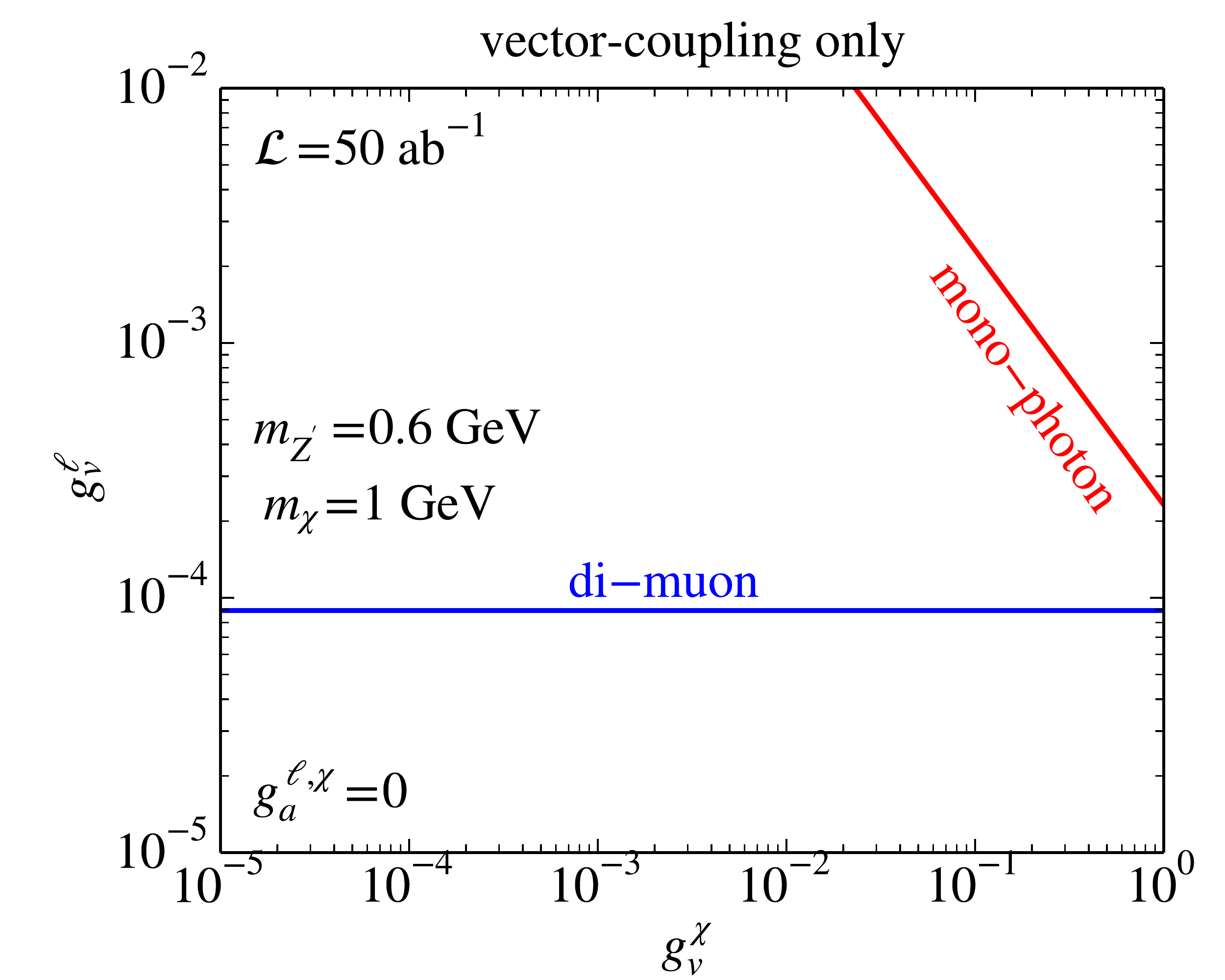}
\caption{Left panel: Expected Belle II {90\%} C.L.\ upper bound on 
$g_v^\ell$ as a function of $g_v^\chi$  
from the {di-muon} channel (blue) 
and from the {mono-photon} channel (red) 
{with 50 $\rm {ab}^{-1}$ integrated luminosity},
$m_{Z^\prime} = 0.6 ~ \rm GeV$, 
$m_\chi = 0.1$ GeV, 
and only vector couplings. 
The parameter space, where the sensitivity 
of the visible channel is the same as 
the invisible channel, is approximated by 
the black dot-dashed line, {which is $g_v^\ell = 2g_v^\chi$}. 
The upper bound in the di-muon channel estimated by 
neglecting the $\gamma-Z'$ interference term is indicated 
by the blue dotted line. 
Right panel: same as the left panel but {with} $m_\chi=1$ GeV.}
\label{fig:Z06gvxgvf}
\end{centering}
\end{figure}

We further compare the 
90\% C.L.\ upper bound on $g_v^\ell$ from the mono-photon channel 
and from the di-muon channel on the $g_v^\chi-g_v^\ell$ plane 
in Fig.\ \eqref{fig:Z06gvxgvf}, 
where we consider the vector-only case 
and take $m_{Z'}= 0.6 ~\rm GeV$. 
For the case where $m_\chi= 0.1 ~\rm GeV$, 
the $Z'$ boson can decay into a pair of 
DM particles. 
For that reason, Belle II can probe a much smaller 
$g_v^\ell$ in the mono-photon channel 
than in the di-muon channel, in the  
range of $g_v^\chi \gtrsim 5 \times 10^{-5}$; 
only for very small $g_v^\chi$ values 
(namely $g_v^\chi \lesssim 5 \times 10^{-5}$), the di-muon channel becomes the better 
channel to constrain the parameter space. 
We further compare the sensitivities from these two channels for all the model points on the 
 $g_v^\ell - g_v^\chi$ plane, and find that the parameter space can be approximately 
 divided by the line 
 $g_v^\ell = 2g_v^\chi$ into two regions:
model points on the left-upper side of 
the line typically receive a stronger constraint from the di-muon 
channel than from the mono-photon channel; 
model points on the right-lower side of the 
line, on the other hand, are better constrained by the mono-photon channel. 
We also estimate the di-muon sensitivity curve by neglecting   
the $\gamma-Z'$ interference term, as indicated by the blue dotted line in the left panel figure 
of Fig.\ \eqref{fig:Z06gvxgvf}.
We find that the $\gamma-Z'$ interference term 
cannot be neglected for the parameter range of $g_v^\chi>0.1$ 
and produces the dominant contribution 
to the di-muon signal for the parameter range of $g_v^\chi \sim 1$ 
(in the vicinity of the sensitivity curve).

For the case where $m_\chi = 1 ~\rm GeV$, 
the $Z'$ boson cannot decay into a pair of 
DM particles. 
For that reason, the sensitivity on $g_v^\ell$ from the di-muon 
channel is always better than the 
mono-photon channel for the entire $g_v^\chi$ range shown 
in the right panel figure of Fig.\ \eqref{fig:Z06gvxgvf}. We also find that 
the di-muon limits in the $m_\chi = 1 ~\rm GeV$ case 
are better than the $m_\chi = 0.1 ~\rm GeV$ case, 
since in the former case the $Z'$ boson can only decay 
into visible final states.

\section{NA64 constraints }
\label{sec:NA64}

Light dark matter that couples to electron can also be searched for 
at the NA64 experiment, an electron {fixed target} %
experiment with a lead target. 
The energy of the incident electron of the NA64 experiment is 100 GeV \cite{Banerjee:2019pds}. 
In this section, 
we compute the constraints on the EFT operators and on the 
light mediator models, by using the  
$2.84 \times 10^{11}$ electrons on target (EOT) data 
accumulated by the NA64 experiment \cite{Banerjee:2019pds}.

A pair of fermionic DM can be produced at NA64 via a 2-to-4 process
\be
e^- (p) + N(P_i) \to  e^- (p') +  N (P_f) +  \chi (k_1) + \bar{\chi }(k_2), 
\ee
where $N$ is the Pb nucleus, 
and we have specified the momentum for each particle 
in the parenthesis. The DM signature is a large missing energy
carried away by the $\chi \bar \chi$ pair. 
The Feynman diagrams of the 2-to-4 process 
for the EFT operators and for the light mediator 
models are shown in Fig.\ (\ref{fig:DM:Na64}).

\begin{figure}[htbp]
\begin{centering}
\includegraphics[width=0.35 \textwidth]{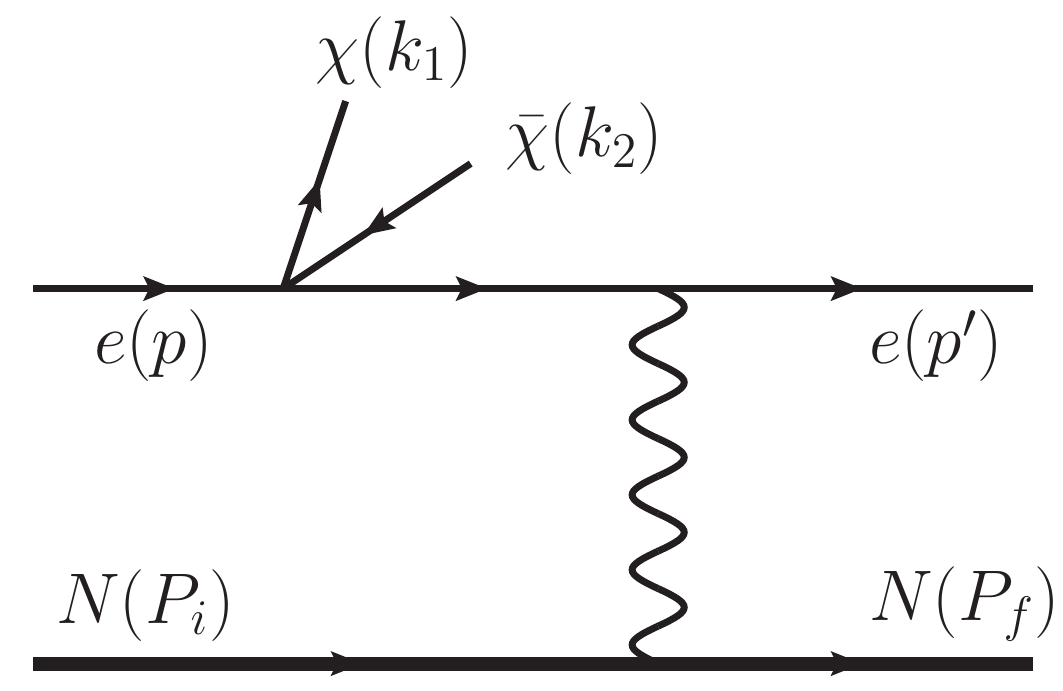}
\includegraphics[width=0.35 \textwidth]{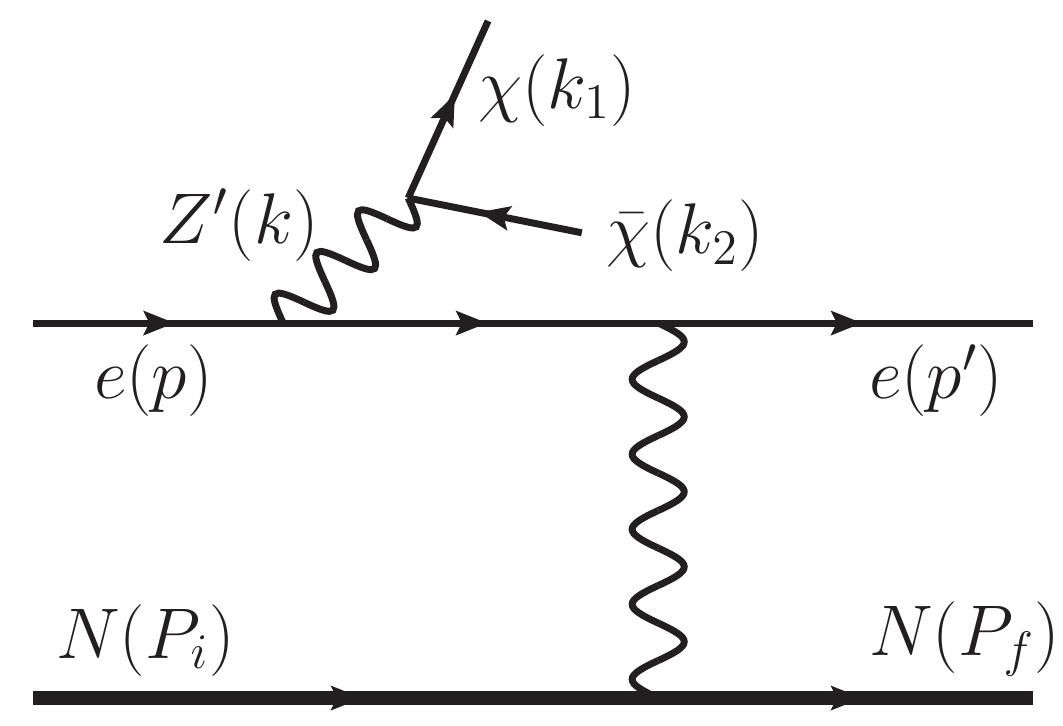}
\caption{Diagrams for the DM 2-to-4 production processes  
at NA64 for the EFT operators 
(left) and for the light mediator models (right).
Only the initial state radiation processes are shown here. 
DM can also be radiated from the final state electron. 
}
\label{fig:DM:Na64}
\end{centering}
\end{figure}

The differential cross section of the 2-to-4 process is computed by 
\be
d \sigma(e N \rightarrow e N \chi \bar{\chi})=
\frac{1}{4 E_e E_N v_{\rm rel}} \overline{|\mathcal{M}| ^{2}} d \Phi^{(4)},
\ee
where $E_e$ $(E_N)$ is the energy of the initial state electron (nucleus),
$v_{\rm rel}$ is the relative velocity of the initial state electron and the 
initial state nucleus,
$d\Phi^{(4)}$ is the four-body phase space for $p'$, $P_f$, $k_1$, and $k_2$, 
and $\overline{|\mathcal{M}| ^{2}}$ is the usual matrix element square 
summed over final spins and averaged over initial spins.
To compute the cross section of the 2-to-4 process, 
we decompose the 4-body phase space 
into a 2-body phase space (for $\chi \bar \chi$) and a 3-body phase space as follows, 
\be
d\Phi^{(4)} (P_f, p', k_1, k_2) = 
{dk^2 \over 2\pi} \, d\Phi^{(3)}(P_f, p', k) \times d\Phi^{(2)}(k_1,k_2), 
\ee
where $k=k_1+k_2$. 
The matrix element of the 2-to-4 process can also be decomposed 
as follows 
\be
\mathcal{M} = \sum_i  \mathcal{M}_{i,A}(p,p',P_i,P_f,k) \times J_{i}^A (k_1,k_2),
\ee
where 
\be
J_{i}^A (k_1,k_2) \equiv \bar{u}(k_1)\Gamma_i^A v(k_2), 
\ee 
with 
$\Gamma_{i}^A = \{1, \gamma^5, \gamma^\mu, \gamma^\mu \gamma^5, \sigma^{\mu\nu}/ \sqrt{2} \}$ 
and $\sigma^{\mu\nu}=\frac{i}{2}[\gamma^\mu,\gamma^\nu]$. 
Here $i$ denotes the interaction type, and 
$A$ denotes the corresponding Lorentz indices. 
Therefore, $J_{i}^A (k_1,k_2)$ are 
$J_1^A$, $J_3^A$, and $J_4^A$ 
for $O_s$, $O_V$, and $O_A$ respectively. 
For the $O_t$ operators, one can use the 
Fierz identity to re-arrange the fermionic fields as follows 
\be \bar{\chi} \ell \bar{\ell}\chi=\sum_{i=1}^5 \lambda_{i}
\bar{\ell}  \Gamma_i \ell
\bar{\chi} \Gamma_i \chi,  
\ee
where $\lambda_i=\frac{1}{4}\{1,1,1,-1,1\}$. 
Thus, all the five $J_{i}^A$'s are needed for the 
$O_t$ case. 
We first integrate out the 2-body phase space 
\be
\chi^{A B}_{ij} (k) \equiv \int \, d \Phi^{(2)} \, \sum_{\rm s_\chi, s_{\bar{\chi}}} J_i^A (J_j^B)^\dagger.
\label{eq:DMTen}
\ee
The expressions of the various $\chi^{A B}_{ij}$'s 
are given in Appendix \ref{sec:dxsec}.

Thus, the differential 
cross section of the 2-to-4 process is given by
\be
d \sigma(eN \rightarrow eN \chi \bar{\chi})
=\frac{d k^2}{2 \pi} {\chi^{A B}_{ij} (k) \over k^4} 
\left[ k^4 d \Phi^{(3)}\frac{\overline{ \mathcal{M}_{i,A} \mathcal{M}_{j,B}^{\dagger} }  
}{4 E_e E_N v_{\rm rel}} \right] 
\equiv  \frac{d k^2}{2 \pi} {\chi^{A B}_{ij} (k) \over k^4}  \times d{\sigma}_{ij,AB}, 
\label{eq:chiABdAB}
\ee 
where the sum with repeated indices $i$ and $j$ is implicit,  
and $d{\sigma}^{AB}_{ij}$ is defined in such a way 
that it has the dimension of cross section.
To compute $d{\sigma}^{AB}_{ij}$, 
we use the 
WWA \cite{Williams:1935dka,vonWeizsacker:1934nji}, 
in which the photon vertex with the lead nucleus 
can be replaced by an effective photon flux function. 
Thus, in the lab frame, 
one has \cite{Liu:2017htz} 
\be
\frac{d{\sigma}_{ij,AB}}{d x}
\simeq k^4
\frac{\alpha \zeta}{16 \pi^2}
(1-x)\beta_k
\int_{\tilde{u}_{\min}}^{\tilde{u}_{\max }}  
{d \tilde{u} \over \tilde{u}^2} \, 
\overline{ {\tilde{\mathcal{M}}}_{i,A} {\tilde{\mathcal{M}}}_{j,B}^{\dagger}} \Bigg|_{t=t_{\min}}, 
\label{eq:WWA}
\ee 
where $\zeta$ is the photon flux, 
$x=k^0/E_e$,
$\beta_k=\sqrt{1-k^2/(x E_e)^2}$,
$t=-(P_f-P_i)^2=-q^2$, 
$t_{\min}=(k^2/2E_e)^2$, 
$\tilde{u}(\theta_k)=-E_e^2\theta_k^2x-k^2(1-x)/x-m_e^2x$, 
with $\theta_k$ being the polar angle of momentum $k$,
$\tilde{u}_{\max}=\tilde{u}(\theta_k=0)$, and
$\tilde{u}_{\min}=\tilde{u}(\theta_k=\pi)$.
\footnote{The limits 
$\tilde{u}_{\min}$ and $\tilde{u}_{\max}$ are determined by the range of $\theta_k$. 
Because the missing energy signature includes dark matter emissions 
with arbitrary $\theta_k$, 
one has $\tilde{u}_{\max}=\tilde{u}(\theta_k=0)$ 
and
$\tilde{u}_{\min}=\tilde{u}(\theta_k=\pi)$.
We note that the usual approximation $\tilde{u}_{\min} \to -\infty$ \cite{Liu:2017htz} can fail for some cases;  
see appendix \ref{sec:dxsec} for more 
detailed discussions.} 
The matrix element $\tilde{\mathcal{M}}_{i,A}$ in Eq.\ \ref{eq:WWA}
corresponds to the diagram 
that is obtained by 
removing the $N$ particles both in the initial state and in final state 
(as well as the $\gamma N N$ vertex) in the diagram of ${\mathcal{M}}_{i,A}$. 
For illustration purposes, we also draw the diagrams for the 
process of $\tilde{\mathcal{M}}_{i,A}$ in Fig.\ \ref{fig:2t2FD}, 
by introducing an imaginary particle $V^A$ with the momentum $k$. 
However, the calculations of the cross section can be carried out 
without introducing the imaginary $V^A$ particle.
The expressions of the integrand in Eq.\ \eqref{eq:WWA} 
contracted with $\chi_{ij}^{A B}$ for various models 
are given in Appendix \ref{sec:dxsec}.

\begin{figure}[htbp]
\begin{centering}
\includegraphics[width=0.35 \textwidth]{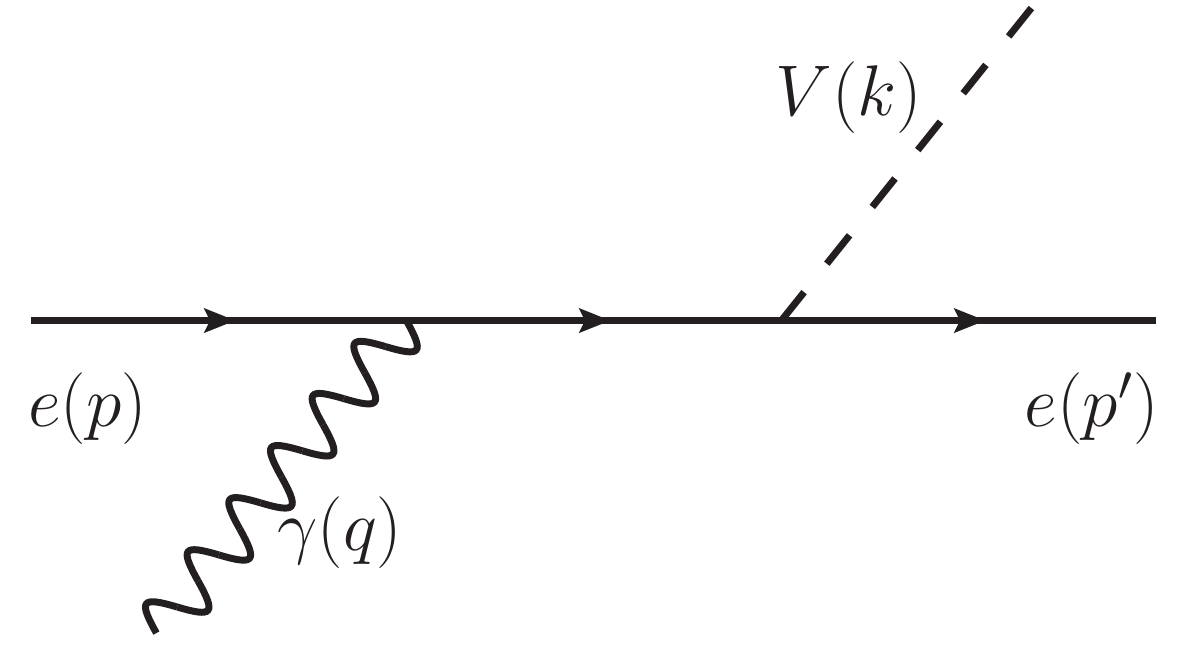}
\includegraphics[width=0.35 \textwidth]{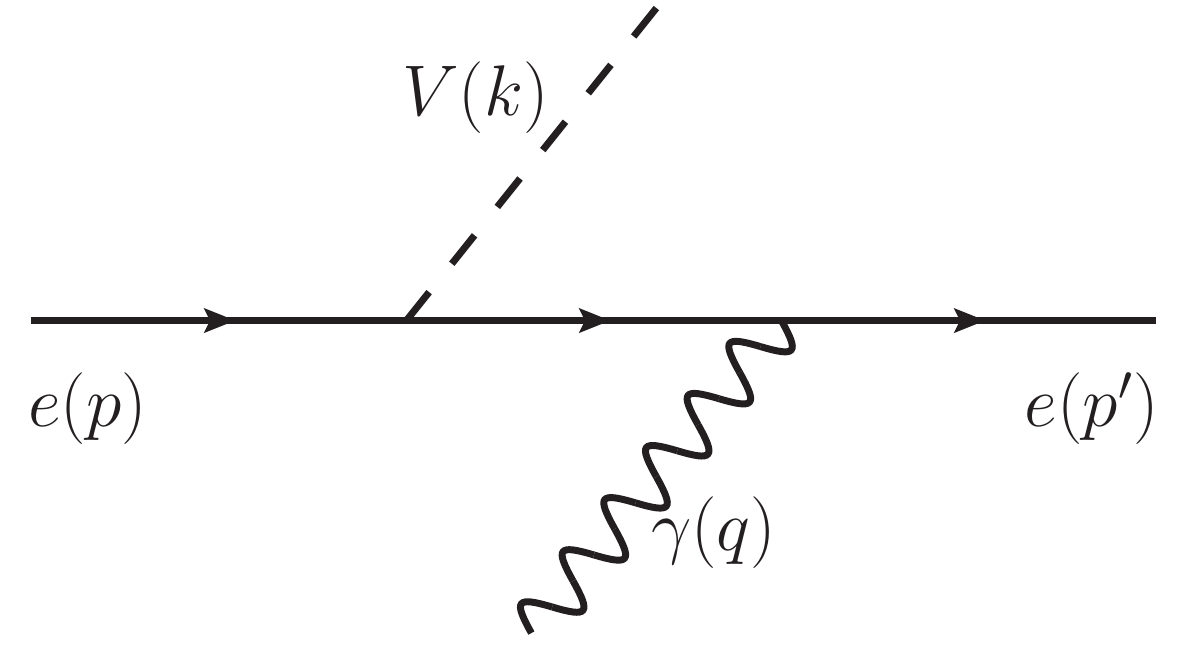}
\caption{Feynman diagrams of  
$e^- (p) + \gamma (q) \to  e^- (p') +  V (k)$, 
where $V(k)$ represents the imaginary particle with momentum $k$, 
which carries the (multiple) Lorentz indices $A$. 
Note that because the 
matrix element $\tilde{\cal M}_{i,A}$ (as well as ${\cal M}_{i,A}$) 
contains the explicit Lorentz indices $A$,    
the corresponding ``polarization vectors'' of $V$ are not included 
in the calculations of $\tilde{\cal M}_{i,A}$.  }
\label{fig:2t2FD}
\end{centering}
\end{figure}

The effective  photon flux $\zeta$ is given by \cite{Bjorken:2009mm,Liu:2017htz}
\be
\zeta =\int_{t_{\rm min}}^{t_{\rm max}} dt \, 
\frac{t-t_{\rm min}}{t^2} \, G_2^{\rm el} (t)
=\int_{t_{\rm min}}^{t_{\rm max}} dt \, \frac{t-t_{\rm min}}{t^2} 
\Bigg[ {Z a^2 t \over (1+ a^2 t) (1+ t/d)} \Bigg]^2,
\ee
where 
$t_{\rm max}=k^2+m_e^2$,
$G_2^{\rm el} (t)$ is the elastic form-factor of the 
lead nucleus,\footnote{We neglect the inelastic form factor, 
which is usually much smaller than the elastic form factor 
for high $Z$ targets \cite{Bjorken:2009mm}.}
$a= 111 m_e^{-1} Z^{-1/3}$, 
$d=0.164 A^{-2/3}$ GeV$^2$, 
and we use $Z=82$ and 
$A=207.2$ for Pb.

Thus, the signal cross section as a function of the missing energy 
is given by 
}\be
{ d \sigma \over dx } 
(eN \rightarrow eN \chi \bar{\chi})
= \int_{4 m_\chi^2}^{x^2 E_e^2} 
d k^2 
\left[ 
\frac{\chi_{ij}^{A B} (k)}{2 \pi k^4}
{d{\sigma}_{ij,AB}\over dx} 
 \right]
\equiv  \int_{4 m_\chi^2}^{x^2 E_e^2} 
dk^2 \frac{d\sigma}{dx dk^2}. 
\label{eq:NA64xsec}
\ee 
The expressions of 
$d\sigma/dx dk^2$
for various models 
are given in Appendix \ref{sec:dxsec}. 
The signal region used by the NA64 collaboration is 
$E_{\rm miss}> 0.5 E_e$  \cite{Banerjee:2019pds},
where $E_e$ is the energy of electron beam and
$E_{\rm miss}$ is the missing energy. 
Thus, the number of the signal events is computed via
\begin{equation}
N_{s}=N_{\mathrm{EOT}} 
\times \frac{\rho_N}{m_N}
\times L_{\rm eff} \times \epsilon_d \times 
\int_{x_{\rm min}}^{1-{m_e\over E_e}} d x  { d \sigma \over dx } 
(eN \rightarrow eN \chi \bar{\chi}), 
\end{equation}
where $x_{\rm min} = 0.5$, 
$N_{\rm EOT}=2.84 \times 10^{11}$ is the total 
number of electron on target, 
$\rho_N=11.34 ~\rm g/cm^3$ is the mass density of lead, 
$m_N=A m_p$ is the mass of the lead nucleus 
with $A=207.2$ and $m_p \simeq 0.93$ GeV,
$\epsilon_d \simeq 0.5$ is the detection efficiency \cite{Banerjee:2019pds}, 
$L_{\rm eff}$ is the effective length of the lead target for electron collision.
We use the radiation length as the effective length,
namely $L_{\rm eff}\simeq X_{0}^{(e)}\simeq 0.5~\rm cm $, 
since the lead target in NA64 is a thick target for electrons \cite{Gninenko:2018ter}. 
We then compute the {90\%} C.L.\ 
limits on the light mediator models and EFT operators 
by using the criterion {$N_s=2.3$} 
based on 
the null background assumption \cite{Gninenko:2017yus}.

 \begin{figure}[htbp]
\begin{centering}
\includegraphics[width=0.49 \textwidth]{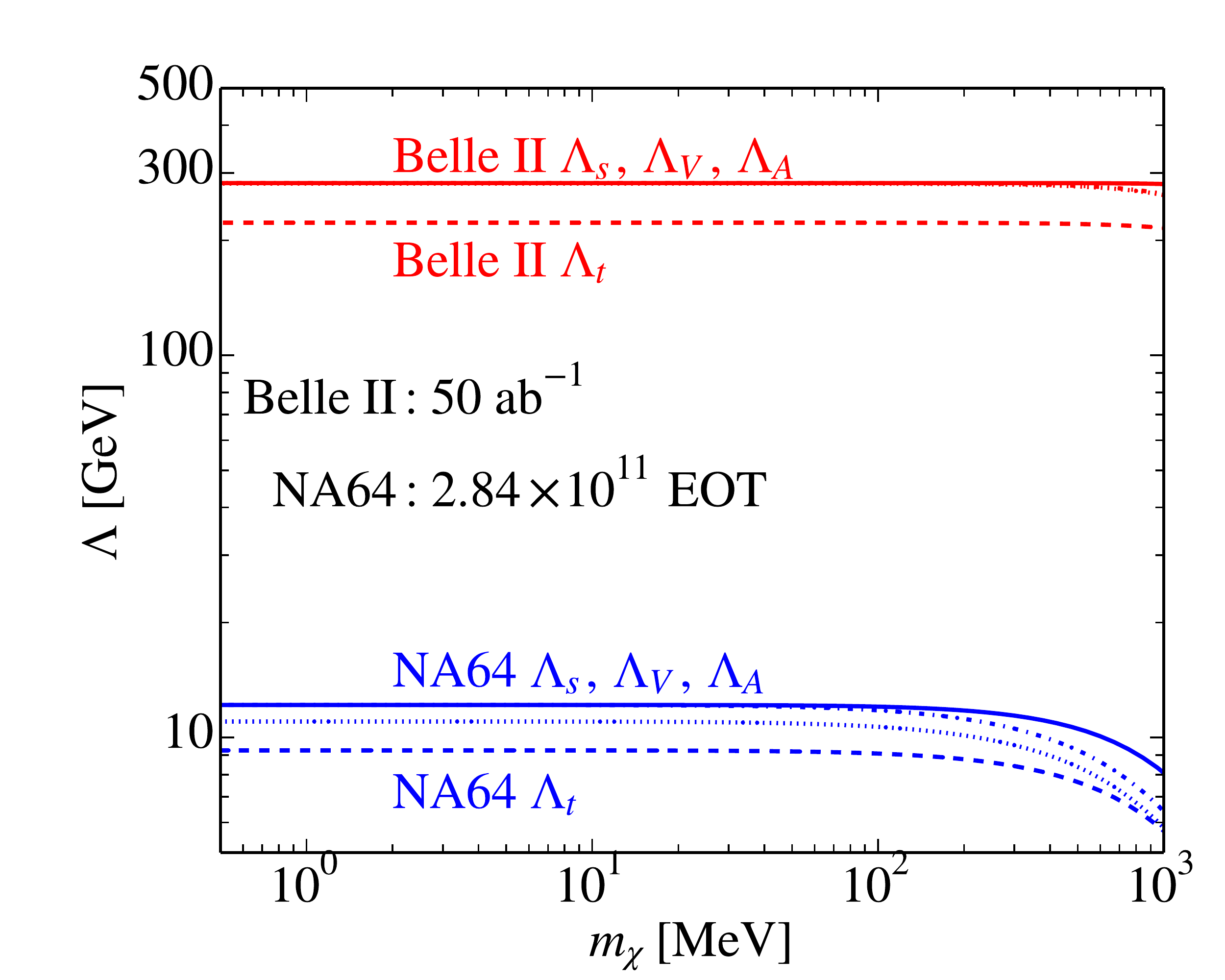}
\caption{NA64 90\% C.L.\ upper bond on the 
EFT operators:
$\Lambda_A$ (dot-dashed), $\Lambda_V$ (solid), 
$\Lambda_s$ (dotted), and  $\Lambda_t$ (dashed). 
The Belle II sensitivities are the same as Fig.\ \ref{fig:monoEFT}.} 
\label{fig:NA64EFT}
\end{centering}
\end{figure}

\begin{figure}[htbp]
\begin{centering}
\includegraphics[width=0.49 \textwidth]{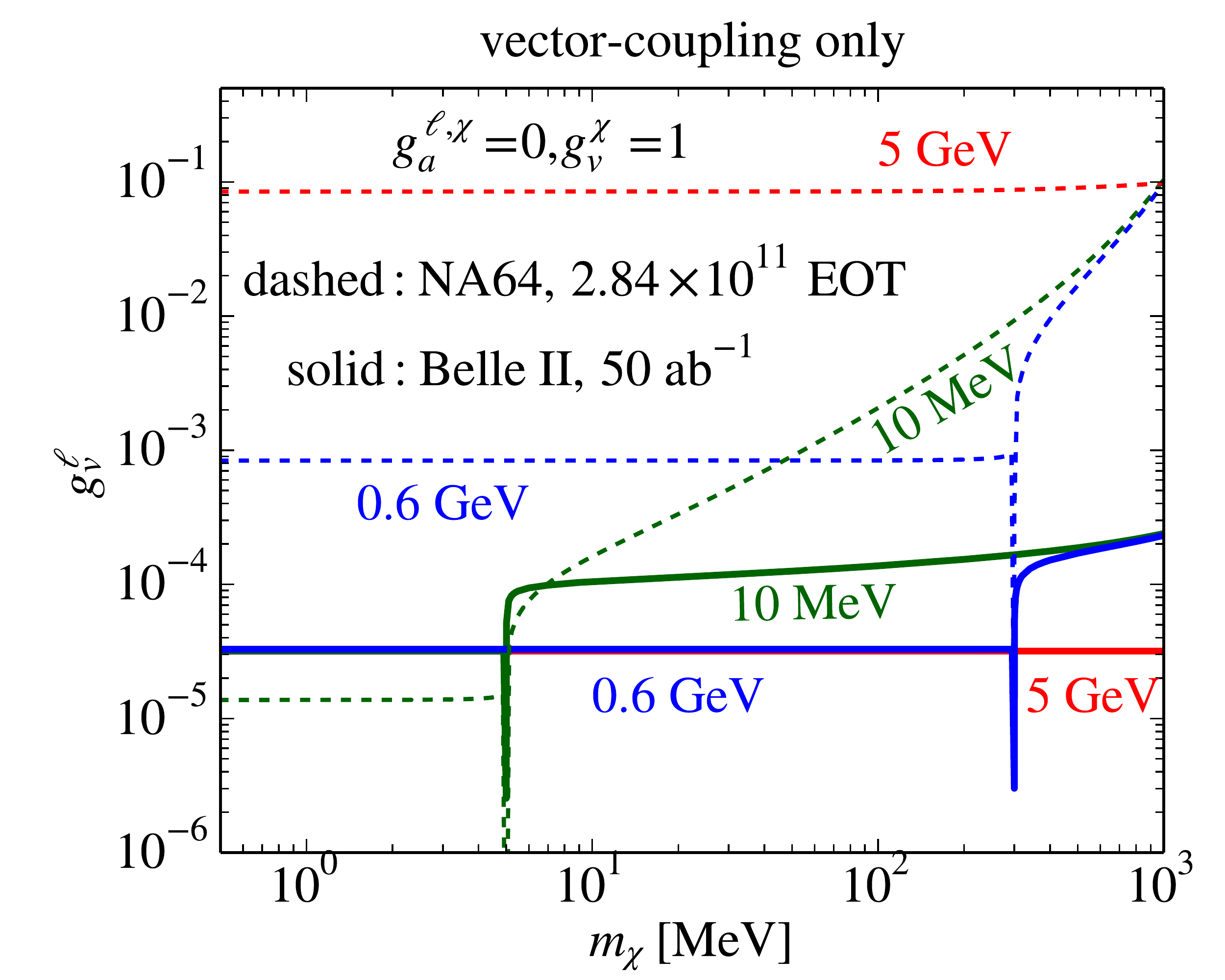}
\includegraphics[width=0.49 \textwidth]{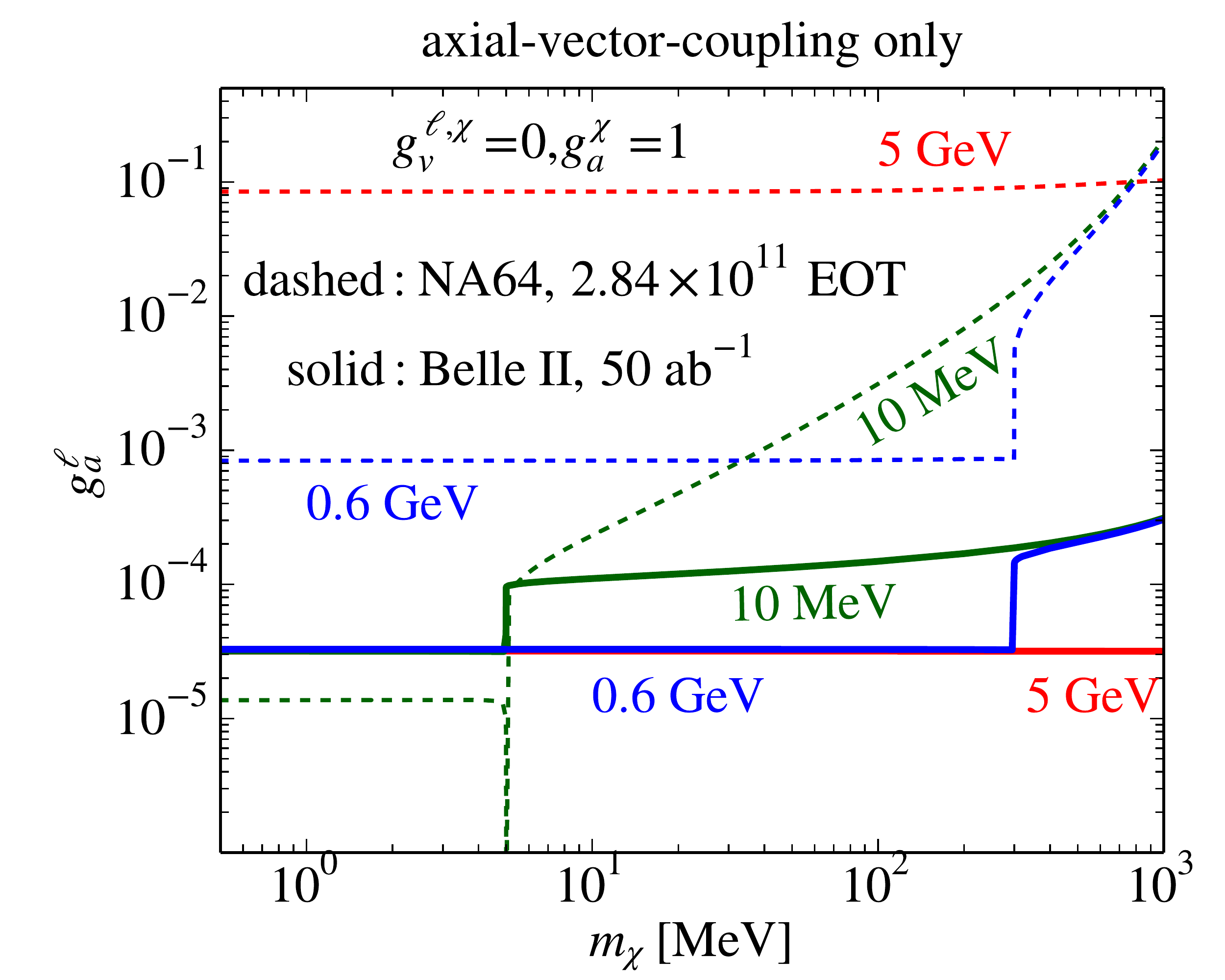}
\caption{NA64 90\% C.L.\ upper bond (dashed lines) on the light mediator models with
vector couplings (left) and axial-vector couplings (right).  
The green, blue, and red lines 
show the constraints {for} $m_{Z'}=10~\rm MeV$, $0.6~\rm GeV$,
and $5~\rm GeV$ respectively.
The Belle II sensitivities (solid lines) 
are the same as %
Fig.\ \ref{fig:monoZV} and Fig.\ \ref{fig:monoZA}.}
\label{fig:NA64LM}
\end{centering}
\end{figure}

The constraints on EFT operators are shown in Fig.\ \ref{fig:NA64EFT}. 
The NA64 constraints with the current EOT data 
on various $\Lambda$'s  
are about one order of magnitude 
smaller than the Belle II expected limits 
with 50 ab$^{-1}$ data. 
The constraints on the light mediators with only vector or axial-vector couplings  
are shown in Fig.\ \ref{fig:NA64LM}. 
For the three different $m_{Z'}$ cases, 
the NA64 constraints with {the current} EOT data are weaker than the 
Belle II expected limits 
with 50 ab$^{-1}$ data, 
except the parameter space where $m_{Z'}=10 ~\rm MeV$ 
and $m_{Z'}>2m_\chi$.

\section{Dark matter relic density}
\label{sec:DMrelic}

In this section, we compare the parameter space of 
the DM operators/models in which the DM relic density (RD)  
is generated by the thermal freeze out 
mechanism, with that probed by the Belle II experiment. 
We compute the thermally averaged DM annihilation cross section 
$\langle \sigma_{{\rm ann}} v_{\rm rel} \rangle$ via \cite{Gondolo:1990dk}
\begin{equation}
\left\langle \sigma_{{\rm ann}} v_{\mathrm{rel}}\right\rangle=
\frac{1}{8 m_\chi^{4} T_F K_{2}^{2}(m_\chi / T_F)} 
\int_{4 m_\chi^{2}}^{\infty} ds \, \sigma_{{\rm ann}} \, 
 K_{1}\left( \frac{\sqrt{s}}{T_F} \right) \left(s-4 m_\chi^{2}\right) 
\sqrt{s} ,
\end{equation}
where 
$\sigma_{{\rm ann}}$ is the total DM annihilation cross section 
as a function of $s$, 
$T$ is the temperature, 
and $K_i$ are the modified Bessel functions of order $i$. 
For the EFT operators, the total DM annihilation cross section 
is $\sigma_{{\rm ann}} = 
\sum_{\ell} \sigma(\bar \chi \chi \to \ell^+ \ell^-)$; 
for the light mediator models, the total DM annihilation cross section 
is 
$\sigma_{{\rm ann}} =  
\sigma (\bar \chi \chi \to Z' Z')
+\sum_{\ell} \sigma  (\bar \chi \chi \to \ell^+ \ell^-)$, 
if kinematically allowed. 
The DM annihilation cross sections for various 
EFT operators and light mediator models 
are given in Appendix \ref{sec:relic}. 
We solve the  freeze-out temperature $T_F$ via 
\cite{Kolb:1990vq, Griest:1990kh, Busoni:2014gta} 
\be
e^{x_F}=\sqrt{45 \over 8} { g_{\chi} m_\chi M_{\rm Pl} \, c (c+2) \langle\sigma_{\rm ann} v_{\rm rel} \rangle \over 2\pi^3 \sqrt{g_*} \sqrt{x_F}},
\label{eq:xF}
\ee
where $x_F = m_\chi/T_F$, $g_\chi$ is the  
degrees of
freedom of the DM, 
$c=1/2$ is a matching constant, 
$M_{\rm Pl}$ is the 
Planck mass, $g_*$ is the 
relativistic degrees of freedom 
in the thermal bath \cite{Steigman:2012nb}.

\begin{figure}[htbp]
\begin{centering}
\includegraphics[width=0.49 \textwidth]{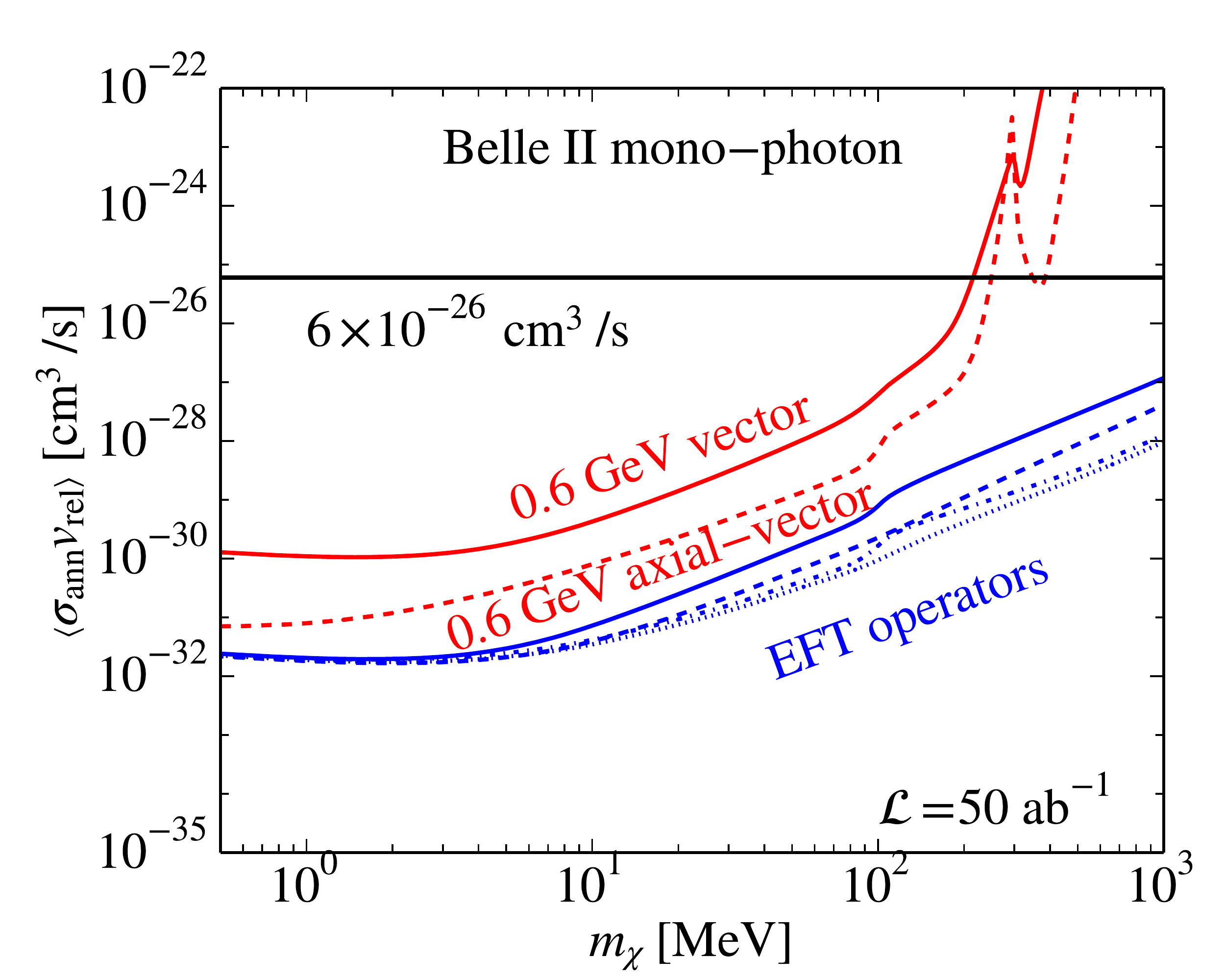}
\includegraphics[width=0.49 \textwidth]{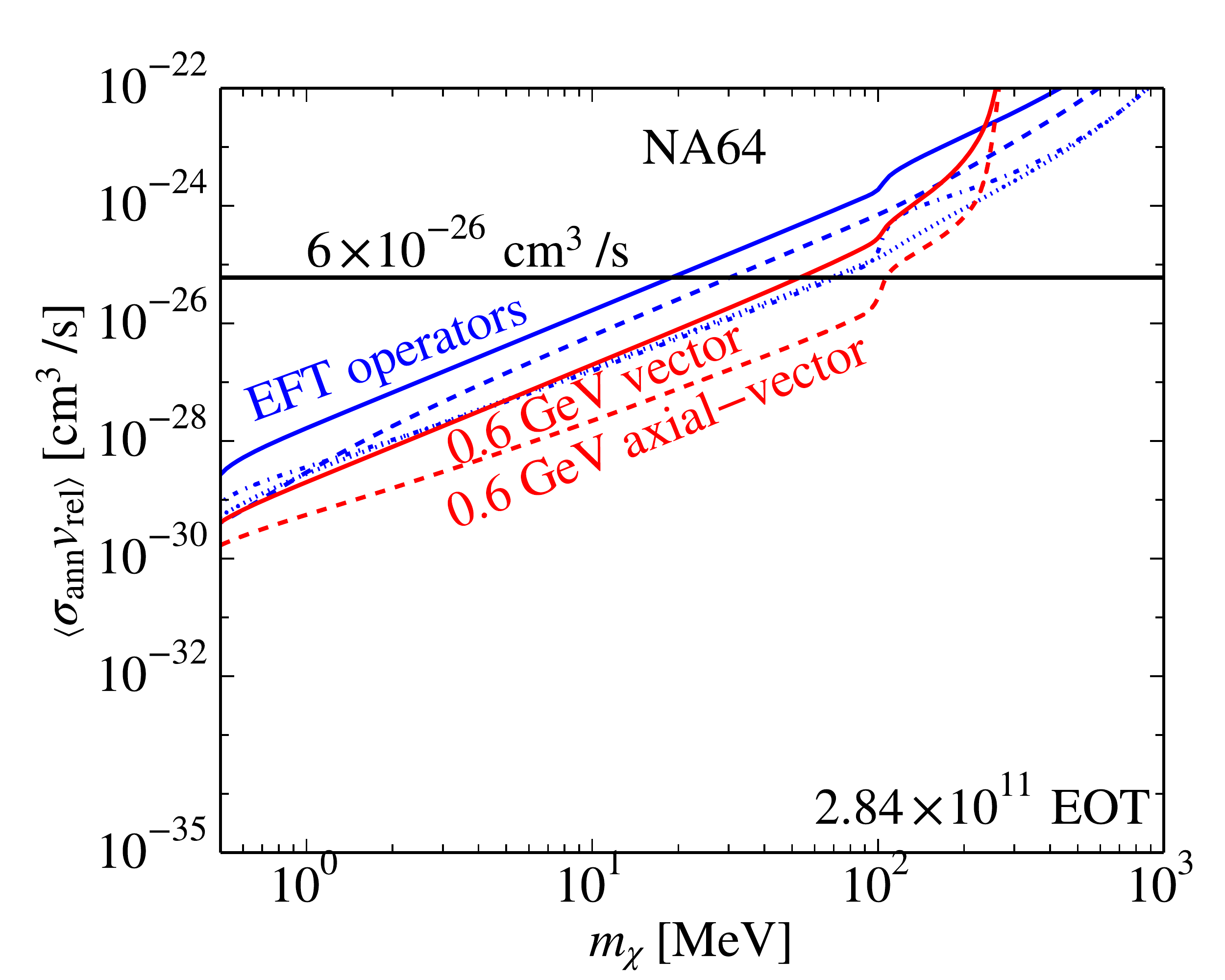}
\caption{Left panel: Belle II 90\% C.L.\ upper bound 
from the mono-photon channel 
with 50 ab$^{-1}$ data on 
the thermally averaged DM annihilation cross section
$\left \langle\sigma v_{\rm rel}\right\rangle$ 
evaluated at the freeze-out temperature.
The blue lines correspond to the limits in Fig.\ (\ref{fig:monoEFT}) 
on EFT operators: 
$O_A$ (dot-dashed),
$O_V$ (solid), 
$O_s$ (dotted), 
and $O_t$ (dashed). 
The red solid and dashed lines 
correspond to the $m_{Z^\prime}=0.6$ GeV models  
in Fig.\ (\ref{fig:monoZV}) and (\ref{fig:monoZA}) respectively. 
Right panel: NA64 90\% C.L. upper bound with $2.84\times 10^{11}$ EOT 
data on the thermally averaged DM annihilation cross section 
evaluated at the freeze-out temperature.
The blue lines correspond to the limits in Fig.\ (\ref{fig:NA64EFT}) on EFT operators. The red solid (dashed) lines 
correspond to the $m_{Z^\prime}=0.6$ GeV 
vector-only (axial-vector-only) model in Fig.\ (\ref{fig:NA64LM}).
The black solid line indicates the thermal cross section for the Dirac DM,  
$6\times 10^{-26}\ \rm cm^3/s$ \cite{Steigman:2012nb}.
}
\label{fig:relic}
\end{centering}
\end{figure}

Fig.\ (\ref{fig:relic}) shows the Belle II sensitivities
(from the mono-photon channel) 
and NA64 constraints on the thermally averaged  
DM annihilation cross section 
$\langle \sigma_{{\rm ann}} v_{\rm rel} \rangle$ 
evaluated at the freeze-out temperature $T_F$, 
for the four EFT operators 
and for two light mediator models. 
The canonical thermal cross section 
($\simeq 6\times 10^{-26}\ \rm cm^3/s$ for Dirac DM \cite{Steigman:2012nb})
in the mass range $m_\chi \lesssim$ GeV 
can be probed by Belle II with 50 ab$^{-1}$ for all the four EFT operators. 
For the two light vector mediator models with 
$m_{Z'} = 0.6$ GeV, the canonical thermal cross section 
in the mass range $m_\chi \lesssim 0.2$ GeV 
can be probed by Belle II with 50 ab$^{-1}$. 
We also find that 
the $2.84\times 10^{11}$ EOT data accumulated at NA64 have already  
probed the canonical DM 
annihilation cross section with  
$m_\chi \lesssim 0.02$ GeV for the EFT operators, 
and with $m_\chi \lesssim 0.04$ GeV for 
the two light vector mediator models with 
$m_{Z'} = 0.6$ GeV.

\section{Summary}
\label{sec:summary}

We investigate the capability of 
the Belle II and the NA64 experiments in probing
the parameter space of the DM models in which 
DM only interacts with charged leptons in the SM.
Our analyses focus on the sub-GeV Dirac DM, which
is less constrained than WIMPs by the current DMDD experiments. 
We consider two different mechanisms to mediate the interactions 
between DM and charged leptons: EFT operators 
and light vector mediators in the MeV-GeV scale.

We compute the
Belle II sensitivities in the mono-photon channel on the EFT operators.
Our analysis shows that  
$\Lambda_{t} \lesssim $ 220 GeV 
can be probed by Belle II with 50 ab$^{-1}$ data, 
and $\Lambda \lesssim $ 280 GeV can be probed 
for $\Lambda_{V}$, $\Lambda_{A}$ and $\Lambda_{s}$. 
We find that the expected Belle II limits
{with 50 ab$^{-1}$ data}
on EFT operators
are of similar size to the LEP limits.
The Belle II and LEP limits for sub-GeV DM 
can be several orders of magnitude 
stronger than the current DMDD limits, 
as well as the white dwarf limit.

The light mediator models can be searched for
both in the mono-photon channel 
and in the di-muon channel at Belle II.
We compute the Belle II sensitivities from both channels
on the light mediator models.
The Belle II mono-photon sensitivities are analyzed for
$m_{Z'}= 10$ MeV, $0.6$ GeV and $5$ GeV. 
The gauge coupling $g^\ell \simeq 3 \times 10^{-5}$ 
(both vector and axial-vector) can be probed by the 
Belle II mono-photon data,
when $m_\chi< m_{Z'}/2$ and $g^\chi=1$.
The di-muon channel is complementary to the 
mono-photon channel 
and sometimes can be much better, 
for example in the parameter where $m_\chi > m_{Z'}/2$. 
Unlike the EFT operators, 
the Belle II sensitivities on the light mediator models
(for example the $m_{Z'} \lesssim 5 ~\rm GeV$ model)
can be several orders of magnitude stronger than the LEP limits, 
in the mono-photon channel.
We also find that the collider limits have a rather weak dependence 
on the mediator mass; 
the DMDD cross section, however, is 
inversely proportional to $m_{Z'}^4$, thus leading to vastly different 
DM-electron interaction cross sections over more than 10 orders of 
magnitude in the parameter space considered in this study.

We also find that the Belle II mono-photon channel  can 
probe the canonical DM thermal annihilation cross section 
for the DM mass $\lesssim$ GeV for the EFT operators
and $\sim 0.2$ GeV for the light mediator models considered.
For both the EFT operators and the light mediator models, 
the Belle II sensitivities can be well below the ``neutrino floor''
expected in silicon detectors in DMDD. 
Thus the Belle II collider can probe the parameter space 
which is beyond the capability of  
current DMDD experiments, unless 
the neutrino floor can be mitigated in a satisfactory way.

We compute the NA64 constraints (with $2.84 \times 10^{11}$ EOT data) 
both on the EFT operators 
and on the light vector mediator models. 
We find that the NA64 can probe interesting parameter space 
of DM models. 
For the EFT operators, the NA64 upper bound on the  new physics scales $\Lambda$  
are typically smaller than the Belle II sensitivities. 
However, for the light mediator models 
(for example, the $m_{Z'} = 10$ MeV model with $m_{\chi}<m_{Z'}/2$),  
NA64 can probe the parameter space that is beyond the 
capability of Belle II. 
Thus, the NA64 and Belle II experiments can be complementary 
in probing sub-GeV DM models. 
The analytic expressions of the DM production cross section 
at the NA64 
for the EFT operators and the vector mediator models 
are provided in the appendix.

\section{Acknowledgement}

We thank 
Yong-Heng Xu, 
Li-Gang Xia, 
and Sandra Robles 
for helpful discussions. 
The work is supported in part  
by the National Natural Science 
Foundation of China under Grant No.\ 
11775109.

\appendix

\section{DM production cross sections at NA64}
\label{sec:dxsec}

We provide the results of the 2-body  
phase space integral for the DM currents and 
the DM production  cross sections  
at NA64 in the WWA in this appendix.
The tensor $\chi^{AB}_{ij}$ for the DM currents can be 
parameterized as follows 
\be
\chi_{i j}^{A B} =  
\sum_{\rm spins} \int d\Phi^{(2)} 
\left[ \bar{u}(k_1) \Gamma_i^{A} v(k_2) \right]
\left[ \bar{u}(k_1) \Gamma_j^{B} v(k_2) \right]^\dagger
\equiv { 1\over 12\pi} \beta_\chi 
T_{ij}^{AB}, 
\label{eq:chiABij}
\ee
where 
$\beta_\chi =\sqrt{1-{4 m_\chi^2 / k^2}}$, and
$\Gamma_{i}^A = \{1, \gamma^5, \gamma^\mu, \gamma^\mu \gamma^5, \sigma^{\mu\nu}/ \sqrt{2} \}$ 
with $\sigma^{\mu\nu}=\frac{i}{2}[\gamma^\mu,\gamma^\nu]$. 
\footnote{Note that there is no explicit Lorentz index for 
$ \{1, \gamma^5\}$;
$A=\mu$ for 
$\{\gamma^\mu, \gamma^\mu \gamma^5 \}$; 
$A=\mu\nu$ for 
$\sigma^{\mu\nu}/ \sqrt{2}$.}
The various two-body phase space integrals   
in Eq.\ \eqref{eq:chiABij} can be 
simplified by the following relations
\be
\int d\Phi^{(2)} F(k_1, k_2) = \frac{\beta_\chi}{96\pi} I (k). 
\label{eq:2bInt}
\ee
We have $I(k)=12$ for $F(k_1, k_2)=1$; 
$I(k)=6k^\mu$ for $F(k_1, k_2)=k_1^\mu$ or $k_2^\mu$;
and 
$I(k)=k^2 \beta_\chi^2 g^{\mu\nu} +(3-\beta_\chi^2)k^\mu k^\nu$ 
for $F(k_1, k_2)=k_1^\mu k_2^\nu$ or $k_2^\mu k_1^\nu$. 
Because 
$( {\chi_{ij}^{A B}} )^\dagger
= \chi_{ji}^{BA} $, 
we have $(  {T_{ij}^{A B}} )^\dagger= 
 T_{ji}^{BA}$. 
The non-zero independent 
${T_{ij}^{A B}}$'s are 
\begin{align}
T_{11}^{AB}=&3k^2 \beta_\chi^2,\\
T_{22}^{AB}=&3k^2,\\
T_{33}^{AB}=&(3-\beta_\chi^2)(k^\mu k^\rho-g^{\mu\rho}k^2),\\
T_{44}^{AB}=& (3-\beta_\chi^2) k^\mu k^\rho-2g^{\mu\rho}k^2\beta_\chi^2,\\
T_{55}^{AB}=&\frac{(3-\beta_\chi^2)}{2}\left(g^{\mu\sigma}k^\nu k^\rho-g^{\mu\rho}k^\sigma k^\nu +g^{\nu\rho}k^\mu k^\sigma-g^{\nu\sigma}k^\mu k^\rho \right) \nonumber \\ 
&
- \frac{k^2\beta_\chi^2}{2}\left(g^{\mu\sigma} g^{\nu\rho}-g^{\nu\sigma} g^{\mu\rho}\right),\\
T_{24}^{AB}=&3\sqrt{k^2(1-\beta_\chi^2)} k^\rho,\\
T_{35}^{AB}=&\frac{3i}{\sqrt{2}}\sqrt{k^2(1-\beta_\chi^2)} (k^\rho g^{\sigma \mu}-k^\sigma g^{\rho \mu}),
\label{eq:Tij}
\end{align}
where we have always used Lorentz indices ``$\mu\nu$'' for ``$A$'' 
and ``$\rho \sigma$'' for ``$B$''. 
The {$\chi^{AB}_{ij}$} 
of $O_s$,  $O_V$, and $O_A$,  
are 
$\chi_{1 1}^{A B}$, $\chi_{3 3}^{A B}$, and $\chi_{4 4}^{A B}$ respectively.
The {$\chi^{AB}_{ij}$} for the light mediator model with the vector (axial-vector) couplings 
is $\chi_{3 3}^{A B}$ ($\chi_{4 4}^{A B}$). 
For the $O_t$ operator, one has to consider all the 
$\chi_{ij}^{AB}$'s combined with the coefficients in the Fierz 
transformation.

The DM differential cross section in NA64 
can be computed in WWA. 
The relevant integrand in Eq.\ \eqref{eq:WWA} 
contracted with $\chi_{ij}^{A B}$ for the EFT operators 
can be parameterized as 
\be 
\frac{1}{\tilde{u}^2} \sum_{i,j} \chi^{AB}_{ij}
\overline{ {\tilde{\mathcal{M}}}_{{i},A} {\tilde{\mathcal{M}}}_{{j},B}^{\dagger}} \Bigg|_{t=t_{\min}}
=  
 \frac{{e^2} \beta_\chi}{6 \pi \Lambda^4} 
 {\sum_{n=1,2,3,4} \frac{C_n}{\tilde{u}^n}}
\label{eq:chiMM}
\ee
We have $C_1=0$ for the $O_V$, $O_A$, and $O_s$ cases. 
The $C_i$ ($i=2,3,4$) for $O_V$ are 
\begin{align}
C_2&=-\frac{2\left(k^2+2 m_\chi^2\right)(x^2-2x+2) }{ (x-1)},\\
C_3&=4x \left(k^2+2 m_\chi^2\right)\left(2 m_e^2+k^2\right),\\
C_4&=-4 \left(k^2+2 m_\chi^2\right) \left(2 m_e^2+k^2\right)  \left(k^2 (x-1)-x^2 m_e^2\right).
\end{align}
The $C_i$ ($i=2,3,4$) for $O_A$ are 
\begin{align}
C_2&=-\frac{ 4 x^2 m_e^2 \left(k^2+2 m_{\chi }^2\right)+2k^2 \left(x^2-2 x+2\right) \left(k^2-4 m_{\chi }^2\right)}{k^2 (x-1)},\\
C_3&=4x  \left(-4 m_e^2 \left(k^2-7 m_{\chi }^2\right)+k^4-4 k^2 m_{\chi }^2\right),\\
C_4&=-4 (-4 m_e^2 \left(k^2-7 m_{\chi }^2\right)+k^4-4 k^2 m_{\chi }^2) \left(k^2 (x-1)-x^2 m_e^2\right).
\end{align}
The $C_i$ ($i=2,3,4$) for $O_s$ are 
\begin{align}
C_2 &=- \frac{3 x^2 (k^2-4 m_{\chi }^2) }{  2(x-1)}, \\ 
C_3 &=3 x(k^2-4 m_{\chi }^2) (k^2-4 m_e^2), \\
C_4 &= -3 (k^2-4 m_{\chi }^2)(k^2-4 m_e^2)  (k^2 (x-1)-x^2 m_e^2).
\end{align}
The $C_i$ ($i=1,2,3,4$) for $O_t$ are 
\begin{align}
C_1&= { x (2m_\chi^2+k^2) \over 4 k^2 (x-1)},\\
C_2&={\frac{-1}{4 k^2  (x-1)}} 
\left[ 6 k^2 x^2 m_e m_\chi+x^2 m_e^2 (k^2+2 m_\chi^2)+2 k^2 (x^2-x+1) (k^2-m_\chi^2 ) 
\right],\\
C_3&=x (6 k^2 m_e m_\chi-m_e^2 (k^2-16 m_\chi^2 )+k^4-k^2 m_\chi^2),\\
C_4&=-(6 k^2 m_e m_\chi-m_e^2 (k^2-16 m_\chi^2)+k^4-k^2 m_\chi^2) (k^2 (x-1)-x^2 m_e^2).
\end{align}

Therefore, 
the differential DM production cross section at NA64 in the WWA 
(defined in Eq.\ \eqref{eq:NA64xsec}) can be written as  
\be
\frac{ d\sigma (eN\to eN \bar\chi \chi)}{dx dk^2}
=  
\frac{\alpha^2 \zeta (1-x)\beta_k \beta_\chi}{48 \pi^3 \Lambda^4}
\left[ C_1 \ln (\tilde{u})
-\sum_{n=2,3,4}\frac{C_n}{(n-1) \tilde{u}^{n-1}}
\right]
\Bigg|_{\tilde{u}_{\rm min}}^{\tilde{u}_{\rm max}}.
\ee
Because the cross section is usually dominated 
by the small angle emissions $\theta_{k}\ll 1$, 
one can use the approximation 
$\tilde{u}_{\min} \to - \infty$ to 
further simplify the expressions 
\cite{Liu:2017htz} 
if $C_1 =0$. 
For the light mediators with vector and axial-vector couplings, 
one just needs to make the following replacement 
\be
\frac{1}{\Lambda^{4}} \to 
\frac{g_\ell^2 g_\chi^2} { (k^2-m_{Z'}^2)^2 + m_{Z'}^2 \Gamma_{Z'}^2 }.
\ee

\section{Mono-photon cross sections of the EFT operators at Belle II}
\label{sec:xsecEFT}

The differential cross section of $e^+ e^- \to \chi \bar{\chi} \gamma$ for 
the four EFT operators in our analysis
have been computed in Ref.\ \cite{Chae:2012bq}.
Here we collect the expressions of the 
cross section for the 
$e^+ e^- \to \chi \bar{\chi} \gamma$ process given in 
Ref.\ \cite{Chae:2012bq}.
For the vector case, the cross section is  
\be
\frac{d\sigma}{dE_\gamma d\cos\theta_\gamma}=\frac{\alpha\sqrt{s}}{12 \pi^{2}\Lambda_V^{4}} \frac{\left(1-z+2 \mu^{2}\right)}{z \sin^2\theta_\gamma} \sqrt{\frac{1-z-4 \mu^{2}}{1-z}}
{\left[(z-2)^2 + z^2 \cos^2\theta_\gamma \right].}
\label{eq: Vdiffcrosssection}
\ee
For the axial-vector case, the cross section is
\be
\frac{d\sigma}{dE_\gamma d\cos\theta_\gamma}=\frac{\alpha\sqrt{s}}{12 \pi^{2}\Lambda_A^{4}}\frac{(1-z)}{z \sin^2\theta_\gamma}\left(\frac{1-z-4 \mu^{2}}{1-z}\right)^{3 / 2}
{\left[(z-2)^2 + z^2 \cos^2\theta_\gamma \right]}.
\label{eq: Adiffcrosssection}
\ee
For the ``s-channel'' scalar case, the cross section is  
\be
\frac{d\sigma}{dE_\gamma d\cos\theta_\gamma}=\frac{\alpha\sqrt{s}}{8 \pi^{2}\Lambda_s^{4}}\frac{(1-z)}{z\sin^2\theta_\gamma}\left(\frac{1-z-4 \mu^{2}}{1-z}\right)^{3 / 2}\left[2(1-z)+z^{2}\right].
\label{eq: SSdiffcrosssection}
\ee
For the ``t-channel'' scalar case, the cross section is  
\begin{align}
\frac{d\sigma}{dE_\gamma d\cos\theta_\gamma}=
&\frac{\alpha\sqrt{s}}{192 \pi^{2}\Lambda_t^{4}}\frac{1}{z\sin^2\theta_\gamma} \sqrt{\frac{1-z-4 \mu^{2}}{1-z}} \left[\left(2-z+\frac{2 \mu^{2}z}{1-z}\right)\left(3 z^{2}-6 z+4\right)-8\mu^{2}
\right. \nonumber\\
& {\left.+ \left(1-z+2 \mu^{2}\right) \left( 2 (z-2)^2 + \left( 2 z^2 - \frac{1}{1-z} \right) \cos^2\theta_\gamma \right) \right].} 
\label{eq: STdiffcrosssection}
\end{align}
Here $E_\gamma$ and $\theta_\gamma$ are the energy and the 
polar angle (with respect to the direction of the initial electron) 
of the final state photon in the CM frame, 
$s$ is the square of the center of mass energy, $m_\chi$ is the mass of the dark matter,
$z=2E_\gamma/ \sqrt{s}$, and 
$\mu=m_\chi/ \sqrt{s}$.

\section{Matrix elements for EFT operators in DMDD}
\label{app:ME}

The matrix elements of the four EFT operators, given in 
Eqs.\ (\ref{eq:EFT1}, \ref{eq:EFT2}, \ref{eq:EFT3}, \ref{eq:EFT4}), 
in the DMDD experiments are given by 
\begin{eqnarray}
\label{eq:M2EFT1}
\Lambda_s^4 \overline{\left|\mathcal{M}_{\chi e}\right|^{2}}
&=&(t-4 m_e^2)(t-4 m_\chi^2),  \\
\label{eq:M2EFT2}
\Lambda^4_t \overline{\left|\mathcal{M}_{\chi e}\right|^{2}}
&=&((m_e-m_\chi)^2-s)^2,  \\
\label{eq:M2EFT3}
\Lambda^4_V \overline{\left|\mathcal{M}_{\chi e}\right|^{2}}
&=&4 (m_e^2+m_\chi^2-s)^2+4 s t +2 t^2,  \\
\label{eq:M2EFT4}
\Lambda^4_A \overline{\left|\mathcal{M}_{\chi e}\right|^{2}}
&=&2 (2m_e^4 +4 m_e^2 (5 m_\chi^2-s-t)+2 m_\chi^4 -4 m_\chi^2 (s+t)+2 s^2+2 s t +t^2), 
\end{eqnarray}
where $s$ and $t$ are Mandelstam variables. 
{For the non-relativistic dark matter,
one has $s\simeq (m_e+m_\chi)^2$
and $t \simeq -q^2$.
The typical momentum transfer in the DM-electron scattering 
is $q_{\rm typ}  \sim Z_{\rm eff} \, \alpha \, m_e $,  
where $Z_{\rm eff} =12.4$ for the outermost shell electron of Xenon atom 
\cite{Essig:2015cda, Bloch:2020uzh,clementi1967atomic}.
We display the $q$ dependence of the form factor $F_{\rm DM}(q)$ 
for all the four EFT operators considered, 
which shows that the deviation from $F_{\rm DM}(q)=1$ is 
less than  
0.5\% in the range of $q \lesssim Z_{\rm eff} \alpha m_e$ for 
all the EFT operators. 
Thus, it is a good approximation to use 
$F_{\rm DM}(q)=1$ in the DMDD calculation
for these EFT operators.
}

\begin{figure}[htbp]
\begin{centering}
\includegraphics[width=0.49\textwidth]{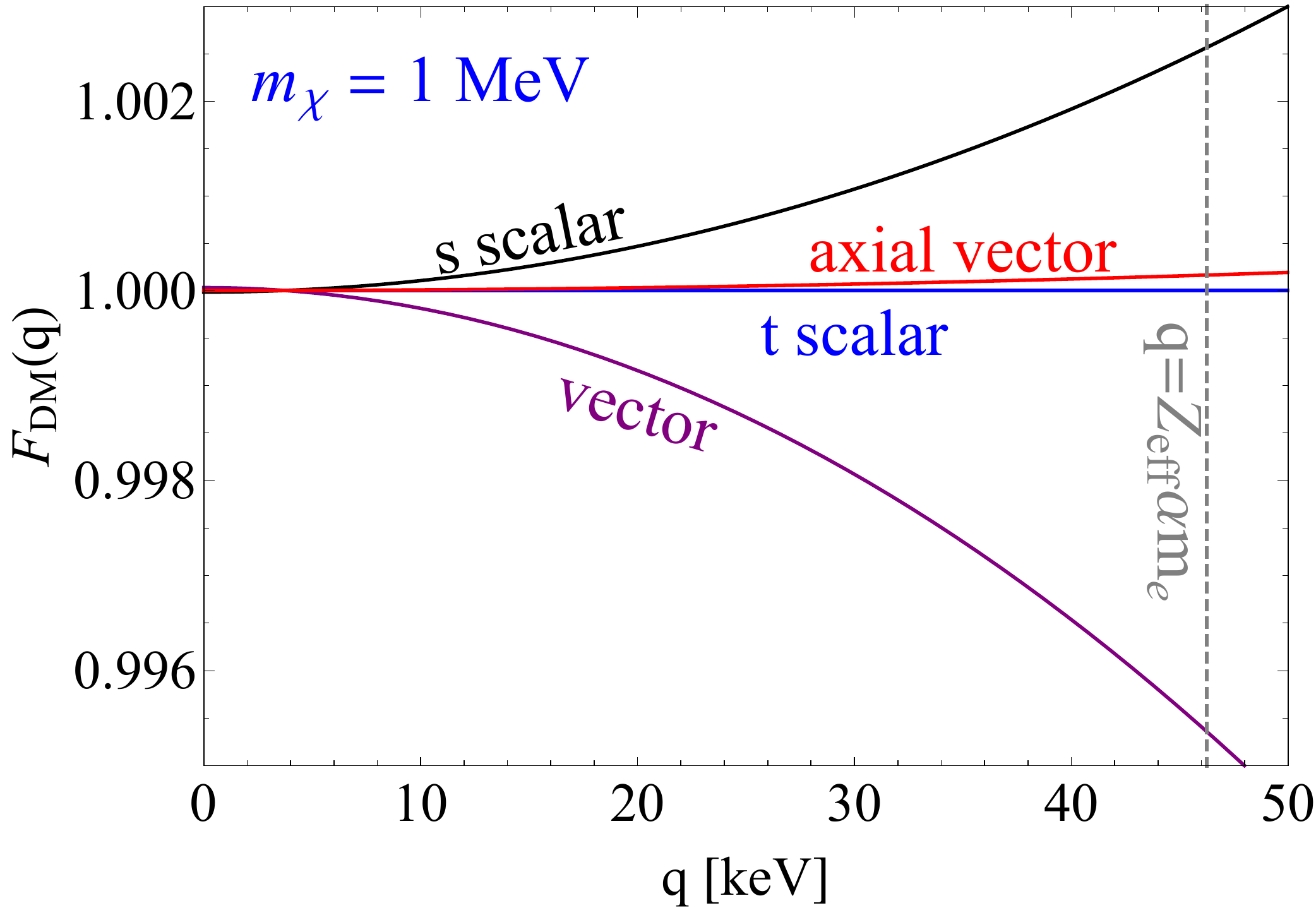}
\caption{{The form factor $F_{\rm DM}(q)$ as a function of $q$ 
for four EFT operators: 
vector (purple), 
axial-vector (red), 
s-channel scalar (black), 
and t-channel scalar (blue). 
The typical $q$ range in DMDD with Xe target is $q \lesssim Z_{\rm eff} \, \alpha \, m_e$
where $Z_{\rm eff}=12.4$. We use $m_\chi=1 ~\rm MeV$ here. 
}}
\label{fig:form}
\end{centering}
\end{figure}

\section{LEP analysis}
\label{sec:LEP}

In this section, we describe our LEP analysis,  
which closely follows the analysis in Ref.\ \cite{Fox:2011fx}. 
To properly take into account the initial state radiation effect, 
we use {\sc CalcHEP}
\cite{Pukhov:2004ca} to generate 
$10^5$ events for each model point for the process 
of $e^+ e^- \to \chi \bar{\chi} \gamma$ 
at $\sqrt{s}=200$ GeV (with 100 GeV for each beam).\footnote{Ref.\ 
\cite{Fox:2011fx} found that using $\sqrt{s}=200$ GeV  
only introduces a small deviation from the full analysis.} 
The DELPHI detector has three main 
electromagnetic calorimeters: 
the Small angle TIle Calorimeter (STIC), 
the Forward ElectroMagnetic Calorimeter (FEMC), 
and the High density Projection Chamber (HPC). 
{We smear the photon events by using 
the gaussian distributions 
with the energy resolutions given in 
Table \ref{tab:DELdetector}.  
Following Ref.\ \cite{Fox:2011fx}, 
an additional Lorentzian energy 
smearing   
\be
L(E) = \frac{1}{\pi}\frac{\Gamma/2}{(E - E_\gamma)^2+(\Gamma/2)^2},
\ee
 where $\Gamma = 0.052 E_\gamma$ is further performed. 
We analyzed the events with the 
preselection cuts shown 
in Table \ref{tab:DELdetector}.

\begin{table}[htbp]
\centering
\begin{tabular}{lclcl} 
\hline 
& $\sigma_{E_\gamma} / E_\gamma$ & Preselection cuts\\
\hline
\multirow{3}*{STIC} & \multirow{3}*{$0.0152 \oplus(0.135 / \sqrt{E_\gamma})$} &(i) $x_\gamma>0.3$\\
~&~& (ii) $\theta_\gamma>9.2^\circ-9^\circ x_\gamma$ when  $3.8^{\circ}<\theta<8^{\circ}$ \\
~&~& $180^\circ -\theta_\gamma>9.2^\circ-9^\circ x_\gamma$ when $172^{\circ}<\theta<176.2^{\circ}$ \\
\hline
\multirow{3}*{FEMC} & \multirow{3}*{$0.03 \oplus(0.12 / \sqrt{E_\gamma}) \oplus(0.11 / E_\gamma)$ } &(i) $x_\gamma>0.1$\\
~&~& (ii) $\theta_\gamma>28^\circ-80^\circ x_\gamma$ when  $12^{\circ}<\theta<32^{\circ}$ \\
~&~& $180-\theta_\gamma>28^\circ-80^\circ x_\gamma$ when $148^{\circ}<\theta<168^{\circ}$\\
\hline
\multirow{2}*{HPC} & \multirow{2}*{ $0.043 \oplus(0.32 / \sqrt{E_\gamma})$ } &(i) $x_\gamma>0.06$\\
~&~& (ii) $45^{\circ}<\theta<135^{\circ}$\\
\hline
\end{tabular}
\caption{Preselection cuts and 
energy resolution for 
the three {sub-detectors} in the 
electromagnetic calorimeters in DELPHI: 
{STIC, FEMC, and HPC \cite{DELPHI:2003dlq}.} 
Here $E_\gamma$ is in unit of GeV, and 
$x_\gamma$=$E_\gamma/E_{\rm beam}$. 
}
\label{tab:DELdetector}
\end{table}

We further take into account other 
efficiency factors beyond the 
detector cuts given in Table \ref{tab:DELdetector}, 
as analyzed in Ref.\ \cite{Fox:2011fx}. 
They include the trigger efficiency, 
the analysis efficiency, 
and an overall factor of 90\%, 
which is found to be necessary for the simulations in Ref.\  \cite{Fox:2011fx} 
to match the simulations in Ref.\ \cite{DELPHI:2003dlq}. 
For HPC, the trigger efficiency 
is a linear interpolation function with 
52\% at $E_\gamma=6~\rm GeV$,
77\% at $E_\gamma=30~\rm GeV$, 
and 84\% at $E_\gamma=100 ~\rm GeV$;
the analysis efficiency
is a linear interpolation function with 
41\% at $E_\gamma=6~\rm GeV$ 
and 78\% at $E_\gamma > 80~\rm GeV$ \cite{Fox:2011fx}.
For FEMC, the trigger efficiency 
is a linear interpolation function with 
93\% at $E_\gamma=10 ~\rm GeV$
and 100\% at $E_\gamma>15 ~\rm GeV$;
the analysis efficiency is a linear interpolation function with 
51\% at $E_\gamma=10 ~\rm GeV$
and 67\% at $E_\gamma=100 ~\rm GeV$ \cite{Fox:2011fx}.
For STIC, the product of {the} trigger efficiency {and the} 
analysis efficiency is 48\% 
{for $E_\gamma>$30 GeV.} \cite{Fox:2011fx}.

We bin the data in 19 bins with $ 0.05<x_\gamma<1$, 
where $x_\gamma=E_\gamma  / E_{\rm beam}$ 
and compute the $\chi^2$ via  
\be
\chi^2=\sum_{i=1}^{19}\frac{(N^s_i+N^b_i-N^{o}_i)^2}{\sigma_i^2}, 
\ee
where 
$N^s_i$ is the number of signal events, 
$N^b_i$ is the number of background events, 
 $N^{o}_i$ is the number of observed data events, 
 and 
 $\sigma_i$ is the uncertainty. 
Here the dominant background process is the 
$e^+ e^- \to \nu \nu \gamma$ process. 
We take $N^b_i$, $N^{ o}_i $, and $\sigma_i$ from 
Refs.\ \cite{Fox:2011fx, DELPHI:2003dlq}.
The LEP limits {at {90\%} CL} on light mediator models
are obtained by $\chi^2/\text{dof} =27.2 /19$.

\section{Confidence Level Limits}
\label{sec:CL}

We provide the derivation for the two confidence level 
(denoted as C.L.\ or CL) limits 
used in our analysis which are based on a non-zero background 
and a null background respectively. 
The CL is usually defined via 
$\rm CL=1-\alpha$ \cite{Lista:2017jsy}.

For the analysis with a non-zero background (e.g., the Belle II limits in our analysis), 
we assume 
a Gaussian distribution for 
the likelihood distribution ${\cal L}(N_s)$
for the signal events 
$N_s$ as follows 
\be
\mathcal{L}\left(N_{s}\right)=\frac{1}{\sqrt{2\pi}\sigma}\exp \left(-\frac{N_{s}^{2}}{2 \sigma^{2}}\right), 
\ee
where $\sigma = \sqrt{N_b}$ with $N_b$ being 
the number of the background events. In this case, 
$\alpha$ is given by \cite{Lista:2017jsy}
\be
\frac{\rm \alpha}{2} =\int_{N_s^{\rm up}}^{+\infty} \mathcal{L}\left(N_{s}\right) d N_{s},
\ee
where 
$N_s^{\rm up}$ is the upper bound on the 
signal events $N_s$ with the given confidence level. 
Therefore, the 90\% CL limit 
in this case
corresponds to $N^{\rm up}_s \simeq \sqrt{2.71} \sqrt{N_b}$.

For the analysis with a null background (e.g., the NA64 limits in our analysis), 
we assume a Poisson distribution for 
${\cal L \left(N_{\rm s}\right)}$, which is 
\be
\mathcal{L}\left(N_{\rm s}\right)=e^{ -{N_s}}.
\ee
In this case, $\alpha$ is given by \cite{Lista:2017jsy}
\be
\alpha =\int_{N_s^{\rm up}}^{+\infty} \mathcal{L}\left(N_{s}\right) d N_{s}. 
\ee
Therefore, the 90\% CL limit in this case 
corresponds to $N^{\rm up}_s =2.3$.

\section{DM annihilation cross sections}
\label{sec:relic}

For the $Z'$ model with only vector couplings, the 
DM annihilation cross sections are
\begin{align}
\sigma(\chi\bar{\chi}\to \ell \ell)=&\frac{(g_v^\ell g_v^\chi)^{2}  
\beta_\ell\left(2 m_\ell^{2}+s\right)\left(2 m_\chi^{2}+s\right)}
{12 \pi s \beta_{d}\left(\beta_{d}^{2} 
m_{Z^\prime}^{2}s+\left(m_{Z^\prime}^{2}-s\right)^{2}\right)},\\
\sigma(\chi\bar{\chi}\to Z^\prime Z^\prime)=&\frac{(g_v^{\chi })^4 
\beta_{Z'}}{4 \pi  s\beta_{d}}\left(-
\frac{\left(2 m_{\chi }^2+m_{Z'}^2\right)^2+\beta_{Z'}^{2}m_{\chi }^2 s+m_{Z'}^4}{\beta_{Z'}^{2}m_{\chi }^2 s+m_{Z'}^4}
\right. \nonumber \\& 
\left. +\frac{2 \left(4 m_{\chi }^2 \left(s-2 m_{Z'}^2-2m_{\chi }^2\right)
+4 m_{Z'}^4+s^2\right)}{\left(s-2 m_{Z'}^2\right) \beta_{Z'}\beta_{d}s}  \text{arccoth} \left(\frac{s-2 m_{Z'}^2}{\beta_{Z'}\beta_{d}s}\right)
\right),
\end{align}
where $m_\ell$ is the lepton mass, and 
\be
\beta_\ell=\sqrt{1-\frac{4m_\ell^2}{s}}, \, \, 
\beta_{d}=\sqrt{1-\frac{4m_\chi^2}{s}}, \, \,
\beta_{Z'}=\sqrt{1-\frac{4m_{Z'}^2}{s}}. 
\ee
For the $Z'$ model with only axial-vector couplings, the 
DM annihilation cross sections are 
\begin{align}
&\sigma(\chi\bar{\chi}\to \ell \ell)=\frac{(g_a^\ell g_a^\chi)^{2}\beta_\ell\left(-4s(m_\ell^2+m_\chi^2)+28m_\ell^2m_\chi^2+s^2\right)}{12 \pi s \beta_{d}\left(\beta_{d}^{2}m_{Z^\prime}^{2}s+\left(m_{Z^\prime}^{2}-s\right)^{2}\right)},\\
&\sigma(\chi\bar{\chi}\to Z^\prime Z^\prime)=\frac{(g_a^{\chi })^4 \beta_{Z'}}{4 \pi  s m_{Z'}^4\beta_{d}}\left(2 \beta_{Z'}^{2}m_{\chi }^2 s-m_{Z'}^4 -\frac{m_{Z'}^4 \left(m_{Z'}^2-4 m_{\chi }^2\right)^2}{\beta_{Z'}^{2}m_{\chi }^2 s+m_{Z'}^4}
\right. \nonumber \\
& \left. +\frac{2 \left(m_{Z'}^4 \left(4 m_{Z'}^4+s^2\right)+4\beta_{Z'}^{2}m_{\chi }^2 s\left(s\left(m_{Z'}^2-m_\chi^2\right)+m_{Z'}^4\right)\right) }{\left(s-2 m_{Z'}^2\right) \beta_{Z'}\beta_{d}s} \text{arccoth} \left(\frac{s-2 m_{Z'}^2}{\beta_{Z'}\beta_{d}s}\right)
\right).
\end{align}
The DM annihilation cross sections for the EFT operators are 
\begin{align}
\sigma_{V}(\chi\bar{\chi}\to \ell \ell)&=\frac{\beta_\ell\left(2 m_\ell^{2}+s\right)\left(2 m_\chi^{2}+s\right)}{12 \pi s \Lambda_V^4  \beta_{d}},\\
\sigma_{A}(\chi\bar{\chi}\to \ell \ell)&=\frac{\beta_\ell \left(-4s(m_\ell^2+m_\chi^2)+28m_\ell^2m_\chi^2+s^2\right)} {12 \pi \Lambda_A^4 s \beta_{d}},\\
 \sigma_s(\chi\bar{\chi}\to \ell\ell)&=\frac{\beta_\ell\beta_{d}s}{16 \pi \Lambda_s^4  },\\
\sigma_t(\chi\bar{\chi}\to \ell \ell)&=\frac{\beta_\ell\left(-s(m_\ell^2+6m_\ell m_\chi+m_\chi^2)+16m_\ell^2m_\chi^2+s^2\right)}{48 \pi \Lambda_t^4  s \beta_{d}}.
\end{align}

\bibliography{ref.bib}{}

\providecommand{\href}[2]{#2}\begingroup\raggedright\begin{thebibliography}{10}

\bibitem{Bertone:2004pz}
G.~Bertone, D.~Hooper, and J.~Silk, ``{Particle dark matter: Evidence,
  candidates and constraints},''
  \href{https://dx.doi.org/10.1016/j.physrep.2004.08.031}{Phys.\  Rept.\
  {\bfseries 405} (2005) 279--390} {\ttfamily
  [\href{https://arxiv.org/abs/hep-ph/0404175}{hep-ph/0404175}]}.

\bibitem{Feng:2010gw}
J.~L.~Feng, ``{Dark Matter Candidates from Particle Physics and Methods of
  Detection},''
  \href{https://dx.doi.org/10.1146/annurev-astro-082708-101659}{Ann.\  Rev.\
  Astron.\  Astrophys.\  {\bfseries 48} (2010) 495--545} {\ttfamily
  [\href{https://arxiv.org/abs/1003.0904}{arXiv:1003.0904}]}.

\bibitem{Roszkowski:2017nbc}
L.~Roszkowski, E.~M.~Sessolo, and S.~Trojanowski, ``{WIMP dark matter
  candidates and searches\textemdash{}current status and future prospects},''
  \href{https://dx.doi.org/10.1088/1361-6633/aab913}{Rept.\  Prog.\  Phys.\
  {\bfseries 81} (2018) 066201} {\ttfamily
  [\href{https://arxiv.org/abs/1707.06277}{arXiv:1707.06277}]}.

\bibitem{Schumann:2019eaa}
M.~Schumann, ``{Direct Detection of WIMP Dark Matter: Concepts and Status},''
  \href{https://dx.doi.org/10.1088/1361-6471/ab2ea5}{J.\  Phys.\  G {\bfseries
  46} (2019) 103003} {\ttfamily
  [\href{https://arxiv.org/abs/1903.03026}{arXiv:1903.03026}]}.

\bibitem{Essig:2017kqs}
R.~Essig, T.~Volansky, and T.-T.~Yu, ``{New Constraints and Prospects for
  sub-GeV Dark Matter Scattering off Electrons in Xenon},''
  \href{https://dx.doi.org/10.1103/PhysRevD.96.043017}{Phys.\  Rev.\  D
  {\bfseries 96} (2017) 043017} {\ttfamily
  [\href{https://arxiv.org/abs/1703.00910}{arXiv:1703.00910}]}.

\bibitem{Aprile:2019xxb}
{\bfseries XENON} Collaboration, ``{Light Dark Matter Search with Ionization
  Signals in XENON1T},''
  \href{https://dx.doi.org/10.1103/PhysRevLett.123.251801}{Phys.\  Rev.\
  Lett.\  {\bfseries 123} (2019) 251801} {\ttfamily
  [\href{https://arxiv.org/abs/1907.11485}{arXiv:1907.11485}]}.

\bibitem{DarkSide:2018ppu}
{\bfseries DarkSide} Collaboration, ``{Constraints on Sub-GeV
  Dark-Matter\textendash{}Electron Scattering from the DarkSide-50
  Experiment},''
  \href{https://dx.doi.org/10.1103/PhysRevLett.121.111303}{Phys.\  Rev.\
  Lett.\  {\bfseries 121} (2018) 111303} {\ttfamily
  [\href{https://arxiv.org/abs/1802.06998}{arXiv:1802.06998}]}.

\bibitem{PandaX-II:2021nsg}
{\bfseries PandaX-II} Collaboration, ``{Search for Light Dark Matter-Electron
  Scatterings in the PandaX-II Experiment},''
  \href{https://dx.doi.org/10.1103/PhysRevLett.126.211803}{Phys.\  Rev.\
  Lett.\  {\bfseries 126} (2021) 211803} {\ttfamily
  [\href{https://arxiv.org/abs/2101.07479}{arXiv:2101.07479}]}.

\bibitem{Barak:2020fql}
{\bfseries SENSEI} Collaboration, ``{SENSEI: Direct-Detection Results on
  sub-GeV Dark Matter from a New Skipper-CCD},''
  \href{https://dx.doi.org/10.1103/PhysRevLett.125.171802}{Phys.\  Rev.\
  Lett.\  {\bfseries 125} (2020) 171802} {\ttfamily
  [\href{https://arxiv.org/abs/2004.11378}{arXiv:2004.11378}]}.

\bibitem{Aguilar-Arevalo:2019wdi}
{\bfseries DAMIC} Collaboration, ``{Constraints on Light Dark Matter Particles
  Interacting with Electrons from DAMIC at SNOLAB},''
  \href{https://dx.doi.org/10.1103/PhysRevLett.123.181802}{Phys.\  Rev.\
  Lett.\  {\bfseries 123} (2019) 181802} {\ttfamily
  [\href{https://arxiv.org/abs/1907.12628}{arXiv:1907.12628}]}.

\bibitem{Arnaud:2020svb}
{\bfseries EDELWEISS} Collaboration, ``{First germanium-based constraints on
  sub-MeV Dark Matter with the EDELWEISS experiment},''
  \href{https://dx.doi.org/10.1103/PhysRevLett.125.141301}{Phys.\  Rev.\
  Lett.\  {\bfseries 125} (2020) 141301} {\ttfamily
  [\href{https://arxiv.org/abs/2003.01046}{arXiv:2003.01046}]}.

\bibitem{Agnese:2018col}
{\bfseries SuperCDMS} Collaboration, ``{First Dark Matter Constraints from a
  SuperCDMS Single-Charge Sensitive Detector},''
  \href{https://dx.doi.org/10.1103/PhysRevLett.121.051301}{Phys.\  Rev.\
  Lett.\  {\bfseries 121} (2018) 051301} {\ttfamily
  [\href{https://arxiv.org/abs/1804.10697}{arXiv:1804.10697}]}. [Erratum:
  Phys.Rev.Lett. 122, 069901 (2019)].

\bibitem{Bell:2021fye}
N.~F.~Bell, G.~Busoni, M.~E.~Ramirez-Quezada, S.~Robles, and M.~Virgato,
  ``{Improved treatment of dark matter capture in white dwarfs},''
  \href{https://dx.doi.org/10.1088/1475-7516/2021/10/083}{JCAP {\bfseries 10}
  (2021) 083} {\ttfamily
  [\href{https://arxiv.org/abs/2104.14367}{arXiv:2104.14367}]}.

\bibitem{Bertone:2007ae}
G.~Bertone and M.~Fairbairn, ``{Compact Stars as Dark Matter Probes},''
  \href{https://dx.doi.org/10.1103/PhysRevD.77.043515}{Phys.\  Rev.\  D
  {\bfseries 77} (2008) 043515} {\ttfamily
  [\href{https://arxiv.org/abs/0709.1485}{arXiv:0709.1485}]}.

\bibitem{McCullough:2010ai}
M.~McCullough and M.~Fairbairn, ``{Capture of Inelastic Dark Matter in White
  Dwarves},'' \href{https://dx.doi.org/10.1103/PhysRevD.81.083520}{Phys.\
  Rev.\  D {\bfseries 81} (2010) 083520} {\ttfamily
  [\href{https://arxiv.org/abs/1001.2737}{arXiv:1001.2737}]}.

\bibitem{Hooper:2010es}
D.~Hooper, D.~Spolyar, A.~Vallinotto, and N.~Y.~Gnedin, ``{Inelastic Dark
  Matter As An Efficient Fuel For Compact Stars},''
  \href{https://dx.doi.org/10.1103/PhysRevD.81.103531}{Phys.\  Rev.\  D
  {\bfseries 81} (2010) 103531} {\ttfamily
  [\href{https://arxiv.org/abs/1002.0005}{arXiv:1002.0005}]}.

\bibitem{Amaro-Seoane:2015uny}
P.~Amaro-Seoane, J.~Casanellas, R.~Sch\"odel, E.~Davidson, and J.~Cuadra,
  ``{Probing dark matter crests with white dwarfs and IMBHs},''
  \href{https://dx.doi.org/10.1093/mnras/stw433}{Mon.\  Not.\  Roy.\  Astron.\
  Soc.\  {\bfseries 459} (2016) 695--700} {\ttfamily
  [\href{https://arxiv.org/abs/1512.00456}{arXiv:1512.00456}]}.

\bibitem{Panotopoulos:2020kuo}
G.~Panotopoulos and I.~Lopes, ``{Constraints on light dark matter particles
  using white dwarf stars},''
  \href{https://dx.doi.org/10.1142/S0218271820500583}{Int.\  J.\  Mod.\  Phys.\
   D {\bfseries 29} (2020) 2050058} {\ttfamily
  [\href{https://arxiv.org/abs/2005.11563}{arXiv:2005.11563}]}.

\bibitem{Ema:2018bih}
Y.~Ema, F.~Sala, and R.~Sato, ``{Light Dark Matter at Neutrino Experiments},''
  \href{https://dx.doi.org/10.1103/PhysRevLett.122.181802}{Phys.\  Rev.\
  Lett.\  {\bfseries 122} (2019) 181802} {\ttfamily
  [\href{https://arxiv.org/abs/1811.00520}{arXiv:1811.00520}]}.

\bibitem{Dent:2020syp}
J.~B.~Dent, B.~Dutta, J.~L.~Newstead, I.~M.~Shoemaker, and N.~T.~Arellano,
  ``{Present and future status of light dark matter models from cosmic-ray
  electron upscattering},''
  \href{https://dx.doi.org/10.1103/PhysRevD.103.095015}{Phys.\  Rev.\  D
  {\bfseries 103} (2021) 095015} {\ttfamily
  [\href{https://arxiv.org/abs/2010.09749}{arXiv:2010.09749}]}.

\bibitem{Cao:2020bwd}
Q.-H.~Cao, R.~Ding, and Q.-F.~Xiang, ``{Searching for sub-MeV boosted dark
  matter from xenon electron direct detection},''
  \href{https://dx.doi.org/10.1088/1674-1137/abe195}{Chin.\  Phys.\  C
  {\bfseries 45} (2021) 045002} {\ttfamily
  [\href{https://arxiv.org/abs/2006.12767}{arXiv:2006.12767}]}.

\bibitem{An:2017ojc}
H.~An, M.~Pospelov, J.~Pradler, and A.~Ritz, ``{Directly Detecting MeV-scale
  Dark Matter via Solar Reflection},''
  \href{https://dx.doi.org/10.1103/PhysRevLett.120.141801}{Phys.\  Rev.\
  Lett.\  {\bfseries 120} (2018) 141801} {\ttfamily
  [\href{https://arxiv.org/abs/1708.03642}{arXiv:1708.03642}]}. [Erratum:
  Phys.Rev.Lett. 121, 259903 (2018)].

\bibitem{Emken:2021lgc}
T.~Emken, ``{Solar reflection of light dark matter with heavy mediators}.''
  {\ttfamily \href{https://arxiv.org/abs/2102.12483}{arXiv:2102.12483}}.

\bibitem{Kou:2018nap}
E.~Kou and P.~Urquijo, eds., ``{The Belle II Physics Book},''
  \href{https://dx.doi.org/10.1093/ptep/ptz106}{PTEP {\bfseries 2019} (2019)
  123C01} {\ttfamily
  [\href{https://arxiv.org/abs/1808.10567}{arXiv:1808.10567}]}. [Erratum: PTEP
  2020, 029201 (2020)].

\bibitem{BaBar:2001yhh}
{\bfseries BaBar} Collaboration, ``{The BaBar detector},''
  \href{https://dx.doi.org/10.1016/S0168-9002(01)02012-5}{Nucl.\  Instrum.\
  Meth.\  A {\bfseries 479} (2002) 1--116} {\ttfamily
  [\href{https://arxiv.org/abs/hep-ex/0105044}{hep-ex/0105044}]}.

\bibitem{Liang:2019zkb}
J.~Liang, Z.~Liu, Y.~Ma, and Y.~Zhang, ``{Millicharged particles at electron
  colliders},'' \href{https://dx.doi.org/10.1103/PhysRevD.102.015002}{Phys.\
  Rev.\  D {\bfseries 102} (2020) 015002} {\ttfamily
  [\href{https://arxiv.org/abs/1909.06847}{arXiv:1909.06847}]}.

\bibitem{Duerr:2019dmv}
M.~Duerr, {\em et al.}, ``{Invisible and displaced dark matter signatures at
  Belle II},'' \href{https://dx.doi.org/10.1007/JHEP02(2020)039}{JHEP
  {\bfseries 02} (2020) 039} {\ttfamily
  [\href{https://arxiv.org/abs/1911.03176}{arXiv:1911.03176}]}.

\bibitem{Duerr:2020muu}
M.~Duerr, T.~Ferber, C.~Garcia-Cely, C.~Hearty, and K.~Schmidt-Hoberg,
  ``{Long-lived Dark Higgs and Inelastic Dark Matter at Belle II},''
  \href{https://dx.doi.org/10.1007/JHEP04(2021)146}{JHEP {\bfseries 04} (2021)
  146} {\ttfamily [\href{https://arxiv.org/abs/2012.08595}{arXiv:2012.08595}]}.

\bibitem{Kang:2021oes}
D.~W.~Kang, P.~Ko, and C.-T.~Lu, ``{Exploring properties of long-lived
  particles in inelastic dark matter models at Belle II},''
  \href{https://dx.doi.org/10.1007/JHEP04(2021)269}{JHEP {\bfseries 04} (2021)
  269} {\ttfamily [\href{https://arxiv.org/abs/2101.02503}{arXiv:2101.02503}]}.

\bibitem{Essig:2013vha}
R.~Essig, J.~Mardon, M.~Papucci, T.~Volansky, and Y.-M.~Zhong, ``{Constraining
  Light Dark Matter with Low-Energy $e^+e^-$ Colliders},''
  \href{https://dx.doi.org/10.1007/JHEP11(2013)167}{JHEP {\bfseries 11} (2013)
  167} {\ttfamily [\href{https://arxiv.org/abs/1309.5084}{arXiv:1309.5084}]}.

\bibitem{Izaguirre:2015zva}
E.~Izaguirre, G.~Krnjaic, and B.~Shuve, ``{Discovering Inelastic Thermal-Relic
  Dark Matter at Colliders},''
  \href{https://dx.doi.org/10.1103/PhysRevD.93.063523}{Phys.\  Rev.\  D
  {\bfseries 93} (2016) 063523} {\ttfamily
  [\href{https://arxiv.org/abs/1508.03050}{arXiv:1508.03050}]}.

\bibitem{Filimonova:2019tuy}
A.~Filimonova, R.~Sch\"afer, and S.~Westhoff, ``{Probing dark sectors with
  long-lived particles at BELLE II},''
  \href{https://dx.doi.org/10.1103/PhysRevD.101.095006}{Phys.\  Rev.\  D
  {\bfseries 101} (2020) 095006} {\ttfamily
  [\href{https://arxiv.org/abs/1911.03490}{arXiv:1911.03490}]}.

\bibitem{Izaguirre:2015yja}
E.~Izaguirre, G.~Krnjaic, P.~Schuster, and N.~Toro, ``{Analyzing the Discovery
  Potential for Light Dark Matter},''
  \href{https://dx.doi.org/10.1103/PhysRevLett.115.251301}{Phys.\  Rev.\
  Lett.\  {\bfseries 115} (2015) 251301} {\ttfamily
  [\href{https://arxiv.org/abs/1505.00011}{arXiv:1505.00011}]}.

\bibitem{Boehm:2020wbt}
C.~Boehm, X.~Chu, J.-L.~Kuo, and J.~Pradler, ``{Scalar dark matter candidates
  revisited},'' \href{https://dx.doi.org/10.1103/PhysRevD.103.075005}{Phys.\
  Rev.\  D {\bfseries 103} (2021) 075005} {\ttfamily
  [\href{https://arxiv.org/abs/2010.02954}{arXiv:2010.02954}]}.

\bibitem{Ellis:2001hv}
J.~R.~Ellis, J.~L.~Feng, A.~Ferstl, K.~T.~Matchev, and K.~A.~Olive,
  ``{Prospects for detecting supersymmetric dark matter at post LEP benchmark
  points},'' \href{https://dx.doi.org/10.1007/s100520200897}{Eur.\  Phys.\  J.\
   C {\bfseries 24} (2002) 311--322} {\ttfamily
  [\href{https://arxiv.org/abs/astro-ph/0110225}{astro-ph/0110225}]}.

\bibitem{Freitas:2014jla}
A.~Freitas and S.~Westhoff, ``{Leptophilic Dark Matter in Lepton Interactions
  at LEP and ILC},'' \href{https://dx.doi.org/10.1007/JHEP10(2014)116}{JHEP
  {\bfseries 10} (2014) 116} {\ttfamily
  [\href{https://arxiv.org/abs/1408.1959}{arXiv:1408.1959}]}.

\bibitem{Primulando:2020rdk}
R.~Primulando, J.~Julio, and P.~Uttayarat, ``{Collider Constraints on a Dark
  Matter Interpretation of the XENON1T Excess},''
  \href{https://dx.doi.org/10.1140/epjc/s10052-020-08652-x}{Eur.\  Phys.\  J.\
  C {\bfseries 80} (2020) 1084} {\ttfamily
  [\href{https://arxiv.org/abs/2006.13161}{arXiv:2006.13161}]}.

\bibitem{Richard:2014vfa}
F.~Richard, G.~Arcadi, and Y.~Mambrini, ``{Searching for dark matter at
  colliders},'' \href{https://dx.doi.org/10.1140/epjc/s10052-015-3379-8}{Eur.\
  Phys.\  J.\  C {\bfseries 75} (2015) 171} {\ttfamily
  [\href{https://arxiv.org/abs/1411.0088}{arXiv:1411.0088}]}.

\bibitem{Fox:2011fx}
P.~J.~Fox, R.~Harnik, J.~Kopp, and Y.~Tsai, ``{LEP Shines Light on Dark
  Matter},'' \href{https://dx.doi.org/10.1103/PhysRevD.84.014028}{Phys.\  Rev.\
   D {\bfseries 84} (2011) 014028} {\ttfamily
  [\href{https://arxiv.org/abs/1103.0240}{arXiv:1103.0240}]}.

\bibitem{Essig:2009nc}
R.~Essig, P.~Schuster, and N.~Toro, ``{Probing Dark Forces and Light Hidden
  Sectors at Low-Energy e+e- Colliders},''
  \href{https://dx.doi.org/10.1103/PhysRevD.80.015003}{Phys.\  Rev.\  D
  {\bfseries 80} (2009) 015003} {\ttfamily
  [\href{https://arxiv.org/abs/0903.3941}{arXiv:0903.3941}]}.

\bibitem{Anastasi:2015qla}
A.~Anastasi {\em et~al.}, ``{Limit on the production of a low-mass vector boson
  in $\mathrm{e}^{+}\mathrm{e}^{-} \to \mathrm{U}\gamma$, $\mathrm{U} \to
  \mathrm{e}^{+}\mathrm{e}^{-}$ with the KLOE experiment},''
  \href{https://dx.doi.org/10.1016/j.physletb.2015.10.003}{Phys.\  Lett.\  B
  {\bfseries 750} (2015) 633--637} {\ttfamily
  [\href{https://arxiv.org/abs/1509.00740}{arXiv:1509.00740}]}.

\bibitem{BaBar:2017tiz}
{\bfseries BaBar} Collaboration, ``{Search for Invisible Decays of a Dark
  Photon Produced in ${e}^{+}{e}^{-}$ Collisions at BaBar},''
  \href{https://dx.doi.org/10.1103/PhysRevLett.119.131804}{Phys.\  Rev.\
  Lett.\  {\bfseries 119} (2017) 131804} {\ttfamily
  [\href{https://arxiv.org/abs/1702.03327}{arXiv:1702.03327}]}.

\bibitem{BaBar:2008aby}
{\bfseries BaBar} Collaboration in {\em {34th International Conference on High
  Energy Physics}}.
\newblock 2008.
\newblock {\ttfamily \href{https://arxiv.org/abs/0808.0017}{arXiv:0808.0017}}.

\bibitem{Liu:2018jdi}
Z.~Liu and Y.~Zhang, ``{Probing millicharge at BESIII via monophoton
  searches},'' \href{https://dx.doi.org/10.1103/PhysRevD.99.015004}{Phys.\
  Rev.\  D {\bfseries 99} (2019) 015004} {\ttfamily
  [\href{https://arxiv.org/abs/1808.00983}{arXiv:1808.00983}]}.

\bibitem{Habermehl:2020njb}
M.~Habermehl, M.~Berggren, and J.~List, ``{WIMP Dark Matter at the
  International Linear Collider},''
  \href{https://dx.doi.org/10.1103/PhysRevD.101.075053}{Phys.\  Rev.\  D
  {\bfseries 101} (2020) 075053} {\ttfamily
  [\href{https://arxiv.org/abs/2001.03011}{arXiv:2001.03011}]}.

\bibitem{Chae:2012bq}
Y.~J.~Chae and M.~Perelstein, ``{Dark Matter Search at a Linear Collider:
  Effective Operator Approach},''
  \href{https://dx.doi.org/10.1007/JHEP05(2013)138}{JHEP {\bfseries 05} (2013)
  138} {\ttfamily [\href{https://arxiv.org/abs/1211.4008}{arXiv:1211.4008}]}.

\bibitem{Dev:2021jrg}
P.~S.~B.~Dev, ``{Leptophilic Dark Matter at Linear Colliders}.'' {\ttfamily
  \href{https://arxiv.org/abs/2111.03024}{arXiv:2111.03024}}.

\bibitem{Kalinowski:2021tyr}
J.~Kalinowski, W.~Kotlarski, K.~Mekala, P.~Sopicki, and A.~F.~Zarnecki,
  ``{Sensitivity of future linear $\hbox {e}^+\hbox {e}^-$ colliders to
  processes of dark matter production with light mediator exchange},''
  \href{https://dx.doi.org/10.1140/epjc/s10052-021-09758-6}{Eur.\  Phys.\  J.\
  C {\bfseries 81} (2021) 955} {\ttfamily
  [\href{https://arxiv.org/abs/2107.11194}{arXiv:2107.11194}]}.

\bibitem{Barman:2021hhg}
B.~Barman, S.~Bhattacharya, S.~Girmohanta, and S.~Jahedi, ``{Catch 'em all:
  Effective Leptophilic WIMPs at the $e^+\,e^-$ Collider}.'' {\ttfamily
  \href{https://arxiv.org/abs/2109.10936}{arXiv:2109.10936}}.

\bibitem{Profumo:2009tb}
S.~Profumo, K.~Sigurdson, and L.~Ubaldi, ``{Can we discover multi-component
  WIMP dark matter?}''
  \href{https://dx.doi.org/10.1088/1475-7516/2009/12/016}{JCAP {\bfseries 12}
  (2009) 016} {\ttfamily
  [\href{https://arxiv.org/abs/0907.4374}{arXiv:0907.4374}]}.

\bibitem{Liu:2019ogn}
Z.~Liu, Y.-H.~Xu, and Y.~Zhang, ``{Probing dark matter particles at CEPC},''
  \href{https://dx.doi.org/10.1007/JHEP06(2019)009}{JHEP {\bfseries 06} (2019)
  009} {\ttfamily [\href{https://arxiv.org/abs/1903.12114}{arXiv:1903.12114}]}.

\bibitem{Xiang:2017yfs}
Q.-F.~Xiang, X.-J.~Bi, P.-F.~Yin, and Z.-H.~Yu, ``{Exploring Fermionic Dark
  Matter via Higgs Boson Precision Measurements at the Circular Electron
  Positron Collider},''
  \href{https://dx.doi.org/10.1103/PhysRevD.97.055004}{Phys.\  Rev.\  D
  {\bfseries 97} (2018) 055004} {\ttfamily
  [\href{https://arxiv.org/abs/1707.03094}{arXiv:1707.03094}]}.

\bibitem{Birkedal:2004xn}
A.~Birkedal, K.~Matchev, and M.~Perelstein, ``{Dark matter at colliders: A
  Model independent approach},''
  \href{https://dx.doi.org/10.1103/PhysRevD.70.077701}{Phys.\  Rev.\  D
  {\bfseries 70} (2004) 077701} {\ttfamily
  [\href{https://arxiv.org/abs/hep-ph/0403004}{hep-ph/0403004}]}.

\bibitem{Yu:2014ula}
Z.-H.~Yu, X.-J.~Bi, Q.-S.~Yan, and P.-F.~Yin, ``{Dark matter searches in the
  mono-$Z$ channel at high energy $e^+e^-$ colliders},''
  \href{https://dx.doi.org/10.1103/PhysRevD.90.055010}{Phys.\  Rev.\  D
  {\bfseries 90} (2014) 055010} {\ttfamily
  [\href{https://arxiv.org/abs/1404.6990}{arXiv:1404.6990}]}.

\bibitem{Hochberg:2017khi}
Y.~Hochberg, E.~Kuflik, and H.~Murayama, ``{Dark spectroscopy at lepton
  colliders},'' \href{https://dx.doi.org/10.1103/PhysRevD.97.055030}{Phys.\
  Rev.\  D {\bfseries 97} (2018) 055030} {\ttfamily
  [\href{https://arxiv.org/abs/1706.05008}{arXiv:1706.05008}]}.

\bibitem{Liu:2017lpo}
J.~Liu, X.-P.~Wang, and F.~Yu, ``{A Tale of Two Portals: Testing Light, Hidden
  New Physics at Future $e^+ e^-$ Colliders},''
  \href{https://dx.doi.org/10.1007/JHEP06(2017)077}{JHEP {\bfseries 06} (2017)
  077} {\ttfamily [\href{https://arxiv.org/abs/1704.00730}{arXiv:1704.00730}]}.

\bibitem{Alikhanov:2017cpy}
I.~Alikhanov and E.~A.~Paschos, ``{Searching for new light gauge bosons at
  $e^+e^-$ colliders},''
  \href{https://dx.doi.org/10.1103/PhysRevD.97.115004}{Phys.\  Rev.\  D
  {\bfseries 97} (2018) 115004} {\ttfamily
  [\href{https://arxiv.org/abs/1710.10131}{arXiv:1710.10131}]}.

\bibitem{Borodatchenkova:2005ct}
N.~Borodatchenkova, D.~Choudhury, and M.~Drees, ``{Probing MeV dark matter at
  low-energy e+e- colliders},''
  \href{https://dx.doi.org/10.1103/PhysRevLett.96.141802}{Phys.\  Rev.\  Lett.\
   {\bfseries 96} (2006) 141802} {\ttfamily
  [\href{https://arxiv.org/abs/hep-ph/0510147}{hep-ph/0510147}]}.

\bibitem{Graham:2021ggy}
M.~Graham, C.~Hearty, and M.~Williams, ``{Searches for dark photons at
  accelerators}.'' {\ttfamily
  \href{https://arxiv.org/abs/2104.10280}{arXiv:2104.10280}}.

\bibitem{Zhang:2019wnz}
Y.~Zhang, {\em et al.}, ``{Probing invisible decay of dark photon at BESIII and
  future STCF via monophoton searches},''
  \href{https://dx.doi.org/10.1103/PhysRevD.100.115016}{Phys.\  Rev.\  D
  {\bfseries 100} (2019) 115016} {\ttfamily
  [\href{https://arxiv.org/abs/1907.07046}{arXiv:1907.07046}]}.

\bibitem{XENON:2020rca}
{\bfseries XENON} Collaboration, ``{Excess electronic recoil events in
  XENON1T},'' \href{https://dx.doi.org/10.1103/PhysRevD.102.072004}{Phys.\
  Rev.\  D {\bfseries 102} (2020) 072004} {\ttfamily
  [\href{https://arxiv.org/abs/2006.09721}{arXiv:2006.09721}]}.

\bibitem{Banerjee:2019pds}
D.~Banerjee {\em et~al.}, ``{Dark matter search in missing energy events with
  NA64},'' \href{https://dx.doi.org/10.1103/PhysRevLett.123.121801}{Phys.\
  Rev.\  Lett.\  {\bfseries 123} (2019) 121801} {\ttfamily
  [\href{https://arxiv.org/abs/1906.00176}{arXiv:1906.00176}]}.

\bibitem{Gninenko:2017yus}
S.~N.~Gninenko, D.~V.~Kirpichnikov, M.~M.~Kirsanov, and N.~V.~Krasnikov, ``{The
  exact tree-level calculation of the dark photon production in high-energy
  electron scattering at the CERN SPS},''
  \href{https://dx.doi.org/10.1016/j.physletb.2018.05.010}{Phys.\  Lett.\  B
  {\bfseries 782} (2018) 406--411} {\ttfamily
  [\href{https://arxiv.org/abs/1712.05706}{arXiv:1712.05706}]}.

\bibitem{Gninenko:2018ter}
S.~N.~Gninenko, D.~V.~Kirpichnikov, and N.~V.~Krasnikov, ``{Probing
  millicharged particles with NA64 experiment at CERN},''
  \href{https://dx.doi.org/10.1103/PhysRevD.100.035003}{Phys.\  Rev.\  D
  {\bfseries 100} (2019) 035003} {\ttfamily
  [\href{https://arxiv.org/abs/1810.06856}{arXiv:1810.06856}]}.

\bibitem{Chu:2018qrm}
X.~Chu, J.~Pradler, and L.~Semmelrock, ``{Light dark states with
  electromagnetic form factors},''
  \href{https://dx.doi.org/10.1103/PhysRevD.99.015040}{Phys.\  Rev.\  D
  {\bfseries 99} (2019) 015040} {\ttfamily
  [\href{https://arxiv.org/abs/1811.04095}{arXiv:1811.04095}]}.

\bibitem{Berlin:2020uwy}
A.~Berlin, P.~deNiverville, A.~Ritz, P.~Schuster, and N.~Toro, ``{Sub-GeV dark
  matter production at fixed-target experiments},''
  \href{https://dx.doi.org/10.1103/PhysRevD.102.095011}{Phys.\  Rev.\  D
  {\bfseries 102} (2020) 095011} {\ttfamily
  [\href{https://arxiv.org/abs/2003.03379}{arXiv:2003.03379}]}.

\bibitem{Williams:1935dka}
E.~J.~Williams, ``{Correlation of certain collision problems with radiation
  theory},'' Kong.\  Dan.\  Vid.\  Sel.\  Mat.\  Fys.\  Med.\  {\bfseries 13N4}
  (1935) 1--50.

\bibitem{vonWeizsacker:1934nji}
C.~F.~von Weizsacker, ``{Radiation emitted in collisions of very fast
  electrons},'' \href{https://dx.doi.org/10.1007/BF01333110}{Z.\  Phys.\
  {\bfseries 88} (1934) 612--625}.

\bibitem{Ma:1978zm}
E.~Ma and J.~Okada, ``{How Many Neutrinos?}''
  \href{https://dx.doi.org/10.1103/PhysRevLett.41.287}{Phys.\  Rev.\  Lett.\
  {\bfseries 41} (1978) 287}. [Erratum: Phys.Rev.Lett. 41, 1759 (1978)].

\bibitem{Gaemers:1978fe}
K.~J.~F.~Gaemers, R.~Gastmans, and F.~M.~Renard, ``{Neutrino Counting in e+ e-
  Collisions},'' \href{https://dx.doi.org/10.1103/PhysRevD.19.1605}{Phys.\
  Rev.\  D {\bfseries 19} (1979) 1605}.

\bibitem{DELPHI:2003dlq}
{\bfseries DELPHI} Collaboration, ``{Photon events with missing energy in e+ e-
  collisions at s**(1/2) = 130-GeV to 209-GeV},''
  \href{https://dx.doi.org/10.1140/epjc/s2004-02051-8}{Eur.\  Phys.\  J.\  C
  {\bfseries 38} (2005) 395--411} {\ttfamily
  [\href{https://arxiv.org/abs/hep-ex/0406019}{hep-ex/0406019}]}.

\bibitem{SuperCDMS:2018mne}
{\bfseries SuperCDMS} Collaboration, ``{First Dark Matter Constraints from a
  SuperCDMS Single-Charge Sensitive Detector},''
  \href{https://dx.doi.org/10.1103/PhysRevLett.121.051301}{Phys.\  Rev.\
  Lett.\  {\bfseries 121} (2018) 051301} {\ttfamily
  [\href{https://arxiv.org/abs/1804.10697}{arXiv:1804.10697}]}. [Erratum:
  Phys.Rev.Lett. 122, 069901 (2019)].

\bibitem{DAMIC:2019dcn}
{\bfseries DAMIC} Collaboration, ``{Constraints on Light Dark Matter Particles
  Interacting with Electrons from DAMIC at SNOLAB},''
  \href{https://dx.doi.org/10.1103/PhysRevLett.123.181802}{Phys.\  Rev.\
  Lett.\  {\bfseries 123} (2019) 181802} {\ttfamily
  [\href{https://arxiv.org/abs/1907.12628}{arXiv:1907.12628}]}.

\bibitem{SENSEI:2020dpa}
{\bfseries SENSEI} Collaboration, ``{SENSEI: Direct-Detection Results on
  sub-GeV Dark Matter from a New Skipper-CCD},''
  \href{https://dx.doi.org/10.1103/PhysRevLett.125.171802}{Phys.\  Rev.\
  Lett.\  {\bfseries 125} (2020) 171802} {\ttfamily
  [\href{https://arxiv.org/abs/2004.11378}{arXiv:2004.11378}]}.

\bibitem{Essig:2018tss}
R.~Essig, M.~Sholapurkar, and T.-T.~Yu, ``{Solar Neutrinos as a Signal and
  Background in Direct-Detection Experiments Searching for Sub-GeV Dark Matter
  With Electron Recoils},''
  \href{https://dx.doi.org/10.1103/PhysRevD.97.095029}{Phys.\  Rev.\  D
  {\bfseries 97} (2018) 095029} {\ttfamily
  [\href{https://arxiv.org/abs/1801.10159}{arXiv:1801.10159}]}.

\bibitem{Essig:2011nj}
R.~Essig, J.~Mardon, and T.~Volansky, ``{Direct Detection of Sub-GeV Dark
  Matter},'' \href{https://dx.doi.org/10.1103/PhysRevD.85.076007}{Phys.\  Rev.\
   D {\bfseries 85} (2012) 076007} {\ttfamily
  [\href{https://arxiv.org/abs/1108.5383}{arXiv:1108.5383}]}.

\bibitem{Essig:2015cda}
R.~Essig, {\em et al.}, ``{Direct Detection of sub-GeV Dark Matter with
  Semiconductor Targets},''
  \href{https://dx.doi.org/10.1007/JHEP05(2016)046}{JHEP {\bfseries 05} (2016)
  046} {\ttfamily [\href{https://arxiv.org/abs/1509.01598}{arXiv:1509.01598}]}.

\bibitem{Sandra:Robles}
S.~Robles, ``private communication.''.

\bibitem{Buen-Abad:2021mvc}
M.~A.~Buen-Abad, R.~Essig, D.~McKeen, and Y.-M.~Zhong, ``{Cosmological
  Constraints on Dark Matter Interactions with Ordinary Matter}.'' {\ttfamily
  \href{https://arxiv.org/abs/2107.12377}{arXiv:2107.12377}}.

\bibitem{Hufnagel:2018bjp}
M.~Hufnagel, K.~Schmidt-Hoberg, and S.~Wild, ``{BBN constraints on MeV-scale
  dark sectors. Part II. Electromagnetic decays},''
  \href{https://dx.doi.org/10.1088/1475-7516/2018/11/032}{JCAP {\bfseries 11}
  (2018) 032} {\ttfamily
  [\href{https://arxiv.org/abs/1808.09324}{arXiv:1808.09324}]}.

\bibitem{Su:2020zny}
L.~Su, W.~Wang, L.~Wu, J.~M.~Yang, and B.~Zhu, ``{Atmospheric Dark Matter and
  Xenon1T Excess},''
  \href{https://dx.doi.org/10.1103/PhysRevD.102.115028}{Phys.\  Rev.\  D
  {\bfseries 102} (2020) 115028} {\ttfamily
  [\href{https://arxiv.org/abs/2006.11837}{arXiv:2006.11837}]}.

\bibitem{Jho:2020sku}
Y.~Jho, J.-C.~Park, S.~C.~Park, and P.-Y.~Tseng, ``{Leptonic New Force and
  Cosmic-ray Boosted Dark Matter for the XENON1T Excess},''
  \href{https://dx.doi.org/10.1016/j.physletb.2020.135863}{Phys.\  Lett.\  B
  {\bfseries 811} (2020) 135863} {\ttfamily
  [\href{https://arxiv.org/abs/2006.13910}{arXiv:2006.13910}]}.

\bibitem{Chen:2020gcl}
Y.~Chen, {\em et al.}, ``{Sun heated MeV-scale dark matter and the XENON1T
  electron recoil excess},''
  \href{https://dx.doi.org/10.1007/JHEP04(2021)282}{JHEP {\bfseries 04} (2021)
  282} {\ttfamily [\href{https://arxiv.org/abs/2006.12447}{arXiv:2006.12447}]}.

\bibitem{Du:2020ybt}
M.~Du, J.~Liang, Z.~Liu, V.~Q.~Tran, and Y.~Xue, ``{On-shell mediator dark
  matter models and the Xenon1T excess},''
  \href{https://dx.doi.org/10.1088/1674-1137/abc244}{Chin.\  Phys.\  C
  {\bfseries 45} (2021) 013114} {\ttfamily
  [\href{https://arxiv.org/abs/2006.11949}{arXiv:2006.11949}]}.

\bibitem{BaBar:2014zli}
{\bfseries BaBar} Collaboration, ``{Search for a Dark Photon in $e^+e^-$
  Collisions at BaBar},''
  \href{https://dx.doi.org/10.1103/PhysRevLett.113.201801}{Phys.\  Rev.\
  Lett.\  {\bfseries 113} (2014) 201801} {\ttfamily
  [\href{https://arxiv.org/abs/1406.2980}{arXiv:1406.2980}]}.

\bibitem{Alwall:2014hca}
J.~Alwall, {\em et al.}, ``{The automated computation of tree-level and
  next-to-leading order differential cross sections, and their matching to
  parton shower simulations},''
  \href{https://dx.doi.org/10.1007/JHEP07(2014)079}{JHEP {\bfseries 07} (2014)
  079} {\ttfamily [\href{https://arxiv.org/abs/1405.0301}{arXiv:1405.0301}]}.

\bibitem{Liu:2017htz}
Y.-S.~Liu and G.~A.~Miller, ``{Validity of the Weizs\"acker-Williams
  approximation and the analysis of beam dump experiments: Production of an
  axion, a dark photon, or a new axial-vector boson},''
  \href{https://dx.doi.org/10.1103/PhysRevD.96.016004}{Phys.\  Rev.\  D
  {\bfseries 96} (2017) 016004} {\ttfamily
  [\href{https://arxiv.org/abs/1705.01633}{arXiv:1705.01633}]}.

\bibitem{Bjorken:2009mm}
J.~D.~Bjorken, R.~Essig, P.~Schuster, and N.~Toro, ``{New Fixed-Target
  Experiments to Search for Dark Gauge Forces},''
  \href{https://dx.doi.org/10.1103/PhysRevD.80.075018}{Phys.\  Rev.\  D
  {\bfseries 80} (2009) 075018} {\ttfamily
  [\href{https://arxiv.org/abs/0906.0580}{arXiv:0906.0580}]}.

\bibitem{Gondolo:1990dk}
P.~Gondolo and G.~Gelmini, ``{Cosmic abundances of stable particles: Improved
  analysis},'' \href{https://dx.doi.org/10.1016/0550-3213(91)90438-4}{Nucl.\
  Phys.\  B {\bfseries 360} (1991) 145--179}.

\bibitem{Kolb:1990vq}
E.~W.~Kolb and M.~S.~Turner,
  \href{https://dx.doi.org/10.1201/9780429492860}{{\em {The Early Universe}}},
  vol.~69.
\newblock 1990.

\bibitem{Griest:1990kh}
K.~Griest and D.~Seckel, ``{Three exceptions in the calculation of relic
  abundances},'' \href{https://dx.doi.org/10.1103/PhysRevD.43.3191}{Phys.\
  Rev.\  D {\bfseries 43} (1991) 3191--3203}.

\bibitem{Busoni:2014gta}
G.~Busoni, A.~De~Simone, T.~Jacques, E.~Morgante, and A.~Riotto, ``{Making the
  Most of the Relic Density for Dark Matter Searches at the LHC 14 TeV Run},''
  \href{https://dx.doi.org/10.1088/1475-7516/2015/03/022}{JCAP {\bfseries 03}
  (2015) 022} {\ttfamily
  [\href{https://arxiv.org/abs/1410.7409}{arXiv:1410.7409}]}.

\bibitem{Steigman:2012nb}
G.~Steigman, B.~Dasgupta, and J.~F.~Beacom, ``{Precise Relic WIMP Abundance and
  its Impact on Searches for Dark Matter Annihilation},''
  \href{https://dx.doi.org/10.1103/PhysRevD.86.023506}{Phys.\  Rev.\  D
  {\bfseries 86} (2012) 023506} {\ttfamily
  [\href{https://arxiv.org/abs/1204.3622}{arXiv:1204.3622}]}.

\bibitem{Bloch:2020uzh}
I.~M.~Bloch, {\em et al.}, ``{Exploring new physics with O(keV) electron
  recoils in direct detection experiments},''
  \href{https://dx.doi.org/10.1007/JHEP01(2021)178}{JHEP {\bfseries 01} (2021)
  178} {\ttfamily [\href{https://arxiv.org/abs/2006.14521}{arXiv:2006.14521}]}.

\bibitem{clementi1967atomic}
E.~Clementi, D.~Raimondi, and W.~P.~Reinhardt, ``Atomic screening constants
  from SCF functions. II. Atoms with 37 to 86 electrons,'' The Journal of
  chemical physics {\bfseries 47} (1967) 1300--1307.

\bibitem{Pukhov:2004ca}
A.~Pukhov, ``{CalcHEP 2.3: MSSM, structure functions, event generation, batchs,
  and generation of matrix elements for other packages}.'' {\ttfamily
  \href{https://arxiv.org/abs/hep-ph/0412191}{hep-ph/0412191}}.

\bibitem{Lista:2017jsy}
L.~Lista, ``{Statistical Methods for Data Analysis in Particle Physics}.''.

\end{thebibliography}\endgroup
\bibliographystyle{utphys28mod}

\end{document}